\documentclass[11pt]{article}
\pdfoutput=1
\usepackage{amsfonts,amsmath}
\usepackage{amssymb}
\usepackage{fancyhdr}
\usepackage{slashed}
\usepackage{graphicx}
\usepackage{subcaption}
\usepackage{color}
\usepackage{cite}

\usepackage{hyperref}
\usepackage[utf8]{inputenc}
\usepackage[titletoc]{appendix}

\def\hybrid{\topmargin -20pt    \oddsidemargin 0pt
	\headheight 0pt \headsep 0pt
	\textwidth 6.25in       % A4 paper
	\textheight 9.25in       % A4 paper
	\marginparwidth .875in
	\parskip 5pt plus 1pt   \jot = 1.5ex}

\hybrid

\def\baselinestretch{1.2}

\catcode`\@=11

\def\marginnote#1{}
\newcount\hour
\newcount\minute
\newtoks\amorpm
\hour=\time\divide\hour by60
\minute=\time{\multiply\hour by60 \global\advance\minute by-\hour}
\edef\standardtime{{\ifnum\hour<12 \global\amorpm={am}%
		\else\global\amorpm={pm}\advance\hour by-12 \fi
		\ifnum\hour=0 \hour=12 \fi
		\number\hour:\ifnum\minute<10 0\fi\number\minute\the\amorpm}}
\edef\militarytime{\number\hour:\ifnum\minute<10 0\fi\number\minute}
\def\draftlabel#1{{\@bsphack\if@filesw {\let\thepage\relax
			\xdef\@gtempa{\write\@auxout{\string
					\newlabel{#1}{{\@currentlabel}{\thepage}}}}}\@gtempa
		\if@nobreak \ifvmode\nobreak\fi\fi\fi\@esphack}
	\gdef\@eqnlabel{#1}}
\def\@eqnlabel{}
\def\@vacuum{}
\def\draftmarginnote#1{\marginpar{\raggedright\scriptsize\tt#1}}

\def\draft{\oddsidemargin -.5truein
		\def\@oddfoot{\sl preliminary draft \hfil
			\rm\thepage\hfil\sl\today\quad\militarytime}
		\let\@evenfoot\@oddfoot \overfullrule 3pt
		\let\label=\draftlabel
		\let\marginnote=\draftmarginnote
		\def\@eqnnum{(\theequation)\rlap{\kern\marginparsep\tt\@eqnlabel}%
			\global\let\@eqnlabel\@vacuum}  }

	\def\preprint{\twocolumn\sloppy\flushbottom\parindent 2em
		\leftmargini 2em\leftmarginv .5em\leftmarginvi .5em
		\oddsidemargin -.5in    \evensidemargin -.5in
		\columnsep .4in \footheight 0pt
		\textwidth 10.in        \topmargin  -.4in
		\headheight 12pt \topskip .4in
		\textheight 6.9in \footskip 0pt
		\def\@oddhead{\thepage\hfil\addtocounter{page}{1}\thepage}
		\let\@evenhead\@oddhead \def\@oddfoot{} \def\@evenfoot{} }

	\def\numberbysection{\@addtoreset{equation}{section}
		\def\theequation{\thesection.\arabic{equation}}}

	\def\underline#1{\relax\ifmmode\@@underline#1\else
		$\@@underline{\hbox{#1}}$\relax\fi}

	\def\titlepage{\@restonecolfalse\if@twocolumn\@restonecoltrue\onecolumn
		\else \newpage \fi \thispagestyle{empty}\c@page\z@
		\def\thefootnote{\fnsymbol{footnote}} }

	\def\endtitlepage{\if@restonecol\twocolumn \else \newpage \fi
		\def\thefootnote{\arabic{footnote}}
		\setcounter{footnote}{0}}  %\c@footnote\z@ }

\catcode`@=12
\relax

\def\figcap{\section*{Figure Captions\markboth
		{FIGURECAPTIONS}{FIGURECAPTIONS}}\list
	{Figure \arabic{enumi}:\hfill}{\settowidth\labelwidth{Figure
			999:}
		\leftmargin\labelwidth
		\advance\leftmargin\labelsep\usecounter{enumi}}}
 \relax
\def\tablecap{\section*{Table Captions\markboth
		{TABLECAPTIONS}{TABLECAPTIONS}}\list
	{Table \arabic{enumi}:\hfill}{\settowidth\labelwidth{Table
			999:}
		\leftmargin\labelwidth
		\advance\leftmargin\labelsep\usecounter{enumi}}}
 \relax
\def\reflist{\section*{References\markboth
		{REFLIST}{REFLIST}}\list
	{[\arabic{enumi}]\hfill}{\settowidth\labelwidth{[999]}
		\leftmargin\labelwidth
		\advance\leftmargin\labelsep\usecounter{enumi}}}
 \relax
	\makeatletter
	\newcounter{pubctr}
	\def\publist{\@ifnextchar[{\@publist}{\@@publist}}
	\def\@publist[#1]{\list
		{[\arabic{pubctr}]\hfill}{\settowidth\labelwidth{[999]}
			\leftmargin\labelwidth
			\advance\leftmargin\labelsep
			\@nmbrlisttrue\def\@listctr{pubctr}
			\setcounter{pubctr}{#1}\addtocounter{pubctr}{-1}}}
	\def\@@publist{\list
		{[\arabic{pubctr}]\hfill}{\settowidth\labelwidth{[999]}
			\leftmargin\labelwidth
			\advance\leftmargin\labelsep
			\@nmbrlisttrue\def\@listctr{pubctr}}}
	 \relax
	\makeatother
	\newskip\humongous \humongous=0pt plus 1000pt minus 1000pt

	\newif\ifdtup

	\relax

	\def\be{\begin{equation}}
		\def\ee{\end{equation}}
	\def\ba{\begin{eqnarray}}
		\def\ea{\end{eqnarray}}

	  \def\cC{{\cal C}}
	  \def\cF{{\cal F}}

	  \def\cO{{\cal O}}
	  
	 \def\cT{{\cal T}}

	\newcommand{\prt}[1]{{\left( {#1} \right)}}
	\newcommand{\prtt}[1]{{\left[ {#1} \right]}}

	\def\no{\noindent}

	\def\pp{\partial}
	
	\newcommand{\ff}{\frac}

	\def\IR{\relax{\rm I\kern-.18em R}}
	\def\IL{\relax{\rm I\kern-.18em L}}

	\def\inv{^{\raise.15ex\hbox{${\scriptscriptstyle -}$}\kern-.05em 1}}

	\def\bea{\begin{eqnarray}}
		\def\eea{\end{eqnarray}}
	\newcommand{\eq}[1]{(\ref{#1})}
	\def\nn{\nonumber}
	
	\newcommand{\la}[1]{\label{#1}}

	\def\a{\alpha}      
	\def\b{\beta}       
	\def\g{\gamma}    
	\def\d{\delta}

	\def\l{\lambda} \def\L{\Lambda}
	\def\m{\mu} \def\n{\nu}

	\def\s{\sigma}  \def\S{\Sigma}
	
	\def\th{\theta}

	\def\n{\nu}

	\definecolor{markcolor2}{rgb}{1,0,0}
	
	\definecolor{markcolor3}{rgb}{0,1,0}

	\newcommand{\revcolor}{\color{blue}}

\begin{document}

		\begin{titlepage}

			\begin{center}

				%\hfill nsysu-xx-yyyy\\
				%\vskip -.1 cm
				%\hfill hep-th/yymmnnn\\

				~
				\vskip .9 cm

				{\large
					\bf Timelike Entanglement Entropy and Renormalization Group Flow Irreversibility}

				\vskip 0.4in
				{\bf Dimitrios Giataganas $^{1,2}$}
				\vskip 0.2in
				{\em
			${}^1$  Department of Physics, National Sun Yat-Sen University,
					Kaohsiung 80424, Taiwan\\
			%		\vskip .1in
            ${}^{2}$ Physics Division, National Center for Theoretical Sciences, Taipei 10617, Taiwan
			%		${}^2$  Center for Theoretical and Computational Physics, \\
			%		National Sun Yat-Sen University				%
			%		\\
			%		Kaohsiung 80424, Taiwan
					\\ \vskip .15in
					{\tt  dimitrios.giataganas@mail.nsysu.edu.tw }
				}

				\vskip .1in
			\end{center}

			\vskip 1.5in

\centerline{\bf Abstract}

We study holographic $c$-theorems based on timelike entanglement entropy and show that timelike $c$-functions capture irreversible renormalization group (RG) flow. We demonstrate that timelike $c$-functions are applicable to both relativistic and nonrelativistic quantum matter in nematic phases with broken rotational symmetry, and that they remain monotonic along holographic anisotropic RG flows under explicit geometric conditions, thereby passing some of the most stringent consistency tests. We derive the exact flow of the composite timelike $c$-function in general homogeneous anisotropic geometries and establish the $c$-theorem for adiabatic flows. Across all classes of scaling backgrounds examined, consistent with the null energy condition and thermodynamic stability, the monotonicity condition universally reduces to a constraint on a running geometric effective dimension. We further construct a second $c$-function purely from the imaginary part of the timelike entanglement entropy and prove its monotonicity under the local effective-dimension condition, without requiring
adiabaticity. For Poincar\'e-invariant flows, this condition follows
directly from the null energy condition. Moreover, we identify a geometric upper bound on the rate of change of the timelike $c$-functions, which constrains how rapidly effective degrees of freedom can be coarse-grained. Our results suggest that capturing RG irreversibility in nonrelativistic flows requires observables that are intrinsically sensitive to the temporal direction. The
applicability of holographic $c$-theorems is thus extended to highly nontrivial RG flows and points toward a new information-theoretic diagnostic of holographic RG dynamics.

		\no
		\end{titlepage}
		\vfill
		\eject

		%\end{center}

		\noindent

		\def\baselinestretch{1.2}
		\baselineskip 19 pt
		\noindent

		%%%%%%%%%%%%%%%

		\setcounter{equation}{0}

\section{Introduction}

The renormalization group (RG) flow is a fundamental framework that establishes relations between theories at different length scales. One powerful way to quantify this relation is by interpreting the central charge of the theory as a measure of the number of degrees of freedom along the RG flow. Since the RG flow proceeds by integrating out high-energy modes \cite{Wilson:1974mb}, this measure typically decreases along the flow from the ultraviolet (UV) to the infrared (IR), thereby  encoding the irreversibility of RG dynamics. This behavior is formalized in the c-theorem, which provides exact  and non-perturbative statements about RG flows, offering valuable insights for strongly coupled systems. The seminal c-theorem \cite{Zamolodchikov:1986gt} states that in two-dimensional quantum field theories (QFTs), there exists a positive real function that decreases monotonically along the RG flow. At RG fixed points, this function takes a value equal to the central charge of the corresponding conformal field theory at each fixed point. In \cite{Cardy:1988cwa}, an analogue of the c-theorem for four-dimensional QFTs was conjectured, suggesting that similar structures should exist in higher even dimensions. In general, the trace of the stress tensor in a conformal theory on curved spacetime receives anomalous contributions. In four dimensions, these contributions involve two trace anomaly coefficients: $c$, associated with the square of the Weyl tensor, and $a$, associated with the topological Euler density. In two dimensions the trace anomaly is governed by the Virasoro central charge; in four dimensions the coefficients
$a$ and $c$ are independent. The $c$-function is identified with the anomaly coefficient associated with the A-type Euler density anomaly.   This conjecture was proven for Lorentz-invariant theories in four dimensions \cite{Komargodski:2011vj}, although a complete proof in higher dimensions using purely field-theoretic methods remains open.

On the other hand, in the strong coupling regime, c-theorems have been reformulated using the gauge/gravity correspondence \cite{Maldacena:1997re,Witten:1998qj}, where monotonic $c$-functions have been proposed and studied in arbitrary dimensions \cite{Girardello:1998pd,Freedman:1999gp,Ryu:2006ef,Myers:2010tj,Myers:2010xs,Casini:2011kv,Liu:2012eea,Myers:2012ed,Chu:2019uoh}. The term $c$-function here is used in a wider sense to describe a fundamental monotonic function along the RG flow. A key tool in these formulations is the entanglement entropy. This is not surprising, as its relation to the central charge at fixed points has been known since early developments. In two-dimensional QFTs, the c-theorem can be extrapolated directly from the entanglement entropy at a fixed point, where the $c$-function takes the form $c=3 L\pp S/\pp L$, with $S(L)$ the entanglement entropy of a spatial orthogonal interval of length $L$. In the holographic framework, entanglement entropy formulations \cite{Ryu:2006bv} generalize naturally to higher dimensions. Moreover, Lorentz symmetry and unitarity in the dual QFT, together with the subadditivity of entanglement entropy, naturally extend to higher dimensions; these properties are crucial, as they underlie the proof of the c-theorem in two dimensions. It has been found that holographic $c$-functions based on entanglement entropy satisfy the c-theorem for Lorentz-invariant theories as long as appropriate energy conditions are imposed in the bulk \cite{Ryu:2006ef,Myers:2010tj,Myers:2012ed}, and there is a rich body of work related to this subject \cite{Myers:2010xs,Casini:2011kv,Liu:2012eea,Hung:2011ta,Chu:2019uoh,Cremonini:2013ipa,Park:2018ebm,Kolekar:2018chf,Hoyos:2021vhl,Jokela:2025qac,Conti:2025wwf}.

Of particular interest are studies of holographic c-theorems in nonrelativistic QFTs and the effort to develop appropriate covariant formalisms for the $c$-function in these theories \cite{Chu:2019uoh,Cremonini:2013ipa}. The RG flow is a fundamental tool for relating theories at different scales and explaining universality in critical phenomena, and one expects this structure to persist even when Lorentz or rotational invariance is broken. It is therefore natural to anticipate the existence of a positive real well-defined $c$-function that decreases monotonically from the UV to the IR in such theories as well.

Nonrelativistic fixed points with broken rotational symmetry are common in quantum matter and many-body systems,  making the formulation of a generalized  holographic c-theorem for such theories a highly relevant endeavor. These phases break spatial rotational symmetry while preserving translations, a pattern of symmetry breaking naturally captured by anisotropic holographic duals. For instance,  electronic nematic phases  arise in strongly correlated systems as states that spontaneously break rotational symmetry while preserving translational invariance \cite{1998Natur393550K,2010ARCMP1153F,PhysRevB.89.155130}.  These anisotropic phases are now widely recognized across a broad range of materials. In high-temperature cuprate superconductors, signatures of electronic nematicity manifest in anisotropic transport and spectroscopic features \cite{PhysRevLett.88.137005,doi:10.1126/science.1138584,doi:10.1126/science.1152309} revealing a tendency toward unidirectional electronic order. In iron-based superconductors, nematic order typically appears above or near the superconducting transition, with strong evidence from scanning tunneling microscopy for nematic electronic structure in the parent compound \cite{doi:10.1126/science.1181083}. Nematic phases also arise in two-dimensional electron gases under high magnetic fields, where quantum Hall nematicity manifests through anisotropic longitudinal transport \cite{PhysRevB.59.8065,Xia_2011}. The ubiquity of nematicity across different material platforms highlights its fundamental role in correlated electron systems and suggests a deep connection between nematic fluctuations and unconventional superconductivity.  Beyond these nematic phases, anisotropic RG flows also describe a variety of other systems, such as strongly coupled plasmas with pressure anisotropy, and various topological quantum phase transitions \cite{yang2014quantum,isobe2016emergent}, for example, between a topologically trivial insulator and a gapless Weyl semimetal, where the critical point can exhibit Lifshitz-like anisotropy. Consequently, understanding the structure of anisotropic RG flows and identifying quantities that directly measure the number and the decimation of degrees of freedom along the flow is of fundamental theoretical importance.

Progress in this direction was made in \cite{Chu:2019uoh}, where a generalized $c$-function based on the entanglement entropy was proposed for anisotropic QFTs, reducing smoothly to earlier known expressions in its isotropic limit. These anisotropic $c$-functions can indeed support a c-theorem in certain classes of theories. Nevertheless, a universally applicable holographic $c$-function for arbitrary anisotropic  systems remains elusive. This is due to a deeper issue: in theories with broken Lorentz invariance, the RG flow can violate the monotonicity of entanglement entropy \cite{Swingle:2013zla}. This is an expected consequence of the definition of holographic entanglement entropy on a fixed time slice. Once Lorentz symmetry is broken, time and space are no longer trivially related, and  entanglement entropy evaluated on a fixed time slice may fail to capture the full RG monotonicity. Hence, the difficulty in defining a proper universal $c$-function based on  entanglement entropy is not surprising. Indeed, this breakdown of a universal c-theorem has also been observed in Lorentz-violating holographic QFTs \cite{Chu:2019uoh,Cremonini:2013ipa}.

These observations motivate the study of timelike entanglement entropy as a probe of RG irreversibility in nonrelativistic and anisotropic theories. A holographic timelike $c$-function was proposed in \cite{Giataganas:2025div}, with applications to Poincar\'e-invariant, Lifshitz and hyperscaling-violating backgrounds. The construction uses extremal surfaces with spacelike and timelike branches,
following the prescription of \cite{Doi:2023zaf,Afrasiar:2024lsi}. Here we derive its exact flow and establish its monotonicity for adiabatic flows, defined by an exact rate condition on the geometry. This identifies the class in which the earlier monotonicity conclusion holds and provides the starting point for studying more general flows.

These results contribute to a rapidly developing body of work demonstrating that timelike entanglement entropy and pseudoentropies, whether defined in the present holographic context or as temporal entanglement measures, encode rich information about quantum field theories and their phase structure. Such quantities have recently become the focus of intense investigation; see, for example,  \cite{Doi:2023zaf, Doi:2022iyj, Narayan:2022afv, Basak:2023otu, Afrasiar:2024lsi, Afrasiar:2024ldn, Chu:2023zah,  Narayan:2023zen, Grieninger:2023knz, Caputa:2024gve, Guo:2024edr, Jena:2024tly, Xu:2024yvf, Chu:2025sjv, Nunez:2025gxq, Nunez:2025ppd, Gong:2025pnu, Nunez:2025puk, Nanda:2025tid, Harper:2025lav,Das:2025fcd}. The developments presented in this work further reinforce the usefulness of these timelike measures as probes of RG structure and quantum dynamics.

In this manuscript, we study in further detail the properties of the holographic timelike $c$-function and the associated c-theorem. In particular, we analyze the form of the $c$-function in homogeneous theories with potentially broken Lorentz and rotational invariance, extending and subjecting the proposal of \cite{Giataganas:2025div} to an even more stringent set of tests. We propose a generalized form of the timelike $c$-function for theories with anisotropic symmetries and identify the precise conditions under which the corresponding c-theorem holds. We derive the exact flow of this $c$-function with respect to the holographic radial direction, and for the equivalent $c_T$-function with respect to the interval $T$, in general homogeneous anisotropic geometries. Its exact flow weights the running effective dimension with a signed kernel, so the local effective-dimension condition alone does not establish monotonicity for general flows; it does within the adiabatic class, where the signed integral is shown to be non-negative.  For adiabatic flows, defined by an exact rate condition on the geometry, the local effective-dimension condition provides a timelike c-theorem for $c_T$ under natural holographic assumptions of the RG flow, which specify the radial orientation and the required convergence and endpoint limits. Outside the adiabatic class, that local condition alone is insufficient. We then construct a second timelike $c$-function $\cC$ from the timelike branch of the extremal surface, built from the imaginary part of the timelike entanglement entropy. Its flow is governed by a positive kernel; the local effective-dimension condition suffices for the monotonicity of the function.

When rotational symmetry is broken, multiple timelike $c$-functions naturally arise, depending on the spatial direction in which the entangling region is localized. A natural question is whether a single covariant quantity can be constructed by combining these into a unique $c$-function. In this work, we adopt the strictest criterion: each individual $c$-function must independently satisfy a c-theorem along the RG flow. Any composite function that preserves the monotonicity of its constituents then automatically obeys a c-theorem. This approach allows us to formulate a general and robust c-theorem in broad classes of holographic backgrounds, relying primarily on geometric properties of the metric.

In particular, in Poincar\'e-invariant holographic flows the entanglement-entropy-based $c$-theorem is guaranteed as long as  the null energy condition (NEC) is satisfied in the bulk \cite{Myers:2012ed,Chu:2019uoh}. For Poincar\'e-invariant flows the local effective-dimension condition is precisely the NEC. The timelike $c$-function $\cC$ is therefore monotonic under the same condition as the entanglement-entropy-based $c$-functions, with ratio of fixed-point values equal to the ratio of the holographic central charges, which serves as a consistency check in the relativistic setting. In two dimensions, $\cC$ is half the magnitude of the imaginary part of the timelike entanglement entropy. For $c$ and $c_T$, adiabaticity provides a sufficient condition for monotonicity, whereas the NEC alone does not guarantee it: along a non-adiabatic flow, such as the one of \cite{Freedman:1999gp} from $\mathcal N=4$ super Yang--Mills theory to the Leigh--Strassler fixed point, which satisfies the NEC, $c_T$ can be not monotonic, since the adiabatic analysis does not apply.

In anisotropic or Lorentz-violating holographic QFTs, however, a fully universal entanglement entropy-based $c$-function has remained elusive \cite{Chu:2019uoh}. Its monotonicity requires additional geometric constraints beyond the NEC, and can fail even in physically acceptable backgrounds, or in a physically acceptable regime of parameters. In contrast, we show that the timelike $c$-function of \cite{Giataganas:2025div} remains monotonic along adiabatic flows in theories with broken Lorentz, scale, or rotational symmetry, under the hypotheses of Section~\ref{sec:adiab}. In the constant-exponent Lifshitz-like and hyperscaling-violating examples, the NEC and positive specific heat establish the orientation of the flow relative to the UV boundary and the convergence properties of the construction. For $\cC$, the directional local conditions fix the sign of the exact flow under its regularity assumptions.

We also discuss the upper bound on the rate of change of the $c$-function. Its upper part, $d\ln c_T/d\L_m<1$, is equivalent to the monotonicity of the boundary time interval as a function of the energy scale, i.e.\ to the assumption that larger intervals probe the IR, rather than an independent constraint.

Our manuscript is organized as follows. In Section~\ref{section2}, we describe the holographic setup for the class of theories we study. We begin by reviewing the Einstein–Maxwell–Dilaton–Axion action, which admits anisotropic solutions belonging to the class of backgrounds considered here. We then present the general formulation of timelike entanglement entropy in homogeneous but anisotropic geometries. In Section~\ref{section3}, we propose a definition of the timelike $c$-function for anisotropic theories, derive its exact flow, and state the timelike c-theorem for adiabatic flows. In Section~\ref{section4}, we relate the flows in the radial and boundary variables and discuss the bound on the rate of change. In Section~\ref{section5}, we compute the NEC and thermodynamic stability conditions for the general class of theories under consideration. In Section~\ref{section:C}, we construct the $c$-function $\cC$ from the timelike branch and prove its monotonicity under the local effective-dimension condition. In Section~\ref{section6}, we apply our formalism to explicit examples: Poincaré-invariant flows, Lifshitz-like anisotropic flows, and hyperscaling-violating Lifshitz-like anisotropic geometries. We conclude with a discussion of our findings and their implications in Section~\ref{section7}. Appendix~\ref{appA} supports the main text with further technicalities, and Appendix~\ref{sec:fgpwsec} discusses the application of the $c$-theorem in the flow of \cite{Freedman:1999gp}.

%%%%%%%%%%%%%%%%%%%%%%%%%%%

\section{Holographic Setup} \label{section2}
\subsection{Einstein-Maxwell-Dilaton-Axion Action and Background Solutions}

In this section, we present the setup of anisotropic geometries with broken Lorentz invariance, broken rotational symmetry, and broken scale invariance. Subclasses of these holographic theories, in particular the Lifshitz-like anisotropic ones,  arise naturally in supergravity by introducing charge distributions along specific spatial directions. A notable example involves linear axion scalars with profiles aligned along a chosen spatial direction, as studied in \cite{Azeyanagi:2009pr, Mateos:2011ix, Mateos:2011tv,Giataganas:2012zy}. The resulting anisotropic solutions exhibit pressure anisotropies along different directions, determined by the constant couplings in the action, leaving in this case no free parameters to tune the anisotropy. This rigidity is a consequence of the Lifshitz-like scaling being encoded directly in the (fixed) coupling constants of the supergravity action. Moreover, it is known that the breaking of rotational symmetry propagates to anisotropic expectation values of observables, as shown in \cite{Giataganas:2012zy}.

More general classes of anisotropic theories, allowing for tunable Lifshitz exponents and/or hyperscaling violation parameters, have been constructed in a variety of models such as those studied in \cite{Giataganas:2017koz,Charmousis:2010zz,Amoretti:2017axe,Baggioli:2024vza,Cremonini:2014pca}, providing a richer structure and enabling more detailed investigations into anisotropic holography.  A  general and flexible framework encompassing these cases is the Einstein-Maxwell-Dilaton-Axion (EMDA) theory, whose action can be written as \cite{Giataganas:2025ing}:
\be\label{action11}
S=\frac{1}{2\kappa^2}\int d^{d+1}x\,\sqrt{-g}\left[R-\frac{1}{2}(\partial \phi )^2+V(\phi )-\frac{1}{2}Z(\phi )(\partial \chi )^2-\frac{1}{4}Y(\phi )F^2\right]\,,
\ee
where the functions $V(\phi)$, $Z(\phi)$ and $Y(\phi)$, can for example be chosen to be exponential of the scalar field $\phi$, given by:
\be\label{potentials}
V(\phi )=V_0 e^{\s \phi}\,,\qquad  Z(\phi )=Z_0 e^{\g\phi}\,,\qquad Y(\phi )=Y_0 e^{\l \phi}\,,
\ee
where $\s,\g,\l$ are coupling constants that characterize the degree of anisotropy and hyperscaling violation in the theory. Depending on the specific models under consideration, some of these constants may be set to zero. Similarly, the coefficients $V_0, Y_0, Z_0$ are constants, which may be set to unity or other appropriate value, or to zero depending on the class of anisotropic solutions we like to study and the properties of the strongly coupled anisotropic theories of interest.  When the axion or Maxwell sector is present, the corresponding kinetic couplings must be positive, $Z(\phi)>0$ and $Y(\phi)>0$; setting $Z_0$ or $Y_0$ to zero removes that sector. Nontrivial RG flows, potentially connecting fixed points with different symmetry properties, can be obtained with more general choices of potentials. In many such models the exponential asymptotics \eqref{potentials} emerge only in the IR.

A representative class of anisotropic scaling geometries arising in EMDA models \cite{Giataganas:2017koz,Charmousis:2010zz,Giataganas:2025ing}, exhibiting anisotropic Lifshitz-like scaling, with two spatial planes  $\prt{\vec{x},\vec{y}}$  is
\bea\la{metr1}
ds^2_{d+1}= r^{-\frac{2\theta}{d-1}}\prt{-r^{2 z_0}dt^2+r^{2z_1} d\vec{x}^2 + r^{2 z_2} d\vec{y}^2+\frac{dr^2}{r^2}}~,
\eea
for suitable relations between the exponents and the couplings in \eqref{potentials}, by appropriate dilaton, axion, and Maxwell field configurations. For instance, when the anisotropy is supported by linear axions, a rotationally invariant anisotropic plane dimension larger than one requires one axion per direction in that plane; other matter content can also support such planes. More intricate anisotropic theories can be realized within the EMDA framework, which are typically known only numerically.

Motivated by the existence of such anisotropic solutions in holography, and in particular their relevance for strongly coupled systems with Lifshitz-like fixed points, we consider a general homogeneous holographic geometry of the form
\be \la{metric1a}
ds^2_{d+1}=-e^{2B(r)} dt^2+e^{2 A_1(r)}d\vec{x}^2+e^{2 A_2(r)} d\vec{y}^2+dr^2~,
\ee
where $d=d_1+d_2+1$, with $d_1$ and $d_2$ denoting the spatial dimensions of the planes spanned by $\vec{x}$ and $\vec{y}$ and $r$ is a proper radial coordinate. Static homogeneous geometries invariant under $SO(d_1)\times SO(d_2)$ can be brought to the diagonal form \eq{metric1a} by a choice of radial coordinate. This covers the anisotropic backgrounds of interest here. 

%%%%%%%%%%%%%%%%%%%%%%%%%%%
\subsection{Timelike Entanglement Entropy}

We consider a boundary time interval defined by $-\frac{T}{2}\le t\le \frac{T}{2}$,  with the entangling region localized at $x_1=0$, while the rest of the coordinates are $x_{i>1}=\s_i$, and $y_j=\s_j$. The corresponding bulk surface is parametrized by $t(r)$. Note that the localization occurs along one of the directions in the $\vec{x}$-plane, however, our analysis and formulas generalize straightforwardly to localizations along the $\vec{y}$-plane as well.

The timelike entanglement entropy  functional reads
\be\la{s1a}
S=2\int dr e^{d_2 A_2(r)+\prt{d_1-1} A_1(r)}\sqrt{1-e^{2 B(r)}t'^2}=2\int dr e^{K(r) -A_1(r)}\sqrt{1-e^{2 B(r)}t'{}^{2}}~,
\ee
where we defined $K(r):=d_1 A_1(r)+ d_2 A_2(r)$. The equation of motion integrates to
\be \la{t1a}
t_{s}=2 \int dr \frac{e^{\L_m-B}}{\sqrt{s  e^{2\L(r)}+  e^{2 \L_m}}}~,
\ee
where $s^2:=1$. The positive sign $s=+1$ corresponds to the spacelike branch of the surface, while the negative sign $s=-1$ to the timelike branch. The definitions of the function $\L(r):=\L_x(r)$, where, for readability,  we omit the subscript $x$ or $y$ in this and other quantities whenever the direction of localization is clear:
\be \la{lamdadef}
\L(r):=B(r)+d_2 A_2(r)+\prt{d_1-1}A_1(r)=K(r)-A_1(r)+B(r)~,\quad \L_m:=\L(r_m)~,
\ee
where $r_m$ denotes the turning point of the symmetric timelike surface for which $t_{-}'(r)$ diverges, and fixes the integration constant. In \eq{t1a} both spacelike and timelike solutions carry the same integration constant, $P_+=P_-=e^{\L_m}$, following the prescription of \cite{Afrasiar:2024lsi}.  In geometries without IR walls the branches are matched asymptotically in the IR, rather than patched at a finite bulk point. Matching the leading IR slopes fixes the relative orientation
of the branches and the combination $T=t_{Im}-t_{Re}$. Their equal momenta are specified by the adopted prescription, rather than determined by leading slope matching alone.  For AdS$_3$, this gives after restoring the gravitational normalization, $S_T=\frac{c}{3}\log\frac{T}{\epsilon}+\frac{i\pi c}{6}$
\cite{Doi:2023zaf,Afrasiar:2024lsi}.  Substituting equation \eq{t1a} into equation \eq{s1a}, we can eliminate the time-interval derivative of the area of the extremal surface to obtain on shell
\be \la{s2a}
S_{s}=2\int dr\ff{e^{2\L-B}}{\sqrt{e^{2\L}+s e^{2 \L_m}}}~,
\ee
where $S_{s}$ represents the corresponding part of the timelike entropy. $S_+$ is real and in general UV divergent, so it may require UV regularization and it corresponds to the spacelike part, while $S_-$ is imaginary with finite magnitude under the stated IR conditions and corresponds to the timelike part of the entropy. Because the theory is anisotropic, the timelike entanglement entropy depends on the localization direction. Localizing along $\vec{x}$ or $\vec{y}$ yields two different functions $\Lambda_x(r)$ and $\Lambda_y(r)$, and hence two distinct timelike entanglement entropies. The directional dependence is encoded in the function $\L(r)$ as defined by \eq{lamdadef}, since here we have localized along $x$. For localization along a $y$-direction, the corresponding function   can be found straightforwardly
\be\la{ly}
\L_y(r)=B(r)+(d_2-1) A_2(r)+d_1 A_1(r)~.
\ee
In what follows we keep the notation $\Lambda(r)$ in general without yet substituting their expressions,  and we work in full generality for the different directions.

The spacelike part of the bulk surface described by the solution $t_{+}(r)$ from \eq{t1a}, contributes the real part of the timelike entanglement entropy, denoted  $S_{+}$ and given by \eq{s2a}. The boundaries of this surface typically lie in the interval $\prt{r_{b},r_{\pp}}$, where  $r_b$ is the deepest point reached in the bulk, while $r_\pp$ is the UV boundary where the surface originates. On the other hand, the timelike surface, described by the solution $t_{-}(r)$,  contributes the imaginary part of the timelike entanglement entropy $S_{-}$. The symmetric timelike surface extends in the interval $\prt{r_m,r_b}$, where $r_m$ is the turning point of the surface in the bulk and $r_b$ is the deepest point that the surfaces reach. The two branches are regulated at a common $r_b$ and joined asymptotically as $r_b\to r_{IR}${}\footnote{$r_{IR}=-\infty$ for Poincar\'e-type geometries in the coordinates \eq{metric1a}.}.

For convenience and presentation reasons we can define 
\be \la{s3}
S_{Re}:=S_+~,\qquad S_{Im}:=i S_{-}~,
\ee
where the $S_\pm$ are given by \eq{s2a}.  The overall factor $V_{d-2}/4G_N$ has been absorbed, so the surface formulas below are normalized areas.   
In later Sec.~\ref{section3} we introduce the real combination $S:=S_{Re}+S_{Im}$, due to its unique property that its derivative with respect to the boundary interval is fixed by the turning point alone. Similarly, we introduce a unified notation for the time intervals as
\be \la{t2}
t_{Re}:=t_{+}=2 s_1 \int_{r_\pp}^{r_b}dr  \frac{e^{\L_m-B}}{\sqrt{  e^{2\L(r)}+  e^{2 \L_m}}}~,\qquad t_{Im}:=t_{-}=2 s_1\int_{r_m}^{r_b}dr  \frac{e^{\L_m-B}}{\sqrt{-  e^{2\L(r)}+  e^{2 \L_m}}}~,
\ee
where we have used \eq{t1a} and we have explicitly stated the integral boundaries in the expressions. The sign convention $s_1=\pm 1$ simply labels the two symmetric branches $t_s(r)$, the final results are independent of this choice, and we could fix either sign without loss of generality. Let us proceed by choosing $s_1=-1$, the right branch with positive boundary time of the solution $t_s(r)$, just for simplicity in presentation. The boundaries of the $S_s$ integrals \eq{s2a} follow straightforwardly from \eq{t2} and we consider them positive.

For geometries without horizons or IR walls, both portions of the surface reach deep into the bulk to the same joining point. For instance, to describe a representative solution,  the spacelike surface $t_{Re}$ initiates from the boundary at $t=T/2$ for $r=r_\pp\to \infty$ and asymptotes to the deep IR for $r=r_b\to r_{IR}$ to the common joining time $t=t_{Im}/2\to \infty$  \cite{Doi:2023zaf,Afrasiar:2024lsi}. On the other hand, the right symmetric branch of the timelike surface $t_{Im}$ begins from the extremal point of the surface at $t=0$ for $r=r_m$ in the bulk, and in theories without horizons or an IR wall, this branch extends to the deep IR  for $r=r_b\to r_{IR}$ where $t=t_{Im}/2\to \infty$. The joining is understood as this asymptotic IR limit.

Since both segments reach the same asymptotic time, the boundary interval is
\be\la{Tdef}
T=t_{Im}-t_{Re}~,
\ee
which is finite although each term diverges in the IR. Throughout we assume the natural relations
\be\la{IRcond}
\int_{r_{IR}}^{r_0}e^{2\L-B}|dr|<\infty~,\qquad \int_{r_{IR}}^{r_0}e^{-B}|dr|\sim\infty~.
\ee 
Here the integrals follow the radial path from the IR endpoint to an interior point $r_0$, with positive measure $|dr|$. The first condition makes $S_{Im}$ and the IR contributions to $S_{Re}$ and $T$ finite; finiteness of $T$ also requires UV convergence. The second makes the separate branch time extents diverge, so both branches reach infinite boundary time.

Reality of the timelike branch profile requires $\L(r)<\L_m$ between the IR and the turning point.
Requiring such branches with regular, nondegenerate turning points at arbitrary $r_m$ therefore enforces $\L$ to increase towards the boundary, with $\L'\neq0$. This allows $\L$ to serve as a radial variable.  The radial coordinate orientation is a convention: when the boundary lies at $r=0$, all statements are understood with $\L$ increasing towards that endpoint.

For monotonic $\L$, the two conditions in \eqref{IRcond} also imply $\L\to-\infty$ in the IR. A finite IR value
would make the two radial integrands proportional up to bounded positive factors, so their convergence properties could not differ.

%%%%%%%%%%%%%%%%%%%%%%%%%%%

\section{Anisotropic Timelike $c$-functions and Monotonicity Conditions}  \label{section3}

A holographic timelike $c$-function was proposed in \cite{Giataganas:2025div} for Poincar\'e-invariant and certain nonrelativistic theories. Here we discuss further its monotonicity using the exact signed flow and providing more details. In this work, we elaborate further on the detailed properties of the timelike $c$-function  and extend this construction to a significantly broader class of theories, namely, those with broken rotational invariance. This constitutes an important and stringent test: despite the existence of anisotropic fixed points in holography, no universally valid $c$-functions have been formulated for such geometries, and in particular entanglement entropy–based constructions are known to fail to remain monotonic in general.

In anisotropic backgrounds the entangling region may be localized along different spatial directions, leading naturally to multiple timelike $c$-functions depending on the spatial direction in which the entangling region is localized. Throughout this work we adopt the strictest possible criterion: each individual directional $c$-function must satisfy a c-theorem along the entire RG flow.  To determine whether such a $c$-function exists and to establish the associated monotonicity conditions, we analyze the dependence of the timelike entanglement entropy on the turning point $r_m$ of the extremal timelike surface, which probes the corresponding energy scale. This requires computing the derivatives of both the time intervals $t_\pm$ and the on-shell entropies $S_\pm$ with respect to $r_m$. These quantities control the response of the real and imaginary parts of the timelike entropy under changes of the radial scale, and hence fully characterize the RG flow of the proposed timelike $c$-function in the most general anisotropic setting.

%%%%%%%%%%%%%%%%%%%%%%%%%%%

\subsection{Derivatives of the Timelike Entanglement Entropy}

In this subsection, we derive analytically the dependence of the timelike entanglement entropy  on the turning point $r_m$ of the extremal surface. Since all anisotropy information is encoded in the function $\Lambda(r)$, the computation follows the structure of the isotropic derivation in \cite{Giataganas:2025div}, and we present here the computation in greater detail and reveal additional properties of the observable.

We first differentiate at fixed radial IR and UV cutoffs. The following timelike expressions are formal regulated combinations: their separate turning-point terms diverge and must be combined before removing the turning-point regulator.  With this understanding, differentiation gives
\be \la{dt1a}
e^{\L_m}\ff{\pp t_{Im}}{\pp r_m}=2 \ff{e^{ 2\L-B}}{\sqrt{e^{2\L_m}-e^{2\L}}}\bigg|_{  r_m}+2 \L_m' e^{2\L_m}  \int_{r_m}^{r_b} dr \ff{e^{2 \L-B}}{\prt{e^{2\L_m}-e^{2 \L}}^{\ff{3}{2}}}~,
\ee
where $\L_m'=\L'(r_m)$. For the imaginary contribution to the area $S_{Im}$, we get
\be \la{ds1a}
\ff{\pp S_{Im}}{\pp r_m}=2\ff{e^{2 \L-B}}{\sqrt{e^{2\L_m}-e^{2\L}}}\bigg|_{  r_m}+
2\L_m'e^{2\L_m}\int_{r_m}^{r_b} dr\ff{e ^{2\L-B}}{\prt{e^{2\L_m}-e^{2 \L}}^{\ff{3}{2}}}~.
\ee
Therefore, the derivative of the timelike part of the area with respect to the time gives
\be \la{derima}
\ff{\pp S_{Im}}{\pp r_m}= e^{\L_m}\ff{\pp t_{Im}}{\pp r_m}\Rightarrow\ff{\pp S_{Im}}{\pp t_{Im}}= e^{\L_m}~.
\ee
Thus the slope of the imaginary part of the entropy with respect to its time interval is fixed universally by the geometry at the turning point.

For the spacelike surface the corresponding equations are:
\be \la{deri2a}
e^{\L_m}\ff{\pp t_{Re}}{\pp r_m}=-2\L_m'e^{2\L_m}\int_{r_\pp}^{r_b}dr\,\frac{ e^{2\L-B}}{\prt{e^{2\L}+e^{2 \L_m}}^{\ff{3}{2}}}~,
\ee
while
\be \la{dt2a}
\ff{\pp S_{Re}}{\pp r_m}=2\L_m' e^{2\L_m} \int_{r_\pp}^{r_b}dr\,\frac{e^{2 \L-B}}{\prt{e^{2\L}+e^{2 \L_m}}^{\ff{3}{2}}}~.
\ee
The right hand sides of \eq{deri2a} and \eq{dt2a} coincide up to a sign and thus the time interval derivative of the area of the spacelike surface reads
\be \la{derrea}
\ff{\pp S_{Re}}{\pp t_{Re}}=- e^{\L_m}~.
\ee
Note that the derivatives \eq{derima} and \eq{derrea} have the same magnitude and differ by a sign, and satisfy the identity
\be \la{equal}
\ff{\pp S_{Re}}{\pp t_{Re}}=-\ff{\pp S_{Im}}{\pp t_{Im}}~.
\ee
This equality is independent of anisotropy and follows solely from the structure of the extremal surface equations, it does not depend on the specific form of $B(r)$ or $A_i(r)$. It plays a key role in the construction of the timelike holographic $c$-function, ensuring that the real and imaginary contributions combine consistently along the RG flow. The boundary terms at $r_m$ in \eqref{dt1a} and \eqref{ds1a} are individually divergent, since the integrands have an integrable singularity at the turning point; they cancel in \eqref{derima}. These identities hold exactly at fixed radial cutoffs, as shown in Appendix~\ref{appA}. They are the Hamilton-Jacobi relations between the on-shell area and the branch time extent,
with conserved momentum of magnitude $e^{\L_m}$.

%%%%%%%%%%%%%%%%%%%%%%%%%%%%%%%%%%%%%%
\subsection{The Timelike $c$-function}

Motivated by \cite{Giataganas:2025div,Chu:2019uoh,Casini:2006es,Myers:2012ed,Ryu:2006ef}, we propose that the timelike holographic $c$-function introduced in \cite{Giataganas:2025div} continues to hold in the presence of spatial anisotropy, provided one uses the appropriate effective dimension $d_x$ (or $d_y$) associated with the direction of localization of the boundary interval, which we explicitly define later. Anisotropy distinguishes the $\vec{x}$ and $\vec{y}$ planes, and there are in general two candidate timelike $c$-functions: $c_x$ and $c_y$, each corresponding to a boundary interval localized along the respective direction.

We consider the branch combination for the $c$-function:
\be \la{cfunctiona} 
c=c_{Re}+c_{Im}~,\qquad c_{(Im,Re)}:=t_{(Im,Re)}^{d_x}\frac{\pp S_{(Im,Re)}}{\pp t_{(Im,Re)}}~.
\ee
Here the computation is presented for $c:=c_x$, where $\L(r)$ is given by \eq{lamdadef}, appropriate for localization along a direction in the $\vec{x}-$plane. Throughout the following we can continue working with $c_x$ for definiteness, without the need to substitute the exact form of the $\L(r)$, our construction remains generic.

For the localization along the $y$-plane, we have $c_y$ and the computation carries on in the same way with
\be \la{cfunctionab}
c_y=c_{y,Re}+c_{y,Im}~,\qquad c_{y,(Im,Re)}:=t_{y,(Im,Re)}^{d_y}\frac{\pp S_{y,(Im,Re)}}{\pp t_{y,(Im,Re)}}~.
\ee
where $\L(r)=\L_y(r)$  given by \eq{ly}, and the exponent $d_y$ is the corresponding effective dimension. The solutions $t_{y}$ and $t_x$ are automatically defined by \eq{t2}, by the appropriate use of $\L_x$ and $\L_y$ in the functional. Thus the anisotropy leads to two independent candidate $c$-functions, each of which must obey monotonicity if a universal timelike c-theorem is to hold.

We first compute the derivatives of the two branch terms:
\be \la{cf00a}
\frac{\pp c_{\prt{Im,Re}}}{\pp r_m}=\pm  t_{\prt{Im,Re}}^{d_x-1}e ^{\L_m} d_x \prt{  \ff{\pp t_{\prt{Im,Re}}}{\pp r_m}+\frac{\L_m'}{d_x} t_{\prt{Im,Re}}}~,
\ee
where we have used \eq{derima} and \eq{derrea}, and the positive (upper) sign corresponds to  $c_{Im}$, while the negative (lower) to $c_{Re}$.

%%%%%%%%%%%%%%%%%%%%%%%%%%%%%%%
\subsection{The $c$-function in Terms of Boundary Data}\label{sec:cbound}

To relate the timelike holographic
$c$-function \eqref{cfunctiona} more directly to boundary QFT observables, it is useful to consider candidate $c$-functions depending on the physical boundary time interval $T$. Absorbing all positive normalization factors into the definition, we introduce the following boundary expressions
\be \la{cf0Ta}
c_{x,T}=T_x^{d_x}\frac{\pp S_x}{\pp T_x}
~\qquad\mbox{and}\qquad
c_{y,T}=T_y^{d_y}\frac{\pp S_y}{\pp T_y}~,
\ee
with $S=S_{Re}+S_{Im}$ and $T=t_{Im}-t_{Re}$ along the relevant directions $(x,y)$ respectively, with the subscript in \eq{cf0Ta} indicating whether the boundary interval is localized along the $x$, or $y$ direction. 
These expressions define candidate $c$-functions purely in terms of boundary quantities and are natural from the perspective of the dual QFT, since they capture how the timelike entanglement entropy scales with the boundary temporal resolution.

Notice that one has to be cautious with the interpretation of the derivative $\pp_T c_T$, since the increase of the boundary time interval $T$ does not always guarantee that lower energy scales are being probed. Here, we assume the standard QFT identification whereby increasing the boundary time interval $T$ probes longer time scales and therefore lower-energy (IR) physics. This orientation must be checked on the geometry. If it changes along the flow, the parametrization by $T$ must be treated on separate monotonic branches.

%%%%%%%%%%%%%%%%%%%%%%%%%%%%%%%%
\subsection{Monotonicity of the Timelike Sector}

We rewrite the time interval \eq{t1a} in the more convenient form
\be \la{tx2a}
t_{s}=2 s_1 \int dr \ff{e^{\ff{\L}{d_x}-B}}{\L'} \cdot F_s~,\qquad F_{s}:=\ff{\L' e^{-\ff{\L}{d_x}+\L_m}}{\sqrt{e^{2 \L_m}+s e^{2\L}}}~.
\ee
Here $\cF_s$ is a primitive of $F_s$. The hypergeometric
parameters are nonsingular for $d_x>1/2$. At $d_x=1/(2n)$,
$n=1,2,\ldots$, use the finite parameter limit after
subtracting an $r$-independent constant:
\be \la{cf}
\cF_s:=\int F_s dr=- d_x e^{-\ff{\L}{d_x}-\L_m}\sqrt{se^{2\L}+ e^{2 \L_m}}~{}_2F_1[1,\ff{1}{2}-\ff{1}{2 d_x},1-\ff{1}{ 2 d_x},-s e^{2 \L-2\L_m}]~.
\ee
Let us work first  with the timelike surface, $s=-1$. We remind the reader that we have chosen the right symmetric branch with $s_1=-1$, as discussed below the equation \eq{t2}. Nevertheless, in this and the next section we demonstrate how the results turn out to be independent of the choice $s_1$ by carrying it explicitly through the computations until obtaining the final expression.  By implementing the integration by parts for the time interval \eq{tx2a}, we obtain
\be \la{im2a}
\frac{t_{Im}}{2 s_1}=\frac{e^{\frac{\L}{d_x}-B}}{\L'}\cF_{-} \bigg|_{r_m}^{r_b}-\int_{r_m}^{r_b}dr\,\frac{ e^{\frac{\L}{d_x}-B}\cF_{-}}{\L'}\prt{\frac{\L'}{d_x}-\frac{\L''}{\L'}-B'}~.
\ee
Its derivative with respect to the turning point $r_m$ after some algebra can be conveniently written as
\bea\la{der2a}
\ff{1}{2 s_1} \frac{\pp t_{Im}}{\pp r_m}=\frac{\pp_{r_m}\cF_{b-}}{\L_b'}e^{\frac{\L_b}{d_x}-B_b}-\frac{\pp_{r_m}\cF_{m-}}{\L_m'}e^{\frac{\L_m}{d_x}-B_m}
-\int_{r_m}^{r_b} dr \frac{e^{\frac{\L}{d_x}-B}\pp_{r_m}\cF_{-} }{\L'} \prt{\frac{\L'}{d_x}-\frac{\L''}{\L'}-B'}~.
\eea
A key simplification comes when we notice that the derivatives of the $\cF$ function can be expressed with the function itself as
\be\la{dera1a}
\frac{\pp_{r_m}\cF_{m-}}{\cF_{m-}}=-\frac{\L_m'}{d_x}~,\qquad  \pp_{r_m}\cF_{-}+\ff{\L_m'\cF_{-} }{d_x}=-\ff{\L_m' e^{\L_m-\frac{\L}{d_x}}}{\sqrt{e^{2 \L_m}-e^{2\L}} }~,
\ee
where in the first expression the function $\cF$ is computed at $r_m$ and then the derivative is taken. Note that for $d_x=1$, $\cF_{m-}$ vanishes identically and the first relation is to be read in its undivided form. 
We note that the derivative of $t_{Im}$ depends on the $t_{Im}$, and  by using the identities \eq{dera1a}, and  equations \eqref{der2a}  with \eqref{im2a}, the algebra collapses dramatically: all dependence on $\cF_-$ reorganizes into a single integral and a boundary term as
\bea\nn
\ff{1}{2 s_1}\prt{ \frac{\pp t_{Im}}{\pp r_m}+\frac{\L_m'}{ d_x}t_{Im}}&=& \frac{e^{\frac{\L_b}{d_x}-B_b}}{\L_b'}  \prt{ \pp_{r_m}\cF_{b-}+\frac{\L_m'\cF_{b-}}{d_x}}\\
&&+\L_m' e^{\L_m} \int_{r_m}^{r_b} dr\frac{ e^{-B}}{\L'\sqrt{e^{2 \L_m}-e^{2\L}}} \prt{\frac{\L'}{d_x}-\frac{\L''}{\L'}-B'}~.\la{der4a}
\eea
The above expression contains a single integral term, since the new integral generated by the use of \eq{dera1a} gives the $t_{Im}$ of the left hand side of the above equation. The left-hand side is proportional to
$\pp_{r_m}c_{Im}$, as follows from \eqref{cf00a}, while the first line of the right hand side can be further simplified by the use of the second equation \eq{dera1a}.
This allows us to write the derivative of the timelike component of the
$c$-function \eq{cf00a} as
\be  
 \frac{\pp c_{Im}}{\pp r_m}=2 s_1 e^{\L_m} t_{Im}^{d_x-1} \L_m' d_x \bigg[\ff{ -e^{\L_m-B_b}}{\L_b'\sqrt{e^{2 \L_m}-e^{2\L_b} }}+ \int_{r_m}^{r_b} dr \frac{e^{\L_m-B}}{\L'\sqrt{e^{2 \L_m}-e^{2\L}}}\prt{\frac{\L'}{d_x}-\frac{\L''}{\L'}-B'}\bigg]~.
\ee
Notice, that   the first term in the brackets is equal to the derivative $t'_{Im\,b}/\L_b'$ of the timelike surface at the deepest point in the bulk $r_b$. Moreover, we can trade $dr$ with $dt$ in the integrand using
\be
dt_{Im}= s_1 \frac{e^{\L_m-B}}{\sqrt{e^{2\L_m}-e^{2 \L}}} dr~.
\ee
Therefore the final compact expression becomes
\bea
\frac{\pp c_{Im}}{\pp r_m}=2  e^{\L_m} t_{Im}^{d_x-1}d_x\L_m'\bigg[ -\frac{t_{Im~b}'}{\L_b'}+\int_{0}^{\frac{t_{Im}}{2}} dt \frac{1}{\L'}\prt{\frac{\L'}{d_x}-\frac{\L''}{\L'}-B'}\bigg]~. \la{cimtota}
\eea
The timelike surface begins at $r=r_m$ when $t=0$,  and asymptotes to the deep IR for $t_{Im}\to\infty$, when choosing the right branch  of $t(r)$, and in geometries without horizons or walls.

%%%%%%%%%%%%%%%%%%%%%%%%%%%
\subsection{Monotonicity of the Spacelike Sector}

We proceed to compute the contribution \eq{cfunctiona} of the spacelike surface $c_{Re}$ to the $c$-function. The time interval $t_{Re}$ is given from \eq{tx2a} for $s=+1$,
and can be written as
\bea
\frac{t_{Re}}{2 s_1}= \frac{e^{\frac{\L}{d_x}-B} }{\L'}\cF_{+}\bigg|_{r_\pp}^{r_b}
-\int_{r_\pp}^{r_b}dr \frac{e^{\frac{\L}{d_x}-B}\cF_+}{\L'}\prt{\frac{\L'}{d_x}-\frac{\L''}{\L'}-B'} ~,\la{tre}
\eea
where $r_\pp$ denotes the UV boundary, and $r_b$ the deep IR endpoint of the spacelike surface.

Taking the derivative with respect to turning point $r_m$ yields
\bea
 \frac{1}{2 s_1}\frac{\pp t_{Re}}{\pp r_m}=\frac{e^{\frac{\L_b}{d_x}-B_b}}{\L_b'}\pp_{r_m}\cF_{b+}-\frac{e^{\frac{\L_\pp}{d_x}-B_\pp}}{\L_\pp'}\pp_{r_m}\cF_{\pp+}-
\int_{r_\pp}^{r_b} dr
\frac{e^{\frac{\L}{d_x}-B}}{\L'}\pp_{r_m}\cF_{+}\prt{\frac{\L'}{d_x}-\frac{\L''}{\L'}-B'}~.\la{tre2a}
\eea
By using
\be
\pp_{r_m}\cF_+=-\frac{\L_m'}{\sqrt{e^{2 \L}+e^{2 \L_m}}}e^{-\frac{\L}{d_x}+\L_m}-\frac{\cF_+}{d_x}\L_m'~,\la{derivfpa}
\ee
we can express the derivative of $t_{Re}$  in terms of $t_{Re}$. Substituting into \eqref{tre2a} and comparing with \eqref{tre}, all terms reorganize cleanly into contributions proportional to $t_{Re}$ itself plus boundary contributions and a single integral. After some algebra  we obtain
\bea \nn
\frac{1}{2 s_1}\prt{\frac{\pp t_{Re}}{\pp r_m}+\frac{\L_m'}{  d_x}t_{Re}}&=&
-\L_m'\bigg[\frac{e^{\L_m-B_b} }{\L_b'\sqrt{e^{2 \L_b}+e^{2\L_m}}}-\frac{ e^{\L_m-B_\pp}}{\L_\pp'\sqrt{e^{2 \L_\pp}+e^{2\L_m}}}\bigg]\\
&&+\L_m'\int_{r_\pp}^{r_b} dr\frac{e^{\L_m-B}}{\L'\sqrt{e^{2 \L}+e^{2\L_m}}}\prt{\frac{\L'}{d_x}-\frac{\L''}{\L'}-B'}~.\la{tre4a}
\eea
The extra integral term generated by the use of \eq{derivfpa} conveniently gives the term of $t_{Re}$ of the left hand side. Interestingly,   the first two terms in the first line of \eq{tre4a} give  precisely $t_{Re~b}'$ and $t_{Re~\pp}'$,  the derivatives of the spacelike surface at the IR and UV endpoints respectively. Moreover, $dr$ is expressed in terms of the time differential
\be  \la{dtre}
dt_{Re}=s_1\frac{e^{\L_m-B}}{\sqrt{e^{2\L}+e^{2 \L_m}}} dr~.
\ee
From \eq{cfunctiona}, and by combining all the above results, we obtain the final expression:
\bea
\frac{\pp c_{Re}}{\pp r_m}&=& -2 e^{\L_m} t_{Re}^{d_x-1}d_x\L_m'\bigg[\prt{-\frac{t_{Re~b}'}{\L_b'}+\frac{t_{Re~\pp}'}{\L_\pp'}}
+\int_{\frac{T}{2}}^{\frac{t_{Im}}{2}} dt \frac{1}{\L'}\prt{\frac{\L'}{d_x}-\frac{\L''}{\L'}-B'}\bigg]~.\la{cretota}
\eea
The positive symmetric spacelike branch $t(r)$ begins at the boundary of the theory $r=r_\pp$ where $t=T/2$, and extends toward the IR, where the common joining time $t_{Im}/2$ diverges as $r_b\to r_{IR}$.

%%%%%%%%%%%%%%%%%%%%%%%%%%%%%%%%%%%%%%%
\subsection{Asymptotic Conditions of the Extremal Surfaces}\label{sec:asym}

The unnormalized gradient normal vector field $g^{\a\b}\pp_\b \S$ satisfies  the condition
\be \la{normalva}
\nabla_\a\pp_\b\S-\nabla_\b\pp_\a\S=0~.
\ee
We define the vector
\begin{align}
\cT_\a=\partial_\a\Sigma=\left(1,0,\mathbf{0}_{d_1-1},\mathbf{0}_{d_2},-t^\prime(r)\right)~,\label{zeroseta}
\end{align}
where its norm, on the spacelike and timelike branches, is given by \cite{Afrasiar:2024ldn}
\be \label{normt}
|\cT|^2=\cT_\a~g^{\a\b}~\cT_\b=-\frac{e^{2 A_1\prt{d_1-1}+2 A_2 d_2}}{s e^{2\L_m}+e^{2 \L}}~.
\ee
The norm is positive on the timelike branch ($s=-1$) and negative ($s=+1$) on the spacelike branch. Its unnormalized magnitude need not vanish in the IR. In the IR, \eqref{normt} gives
\be
|\cT|^2\sim-s^{-1}e^{2A_1(d_1-1)+2A_2d_2-2\L_m}~.
\ee
Since
$e^{2B}t_s'^2=[1+s e^{2(\L-\L_m)}]^{-1}\to 1$, both branches approach null directions in a local orthonormal frame. For the common orientation of the chosen symmetric half-surface,
\be\la{conddera}
\frac{t_{Re~b}'{}^2}{t_{Im~b}'{}^2}\longrightarrow 1~.
\ee
This leading agreement holds for arbitrary nonzero branch momenta. 

For the common momentum and the chosen orientation,
\be\label{ts12}
t'_{Re\,b}-t'_{Im\,b}=\mathcal O\!\left(e^{2\L_b-2\L_m-B_b}\right)~,
\ee
so that $(t'_{Re\,b}-t'_{Im\,b})/\L_b'$ vanishes in the IR under the endpoint condition \eqref{IRendpoint} below.

%%%%%%%%%%%%%%%%%%%%%%%%%%%%%%%%%%%%%

\subsection{The Derivative of the $c$-function}\label{sec:derivc}

Using the asymptotic behavior of the extremal surfaces derived in the previous subsection, we now combine the timelike and spacelike contributions \eqref{cimtota} and \eqref{cretota} for the $c$-function; we remind the reader that we use the notation $c:=c_x$ of \eqref{cfunctiona}, to avoid carrying obvious subindices.  At a common finite IR cutoff, $T>0$ and the common joining-time coordinate is $T_\infty/2$, with $T_\infty:=t_{Im}$. The entropy slopes entering \eqref{cfunctiona} are evaluated at fixed radial cutoffs before choosing the regulated family. The analysis of $c_y$ is straightforward and we comment on it when the derivation completes.  

First, we must specify how the IR cutoff is varied when the turning point changes. To distinguish the RG evolution from regulator-induced dependence, we require the cutoff prescription to respect scaling at matched fixed points. We choose a cutoff that scales with the probe, measuring the regulator time in units of the turning-point scale. At a matched scale-covariant point,
$t_{Re}\mapsto a\,t_{Re}$ and $e^{\L_m}\mapsto a^{-d_x}e^{\L_m}$, so $e^{\L_m/d_x}t_{Re}$ is invariant.
Restricting the regulator label to these two regulated spacelike-branch quantities selects this combination, up to an equivalent monotonic relabelling. We adopt the prescription
\be\la{IRclockprescription}
e^{\L_m/d_x}t_{Re}=\mathcal R~, \qquad \mathcal R>0\quad\text{fixed along the flow}~.
\ee
The prescription measures the regulator in the time unit $e^{-\L_m/d_x}$. By \eqref{derrea}, 
$c_{Re}=-\mathcal R^{d_x}$ remains fixed, while allowing the full $c$ to vary with $r_m$ along an interpolating flow, so that the spacelike branch enters through the physical interval $T$, with its divergent part canceled against that of $c_{Im}$. This distinguishes the construction from the function $\cal C$  we introduce in later sections which uses only the imaginary part of the timelike entanglement entropy.
We adopt this regularization scheme for the additive branch
$c$-function \eqref{cfunctiona}; it yields the monotonicity comparison
with the boundary function $c_T$ of \eqref{cf0Ta} derived below.
The boundary construction itself does not require this prescription.

In pure AdS$_{d+1}$ with $d_x=d-1$, it preserves the stationarity expected at a fixed point. A fixed radial cutoff would instead produce, for $d>2$, the leading regulator-induced dependence $c\propto e^{(d-2)r_m/L}$; for $d=2$ this leading term is constant.

We have $\mathcal R$ sufficiently large as the cutoff lies on the IR side $r_b \to r_{IR}$. At finite regulator, $T>0$ denotes the regulated boundary interval. The cutoff moves with $r_m$ at fixed $\mathcal R$, while $T_\infty=t_{Im}$ is allowed to vary. Combining the interval derivatives entering \eqref{cimtota} and \eqref{cretota},   with the IR cutoff varied according to \eqref{IRclockprescription}, yields
\begin{align}\la{ctot1}
\frac{dc}{dr_m}
&=2e^{\L_m}T_\infty^{d_x-1}d_x\L_m' \bigg[\varepsilon_{\mathcal R}-\frac{t'_{Re\,\pp}}{\L'_\pp}\bigg]
\\\nn
&{}+ 2e^{\L_m}T_\infty^{d_x-1}d_x\L_m' \bigg[\int_0^{\frac{T_\infty}{2}}dt\frac{1}{\L'}\prt{\frac{\L'}{d_x}-\frac{\L''}{\L'}-B'}\Big|_{Im} -\int_{\frac T2}^{\frac{T_\infty}{2}}dt\frac{1}{\L'} \prt{\frac{\L'}{d_x}-\frac{\L''}{\L'}-B'}\Big|_{Re}\bigg]~.
\end{align}
The derivative is taken at fixed $\mathcal R$.
The spacelike surface approaches the UV boundary $e^{A_i}\to \infty$
orthogonally. For the UV asymptotics considered here,
its weighted endpoint term also vanishes:
\be\la{boundarypp}
\frac{t'_{Re\,\pp}}{\L'_\pp}\longrightarrow 0~.
\ee
This weighted limit follows naturally  when $e^B|\L'|\to\infty$ towards the UV.  We also require
UV convergence of the time integrals and their differentiated forms.

The branch identities \eqref{derima} and \eqref{derrea} then give $c=e^{\L_m}(t_{Im}^{d_x}-t_{Re}^{d_x})$. At fixed $\mathcal R$, \eqref{IRclockprescription}  implies $dt_{Re}/d\L_m=-t_{Re}/d_x$, then the  IR endpoint correction in \eq{ctot1} reads
\be\la{IRclockcorrection}
\varepsilon_{\mathcal R}
=\frac{t'_{Re\,b}-t'_{Im\,b}}{\L_b'}
\prt{1-\frac{\L_b'}{\L_m'}\frac{dr_b}{dr_m}}= \cO(\mathcal R^{1-2d_*}) \to 0~.
\ee
The vanishing limit follows since the factor $(t'_{Re\,b}-t'_{Im\,b})/\L_b'$ vanishes by \eqref{ts12} under the natural endpoint condition in backgrounds without IR walls,
\be\la{IRendpoint}
\frac{e^{2\L-B}}{|\L'|}\longrightarrow 0 \qquad\text{in the IR}~,
\ee
together with finiteness of the imaginary area integrals, while the bracket remains bounded, since $d\L_b/d\L_m\to d_*/d_x$ for a regular scaling IR with \eq{IRcond}, where $d_*>1/2$ is the effective dimension of the IR scaling geometry, i.e.\ the IR value of the running dimension $d_{x_r}$ defined below in \eqref{dxr1a};  the correction is then $\mathcal O(\mathcal R^{1-2d_*})$  (Appendix~\ref{appA}). In an exactly scaling geometry with the matching reference exponent and the UV cutoff removed, it vanishes identically at finite $\mathcal R$. The endpoint condition \eqref{IRendpoint} holds for each localization direction in all scaling examples and parameter regions of Section~\ref{section6}. For example, if the entire geometry is exactly scale covariant and $d_x=d_*$, the regulator condition \eq{IRclockprescription} instead fixes $u_b:=\L_b-\L_m$. Hence $d\L_b/d\L_m=1$: the bracket in \eqref{IRclockcorrection} vanishes and $\varepsilon_{\mathcal R}=0$ at every finite regulator, with the UV cutoff removed.

We remark that in \eqref{ctot1} each branch term separately carries a positive measure: by \eqref{cimtota} and
\eqref{cretota}, at fixed radial cutoffs, with $\L'>0$, $t'_{Im\,b},t'_{Re\,b}<0$ and \eqref{boundarypp}, the local condition $\L'{}^{-1} \prt{\L'/d_x- \L''/\L'-B'}  \ge 0$ defined in \eqref{dxrdef} below as $w\ge0$, along the timelike branch gives $\pp_{r_m}c_{Im}\ge0$, and along the spacelike branch gives $\pp_{r_m}(-c_{Re})\ge0$. The criterion of \cite{Giataganas:2025div} therefore applies verbatim to the
timelike-branch term and to the magnitude of the spacelike-branch term separately; it is the relative sign with which the spacelike branch enters $T$ that produces the signed measure below.

We abbreviate the combination of the two branch integrals using the signed measure, where the integrand is evaluated on its own branch,
\be\la{ctsignedmeasure}
\int_0^{T/2}d\tau\,f:=\int_0^{T_\infty/2}dt\,f\big|_{Im}+\int_{T_\infty/2}^{T/2}dt\,f\big|_{Re}~.
\ee
Eventually the final expression \eq{ctot1} becomes
\bea\la{c_tot2a00}
\frac{dc}{dr_m}=2e^{\L_m}T_\infty^{d_x-1}d_x\L_m'
\prtt{\int_0^{\frac T2}d\tau\,\frac1{\L'}\prt{\frac{\L'}{d_x}-\frac{\L''}{\L'}-B'}+\varepsilon_{\mathcal R}}~.
\eea
This expression is exact at finite regulator.  To express the geometric conditions compactly, we introduce the following very useful quantities for the integrand of \eq{c_tot2a00}
\be\la{dxrdef}  
w(r):=\frac{1}{\L'}\prt{\frac{\L'}{d_x}-\frac{\L''}{\L'}-B'}~.
\ee
We refer to 
\be \la{localeff}
w\ge0
\ee
as the local effective-dimension condition for reasons that become clear in Section~\ref{sec:dx}.

The measure $d\tau$  of \eq{c_tot2a00} is the signed branch measure, and it remains signed after removal of the IR regulator, so the pointwise condition $w\ge0$ alone does not fix the sign. In our geometries the IR endpoint correction \eq{IRclockcorrection} vanishes $\varepsilon_{\mathcal R}\to0$. Then \eqref{c_tot2a00}  becomes
\bea\la{c_tot2a}
\frac{dc}{dr_m}
\simeq 2e^{\L_m}T_\infty^{d_x-1}d_x\L_m'\prtt{\int_0^{\frac T2}d\tau\,\frac1{\L'}\prt{\frac{\L'}{d_x}-\frac{\L''}{\L'}-B'}}~.
\eea
The omitted endpoint correction tends to zero inside the flow bracket and drops out in the appropriate normalized limits, under the stated derivative-convergence assumptions we discuss below. At finite regulator, neglecting it requires control relative to the signed integral. The compact form of \eq{c_tot2a} makes explicit that the flow is determined by the complete signed geometric integral and the regulated endpoint correction. A definite sign of the integrand alone is not sufficient. Here $T_\infty=t_{Im}\simeq\mathcal R e^{-\L_m/d_x}$ depends on the turning point and diverges as $\mathcal R\to\infty$. It is positive, so its prefactor does not affect the sign.

Therefore, under our assumptions, the condition for the monotonicity of the $c$-function is
\be
\frac{dc}{dr_m}\ge0
\ee
for a boundary located at infinity in the holographic spacetime, with $\L_m'>0$. For either radial orientation, the UV-oriented criterion is $dc/d\L_m\ge0$. At finite cutoff the sign is determined by the complete bracket in \eqref{c_tot2a00}.

In anisotropic theories where the spatial rotational invariance $SO(d_1+d_2)$ is broken to $SO_x(d_1)$ and $SO_y(d_2)$, two distinct timelike $c$-functions $c_x$ and $c_y$ naturally arise, each corresponding to one of the two subplanes. In theories in which the symmetry is broken further, the number of $c$-functions equals the number of distinct subplanes. It may be also possible to construct a single covariant unique $c$-function by combining these into a single function. Here we adopt the strictest criterion: each individual $c$-function must satisfy a c-theorem along the RG flow. Therefore, any covariant combination of them that preserves monotonicity would then define a global anisotropic $c$-function with the same monotonicity.

For clarity, we state the directional flows explicitly. With the UV cutoff removed and the IR prescription
\eqref{IRclockprescription} imposed separately in each direction, the function $c_x$ satisfies
\bea\la{c_tot2ax}
\frac{dc_x}{dr_m}\simeq 2e^{\L_{m,x}}T_{\infty,x}^{d_x-1}d_x\L_{m,x}'
\prtt{\int_0^{T_x/2}d\tau_x\,\frac{1}{\L_x'}\prt{\frac{\L_x'}{d_x}-\frac{\L_x''}{\L_x'}-B'}}~,
\eea
where $\L_x(r)$ is given by \eqref{lamdadef} and
$\L_{m,x}':=\L_x'(r_m)$. Similarly,
\bea\la{c_tot2ay}
\frac{dc_y}{dr_m} \simeq 2e^{\L_{m,y}}T_{\infty,y}^{d_y-1}d_y\L_{m,y}' \prtt{\int_0^{T_y/2}d\tau_y\,\frac{1}{\L_y'}\prt{\frac{\L_y'}{d_y}-\frac{\L_y''}{\L_y'}-B'}}~,
\eea
with $\L_y(r)$ given by \eqref{ly}. We have already used the stated endpoint and convergence assumptions,
$\varepsilon_{\mathcal R_i}\to0$ as the IR regulator is removed. At finite regulator, the correction may
be omitted when it is negligible compared with the signed integral. The positive factors $T_{\infty,x}^{d_x-1}$ and $T_{\infty,y}^{d_y-1}$ depend on the turning point and remain in the unnormalized flows. Each signed integral is defined by \eqref{ctsignedmeasure} for the corresponding direction, and each derivative is taken at fixed $\mathcal R_i$.

Let us make another remark in this section. Following our formalism one  may consider an alternative branch definition of a $c$-function that combines the positive $d_x$-th roots:  $c_{\rm alt}:=[c_{Im}^{1/d_x}-(-c_{Re})^{1/d_x}]^{d_x}.$ The branch identities give $c_{\rm alt}=e^{\L_m}T^{d_x}$. This construction requires no prescription fixing $c_{Re}$.  The same idea applies to the $y$-direction; here we retain the additive definition \eqref{cfunctiona} and the prescription \eqref{IRclockprescription}.

%%%%%%%%%%%%%%%%%%%%%%%%%%%%%%
\subsection{Monotonicity of $c$-function in Terms of Boundary Interval \texorpdfstring{$T$}{T}}
\label{section::ct}

We now compare the monotonicity of the branch $c$-function \eqref{cfunctiona} with that of the boundary expression \eqref{cf0Ta}, under the IR prescription \eqref{IRclockprescription}. We continue working in the
$x$-direction, with $c:=c_x$, and write $c_T:=c_{x,T}$, $T:=T_x$ and $S:=S_x$. The $y$-direction follows analogously.

We analyze the monotonicity of
\be\label{cttt}
c_T=T^{d_x}\frac{\pp S}{\pp T},
\ee
by taking the derivative with respect to the turning point $r_m$.  After removing the cutoffs under the
stated endpoint and convergence assumptions, the branch identities \eqref{derima} and \eqref{derrea} give
$\pp S/\pp T=e^{\L_m}$. Thus
\be\la{cf0T1a}
\frac{\pp c_T}{\pp r_m}=T^{d_x-1}e^{\L_m}\prtt{d_x\frac{\pp T}{\pp r_m}+T\L_m'}~.
\ee

We work with the right half of the symmetric surface, as in the preceding subsections. The branch time extents are positive, whereas $t'_{Im}$ and $t'_{Re}$ denote signed radial slopes. All conventional positive normalization factors have been absorbed into $c_T$ and do not affect monotonicity.
The final result is independent of the choice of symmetric half-surface.

At finite IR cutoff we distinguish the regulated interval $T_{\mathcal R}:=t_{Im}-t_{Re}$ from its limit
$T=\lim_{\mathcal R\to\infty}T_{\mathcal R}$, and define the regulated boundary function algebraically,
\be\la{cbranchrelation}
c_{T,\mathcal R}:=e^{\L_m}T_{\mathcal R}^{d_x}
\;\xrightarrow{\;\mathcal R\to\infty\;}\;e^{\L_m}T^{d_x}=c_T~,
\ee
rather than through the entropy slope along the moving-cutoff family, which would contain endpoint contributions. As before, $T_\infty:=t_{Im}$ varies with the turning point.

The interval derivative is computed by combining the contributions from $t_{Im}$ and $t_{Re}$.
From \eqref{der4a} we have
\bea\la{der4b}
\frac{\pp t_{Im}}{\pp r_m}
=-\frac{\L_m'}{d_x}t_{Im}-2\L_m'\bigg[\frac{t'_{Im\,b}}{\L_b'}-\int_0^{\frac{T_\infty}{2}}dt\,
\frac{1}{\L'}\prt{\frac{\L'}{d_x}-\frac{\L''}{\L'}-B'}\Big|_{Im}\bigg]~.
\eea
From \eqref{tre4a}, the spacelike contribution reads
\bea\la{tre4b}
 \frac{\pp t_{Re}}{\pp r_m} =-\frac{\L_m'}{d_x}t_{Re}-2\L_m'\bigg[
\frac{t'_{Re\,b}}{\L_b'}-\frac{t'_{Re\,\pp}}{\L_\pp'}-\int_{\frac{T_{\mathcal R}}2}^{\frac{T_\infty}2}dt\,
\frac{1}{\L'}\prt{\frac{\L'}{d_x}-\frac{\L''}{\L'}-B'}\Big|_{Re}
\bigg]~.
\eea
These partial derivatives hold the radial cutoffs fixed, and each integral is evaluated along its own
branch. They can be written in more compact forms using  $w$, defined in \eqref{dxrdef}.

To retain the compact notation of the branch flow, we use the signed integral \eq{ctsignedmeasure}.  In particular, $\int d\tau=T_{\mathcal R}/2$. The limits indicate the net time extent of this signed
path; the integral samples both branches and does not identify their geometries at equal time.

Subtracting \eqref{tre4b} from \eqref{der4b} and including the IR cutoff motion gives
\be\la{dtrmaregulated}
\frac{dT_{\mathcal R}}{dr_m}
=2\L_m'\bigg[-\frac{T_{\mathcal R}}{2d_x}+\int_0^{\frac{T_{\mathcal R}}2}d\tau\,\frac{1}{\L'}\prt{\frac{\L'}{d_x}-\frac{\L''}{\L'}-B'}+\varepsilon_{\mathcal R}-\frac{t'_{Re\,\pp}}{\L_\pp'}
\bigg]~.
\ee
Here the total derivative is taken at fixed $\mathcal R$ and fixed radial UV cutoff. The IR correction $\varepsilon_{\mathcal R}$ is given by \eqref{IRclockcorrection}. The UV endpoint term vanishes under \eqref{boundarypp}.

Under the convergence, endpoint and cutoff-motion conditions of Section~\ref{sec:derivc}, $\varepsilon_{\mathcal R}\to0$, and $T_{\mathcal R}$ and its first flow derivative converge. We therefore obtain
\bea\la{dtrma}
\frac{\pp T}{\pp r_m}
=2\L_m'\prtt{-\frac{T}{2d_x}+\int_0^{\frac T2}d\tau\,\frac{1}{\L'}\prt{\frac{\L'}{d_x}-\frac{\L''}{\L'}-B'}}
%=-2\L_m'\int_0^{\frac T2}\frac{d\tau}{d_{x_r}}~.
\eea
Substituting \eqref{dtrma} into \eqref{cf0T1a}, we obtain
\bea\la{equivct}\la{exactflow}
\frac{\pp c_T}{\pp r_m}
&=&\frac{T^{d_x-1}}{T_\infty^{d_x-1}}\bigg[2e^{\L_m}T_\infty^{d_x-1}d_x\L_m' \int_0^{\frac T2}d\tau\,\frac{1}{\L'}\prt{\frac{\L'}{d_x}-\frac{\L''}{\L'}-B'}
\bigg]~.
\eea
The factors involving $T_\infty$ cancel algebraically, leaving the boundary flow governed by the same limiting
signed integral as the branch flow. At finite regulator, with the UV cutoff removed and the
IR endpoint correction retained in both flows, the comparison is exact:
\be\la{cbranchcomparison}
\frac{dc_{T,\mathcal R}}{dr_m}=\left(\frac{T_{\mathcal R}}{T_\infty}\right)^{d_x-1}
\frac{dc_x}{dr_m}~.
\ee
Since $T_{\mathcal R},T_\infty>0$, the two regulated functions have the same monotonicity and stationary points
for every $d_x>0$. The comparison factor depends on the turning point; as the regulator is removed it tends to zero for $d_x>1$, equals one for $d_x=1$ and diverges for $0<d_x<1$, but the comparison itself survives after normalization. As $\mathcal R\to\infty$ at
fixed turning point, $T_{\mathcal R}\to T$ while $t_{Re}=\mathcal R e^{-\L_m/d_x}\to\infty$, and
$c=e^{\L_m}[(t_{Re}+T_{\mathcal R})^{d_x}-t_{Re}^{d_x}]$ gives
\be\la{cxnormlimit}
\mathcal R^{1-d_x}c\longrightarrow d_xe^{\L_m/d_x}T
=d_x\,c_T^{1/d_x}~.
\ee
The normalization $\mathcal R^{1-d_x}$ is positive and independent of $r_m$, so, assuming the first flow derivative converges locally uniformly in $\L_m$, the normalized branch limit and $c_T$ have the same monotonicity and stationary points. In summary, the functions $c$ and $c_T$ have the same monotonicity along the flow.

%%%%%%%%%%%%%%%%%%%%%%%%%%%

\subsection{The $c$-function in Anisotropic Theories with a Number of Invariant Planes}

The entire analysis generalizes to $n$ invariant $SO(d_i)$ planes, labelled by $i=1,\ldots,n$, with
$\vec{x}_i$ spanning the $i$-th plane.  For each directional branch $c$-function $c_{x_i}$,
choose a fixed reference exponent $d_{x_i}>0$ and impose \eqref{IRclockprescription} separately:
$e^{\L_{m,x_i}/d_{x_i}}t_{i,Re}=\mathcal R_i$, with $\mathcal R_i$ fixed along the flow.

After removing the UV cutoff under the corresponding endpoint condition, the regulated flow is
\bea\la{c_tot2axi}
\frac{dc_{x_i}}{dr_m}&=&2e^{\L_{m,x_i}}T_{\infty,i}^{d_{x_i}-1}d_{x_i}\L_{m,x_i}' \bigg[\int_0^{\frac{T_{i,\mathcal R_i}}2}d\tau_i\,\frac{1}{\L_{x_i}'} \prt{\frac{\L_{x_i}'}{d_{x_i}}-\frac{\L_{x_i}''}{\L_{x_i}'}-B'}+\varepsilon_{\mathcal R_i}\bigg]~,
\eea
where
\be
\L_{x_i}(r):=B(r)+(d_i-1)A_{x_i}(r)+\sum_{\substack{j=1\\j\neq i}}^n d_j A_{x_j}(r)~.
\ee
Here $T_{i,\mathcal R_i}:=t_{i,Im}-t_{i,Re}>0$ and $T_{\infty,i}:=t_{i,Im}$.
The signed measure $d\tau_i$ and endpoint correction $\varepsilon_{\mathcal R_i}$ are defined by
\eqref{ctsignedmeasure} and \eqref{IRclockcorrection} using the geometry for the corresponding direction.
The geometric integrand is equal to $w_i$,  defined by \eqref{dxrdef} using $\L_{x_i}$. Under the corresponding convergence, endpoint and cutoff-motion conditions, $\varepsilon_{\mathcal R_i}\to0$.  

For each subspace, the UV-oriented monotonicity condition is  $dc_{x_i}/d\L_{m,x_i}\ge0$, or, with $\L_{m,x_i}'>0$,
\be
\frac{dc_{x_i}}{dr_m}\ge0~.
\ee
Thus, as rotational symmetry is broken further, the number of independent directional $c$-functions
and corresponding monotonicity conditions increases.

 %%%%%%%%%%%%%%%%%%%%%%%%%%%
\subsection{Role of \texorpdfstring{$d_x$}{dx}}\label{sec:dx}

The parameter $d_x$ plays a central role in the monotonicity of the timelike $c$-function. Physically, $d_x$ can be thought of as acting as an effective dimensionality of the QFT degrees of freedom that contribute to the timelike entanglement entropy in the direction where the strip is localized. This effective dimension can be extracted directly from the symmetry structure of the geometry whenever the bulk metric exhibits scaling behavior. At a scale-covariant point, its matching value is determined by the scaling of the timelike entanglement entropy in the localization direction. Along an interpolating flow, $d_x$ is held fixed, while the geometric effective dimension $d_{x_r}(r)$ may vary. 

Additionally, we can interpret $d_{x_r}$ as a local effective scaling exponent. Since $P=e^{\Lambda}$ is the conserved surface momentum and $\mu=e^B|\Lambda'|$ is a local inverse time scale, $d_{x_r}=d\ln P/d\ln\mu$. Thus $d_{x_r}$ measures the local scaling of $P$ with $\mu$;
and it can vary along an interpolation without changing the number
of spatial directions.

To illustrate this, it is convenient to introduce a radial coordinate $\tilde r$ related to the proper radial distance of \eq{metric1a} by $dr^2=\tilde r^{2\d}d\tilde r^2$, i.e.\ $r=\log\tilde r$ for $\d=-1$ and $r=\tilde r^{\d+1}/|\d+1|$  otherwise, and consider a commonly encountered class of scale-covariant geometries characterized by monomial metric elements. Such backgrounds can be viewed as a special subclass of \eq{metric1a} and, in certain cases, describe fixed points. They take the form
\be\la{metricpolyna}
ds^2=-\tilde r^{2\b}dt^2 +\tilde r^{2\a_1}d\vec{x}^{\,2} + \tilde r^{2\a_2}d\vec{y}^{\,2} +\tilde r^{2\d}d\tilde r^2~.
\ee
Under the anisotropic scaling transformation
\be\la{rescalingsa}
t\rightarrow\l^{\tilde\b}t~,
\qquad x\rightarrow\l^{\tilde a_1}x~,
\qquad y\rightarrow\l^{\tilde a_2}y~,
\qquad\tilde r\rightarrow\l^{-\tilde\d}\tilde r~,
\ee
the metric \eqref{metricpolyna} transforms covariantly when $\tilde a_i=\tilde\d\prt{\a_i-1-\d}$ and $\tilde\b=\tilde\d\prt{\b-1-\d}$. Although the power of the radial metric component can always be fixed by a further coordinate redefinition, we retain the general exponent $\delta$ in order to maintain uniform applicability across different holographic backgrounds and coordinate systems.

For a strip localized in the $x$-direction, the time-dependent part of the timelike entanglement entropy scales as
\be
S_x\sim\tilde r_m^{\a_1\prt{d_1-1}+\a_2d_2+\d+1} \sim T_x^{-\frac{\a_1\prt{d_1-1}+\a_2d_2+\d+1}{\b-\d-1}}~.
\ee
These scaling relations describe the interval-dependent part of
$S_x$, after subtracting interval-independent UV terms. When $\a_1(d_1-1)+\a_2d_2+\d+1=0$, 
the matching exponent is $d_x=1$ and this dependence is logarithmic. The imaginary contribution itself
is then independent of the interval. For $\b-\d-1\neq0$, scale invariance of $c_{x,T}$ fixes
\be\la{dxresulta}
d_x=\frac{\b+\a_1(d_1-1)+\a_2d_2}{\b-\d-1}~.
\ee
This provides a physical interpretation for the effective dimension $d_x$ based on the scaling symmetry of the metric. Similarly, for a strip localized in the $y$-plane, we obtain
\be\la{dxresultay}
d_y=\frac{\b+\a_1d_1+\a_2\prt{d_2-1}}{\b-\d-1}~.
\ee
These expressions provide a direct physical interpretation of $d_x$ and $d_y$:  they encode the scaling of the degrees of freedom probed by timelike entanglement at a fixed point of the RG flow.

Alternatively, the running effective dimension  evaluated at a radial scale $r_f$, gives the purely geometric expression
\be\la{dxr1a}
d_{x_r}(r_f):=\frac{\L'(r_f)^2}{\L''(r_f)+B'(r_f)\L'(r_f)}~,
\ee
for which the integrand \eq{c_tot2a} vanishes: $w(r_f)=0$ when $d_{x_r}(r_f)=d_x$. The fixed-point value in \eqref{dxresulta} agrees exactly with \eqref{dxr1a} for metrics of the form \eqref{metricpolyna}. With the matching reference exponent and the IR prescription \eqref{IRclockprescription}, the $c$-function is stationary in these scale-covariant geometries.

Finally, we can simplify the expression \eqref{c_tot2axi} using the geometric notion of the effective dimension:
\bea\la{c_final2a}
\frac{dc_x}{dr_m}=2e^{\L_{m,x}}T_{\infty,x}^{d_x-1}d_x\L_{m,x}' \prtt{\int_0^{\frac{T_{x,\mathcal R_x}}2}d\tau_x\, \prt{\frac1{d_x}-\frac1{d_{x_r}(r)}} +\varepsilon_{\mathcal R_x}}~,
\eea
with the analogous expression for $c_y$ obtained by replacing the $x$-indices by $y$-indices. This is the exact regulated branch flow. Its sign depends on the complete signed integral and the endpoint correction.  The result generalizes straightforwardly to any number of anisotropic subspaces.

Moreover, using \eqref{dxrdef} and \eqref{dxr1a}, we obtain the expression of the integrand $w$ in terms of the effective dimensionalities
\be\la{localdx} 
w=\ff{1}{d_x}-\ff{1}{d_{x_r}(r)}~.
\ee
Along a flow, we choose a fixed reference exponent $d_x>0$, usually matching the effective dimension of one fixed point. Where the running dimension is positive, the local condition $w\ge0$ requires $d_x\le d_{x_r}(r)$ throughout that region. Endpoint values alone do not establish this condition, since the running dimension may become smaller at an intermediate scale.   For the functions $c$ and $c_T$, this condition provides a sufficient criterion within the adiabatic class of Section~\ref{sec:adiab}. For Poincar\'e-invariant flows with
$d_x=d-1$, the local condition follows from the NEC; in general nonrelativistic flows it must be checked separately. Note that in principle, the choice of $d_x$ need not match either endpoint and therefore need not give finite nonzero fixed-point values.

A fixed reference exponent cannot match two unequal endpoint dimensions. At a regular scaling endpoint with
effective dimension $d_x^{\rm(fp)}$ and convergent integrals, $c_T\propto T^{d_x-d_x^{\rm(fp)}}$, $\cC\propto T_{\rm eff}^{d_x-d_x^{\rm(fp)}}$, and the
normalized branch limit scales as $T^{1-d_x^{\rm(fp)}/d_x}$. Finite nonzero endpoint values therefore require matching the reference exponent to the endpoint dimension. For unequal endpoint dimensions the construction may remain monotonic while diverging or vanishing at the unmatched endpoint; a fixed normalization scale cannot remove these powers. Our $c$-theorems in this work primarily aim at providing a function that is monotonic along the RG flow and takes a finite value at a fixed point, which is the standard requirement for holographic $c$-functions; the comparison of its values at two fixed points is an additional statement that requires the same scaling at both ends, as it does for any other entanglement-entropy $c$-function.

 %%%%%%%%%%%%%%%%%%%%%%%%%%%
\subsection{Monotonicity of the $c$-functions for Adiabatic Flows}
\label{sec:adiab}

The local effective-dimension condition $w\ge0$ does not by itself determine the sign of the branch flow,
because the measure in \eqref{c_tot2a} is signed. We now specify an additional rate condition on the
running geometry that makes this integral non-negative throughout the RG flow. This defines the adiabatic
class considered below. The IR prescription \eqref{IRclockprescription} fixes the regularization of the branch $c$-function and  the vanishing of the endpoint correction $\varepsilon_{\mathcal R}$ is independent of adiabaticity.

For a fixed reference exponent $d_x>0$ and a smooth geometry with $\L'\neq0$, define
\be\la{muQ}
\mu:=e^B|\L'|~,\qquad Q:=w\,\mu^{-1}~,
\ee
so that $d\ln\mu/d\L=1/d_{x_r}$. With $u:=\L-\L_m$, changing variables from $r$ to $\L$ in \eqref{t2} turns the signed flow integral of \eqref{c_tot2a} into
\be\la{Kdef}
\int_0^{T/2}d\tau\,w=\int_{-\infty}^{\infty}du\,K(u)\,Q(\L_m+u)~,
\qquad
K(u):=\frac{\Theta(-u)}{\sqrt{1-e^{2u}}}-\frac{1}{\sqrt{1+e^{2u}}}~,
\ee
where the first term is the timelike branch, present only for
$u<0$, and the second the spacelike branch, entering with the
opposite sign, we see that $K\sim e^{2u}$ in the IR, $K\sim-e^{-u}$ in the
UV, and $\int K\,du=0$. The cumulative kernel
\be\la{Mdef}
M(u):=\int_{-\infty}^{u}K(v)\,dv=
\begin{cases}
\displaystyle\ln\frac{1+\sqrt{1+e^{2u}}}{1+\sqrt{1-e^{2u}}},&u<0,  \\
\displaystyle\ln\frac{1+\sqrt{1+e^{2u}}}{e^{u}},&u\ge0,
\end{cases}
\ee
satisfies $M>0$ at finite $u$ and $M(\pm\infty)=0$.

At this point let us summarize the natural assumptions on the holographic RG flow we already have: (H1) $\L$ increases towards the UV, ranges over
$(-\infty,\infty)$, and $\mu\to\infty$ in the UV; the orientation of $\L$ is the one adopted in
Section~\ref{section2} for the reality of the timelike branch at arbitrary turning points.
(H2) The IR conditions \eqref{IRcond} hold. (H3) The endpoint conditions \eqref{IRendpoint} and
\eqref{boundarypp} hold, also after multiplication by $w$, so that $M\mu^{-1}\to0$ and $MQ\to0$ at both ends. The integrals in \eqref{Kdef} and \eqref{dtrma} converge, and differentiation with respect to $\L_m$ at fixed $u$ is valid under the integral sign. These are natural assumptions for holographic RG flows without IR wall.

Let us now introduce the notion of adiabaticity for the RG flow. We call a flow adiabatic if
\be\la{adiabcond}
\pp_\L Q\le0
\qquad\Longleftrightarrow\qquad
\frac{dw}{d\L}\le\frac{w}{d_{x_r}}
\ee
throughout the geometry: the ratio $w/\mu$ must not increase towards the UV. For $\L'>0$ this reads $dQ/dr\le0$, i.e.
\be\la{adiabcond22}
w'\le w\Big(B'+\frac{\L''}{\L'}\Big)~,
\ee
and for $\L'<0$ the radial inequality reverses. Where $w>0$ and $d_{x_r}>0$ it reads $d\ln w/d\ln\mu\le1$, i.e.\ the positive departure of the running dimension from the reference value grows no faster than the probing energy scale $\mu$.

To summarize so far: for adiabatic flows satisfying (H1)-(H3), the local condition $w\ge0$ throughout the geometry is sufficient for monotonicity of the branch $c$-function in the normalized regulator limit; where $d_{x_r}>0$ it is equivalently $d_x\le d_{x_r}$.

Adiabaticity supplies the control needed because the measure $d\tau$ is signed. Integrating \eqref{Kdef} by parts, with the endpoint terms vanishing under (H3),
\be\la{Kparts}
\int_0^{T/2}d\tau\,w=-\int_{-\infty}^{\infty}d\L\,M(\L-\L_m)\,\partial_\L Q\;\ge0~,
\ee
the sign following from $M>0$ and \eqref{adiabcond}. Since the prefactor of \eqref{c_tot2a} divided by $\L_m'$, namely $2e^{\L_m}T_\infty^{d_x-1}d_x$, is positive, this gives $dc/d\L_m\ge0$ in the normalized regulator limit, at every turning point and separately in each localization direction.  The same integral controls the boundary $c$-function,
\be\la{exactflowL}
\frac{d\ln c_T}{d\L_m}=-\frac{2d_x}{T}\int_{-\infty}^{\infty}d\L\,M(\L-\L_m)\,\partial_\L Q\;\ge0~.
\ee
Outside the adiabatic class $w\ge0$ alone does not fix the sign of \eqref{Kparts}.

In summary, for adiabatic RG flows with respect to \eqref{adiabcond}, the monotonicity of the $c$-function is guaranteed as long as the integrand of \eq{c_tot2a}  is non-negative throughout the geometry. Thus, $w\ge0$ along the adiabatic RG flow guarantees a monotonic $c$-function.

 %%%%%%%%%%%%%%%%%%%%%%%%%%%
\section{Additional $c$-function Properties along the RG Flow}
\label{section4}

So far, we have analyzed the evolution of the timelike $c$-function through its derivative with respect to the
turning point $r_m$. An alternative viewpoint is to ask how it changes with the boundary interval
$T$. This describes the same RG flow whenever increasing $T$ probes lower energy scales. 
We work at finite IR regulator, with $T$ denoting the regulated interval and derivatives taken at fixed
$\mathcal R$ under \eqref{IRclockprescription}. The UV cutoff has been removed. 

The two derivatives are related by
\be\la{dcdt}
\frac{\pp c}{\pp T} =\frac{\pp c}{\pp r_m}\frac{\pp r_m}{\pp T}~,
\ee
where the first factor follows from \eqref{ctot1}.
The interval derivative is given by
\eqref{dtrmaregulated}:
\be
\frac{\pp T}{\pp r_m}=2\L_m'\prtt{-\frac{T}{2d_x}+\int_0^{\frac T2}d\tau\, \frac1{\L'}\prt{\frac{\L'}{d_x}-\frac{\L''}{\L'}-B'}+\varepsilon_{\mathcal R}}~.
\ee
The integral has the signed measure \eqref{ctsignedmeasure}, and $\varepsilon_{\mathcal R}$ is the IR endpoint
correction \eqref{IRclockcorrection}.

Combining the exact regulated flow \eqref{c_tot2a00} with the interval derivative, we obtain
\be\la{dettt}
\frac{\pp c}{\pp T}=\frac{d_x\,\pp_{r_m}c}{-T\L_m'+e^{-\L_m}T_\infty^{1-d_x}\pp_{r_m}c}~.
\ee

For $\L_m'>0$, the c-theorem requires $\pp_{r_m}c\ge0$. Increasing $T$ does not, a priori,
guarantee that lower energy scales are being probed. We therefore retain the orientation
$\pp T/\pp r_m<0$. With this orientation, $\pp_{r_m}c\ge0$ is equivalent to $\pp_Tc\le0$.
If $T(r_m)$ changes orientation, parametrization by $T$ must instead be considered on its individual
monotonic branches.

For $d_x>0$, combining radial monotonicity with this interval orientation gives
\be\la{sand}
0\le\pp_{r_m}c < e^{\L_m}T_\infty^{d_x-1}T\L_m'~.
\ee
The lower bound expresses the $c$-theorem. Combining \eqref{c_tot2a00} and \eqref{dtrmaregulated} shows that the upper bound is equivalent to $dT/dr_m<0$: it does not follow from radial monotonicity, but from the requirement that a larger boundary interval probes lower energy scales. Under this physical requirement it is a genuine bound on the rate of change of the $c$-function, with a right-hand side fixed entirely by the bulk geometry, and it holds along any flow with this orientation, whether or not the flow is adiabatic. At upper-bound saturation, $dT/dr_m=0$ and the inverse parametrization in \eqref{dettt} is singular. Upon removing the regulator, the positive normalization \eqref{cxnormlimit} must be applied to both sides. Strictness of the limiting upper bound requires $dT/dr_m<0$ for the physical interval.

Let us  elaborate further on the bound \eq{sand} and assess its sensibility and consistency with the c-theorem. We rewrite \eq{dettt} in terms of the metric data: 
\be 
\frac{\pp c}{\pp T}= T_{\infty}^{d_x-1} e^{\L_m} d_x \prtt{1-\frac{T}{2 d_x \prt{  \int_0^\frac{T}{2} d\tau \frac{1}{\L'}\prt{\frac{\L'}{d_x}-\frac{\L''}{\L'}-B'}+\varepsilon_{\mathcal R}}}}^{-1}~.
\ee
Therefore for positive $d_x$ and wherever $\pp_{r_m}c>0$, we have:
\be \la{up1}
\frac{T}{2d_x \int_0^\frac{T}{2} d\tau \frac{1}{\L'}\prt{\frac{\L'}{d_x}-\frac{\L''}{\L'}-B'}+2d_x \varepsilon_{\mathcal R}}> 1~,
\ee
where the value of the integral is constrained with respect to the rescaled time interval $T/d_x$. We recast inequality \eqref{up1} in terms of the effective dimension $d_{x_r}$, defined in equation \eqref{dxr1a}, for $\varepsilon_{\mathcal R}\to 0$  yielding
\be \la{up1b}
\frac{T}{2  \int_0^\frac{T}{2} d\tau \prt{1-\frac{d_x}{d_{x_r}}}}>1~.
\ee
When the c-theorem holds the integral appearing in the above expression is non-negative and we can write it as:
\be\la{up2}
\int_0^\frac{T}{2} d\tau \prt{1-\frac{d_x}{d_{x_r}}}< \frac{T}{2}~.
\ee
 
In summary, the conditions ensuring monotonicity of $c(r_m)$ ensure the monotonicity of $c(T)$ as well, provided the increasing boundary interval $T$ probes lower energy scales. The two formulations then define the same RG monotone, expressed in bulk-geometric or boundary-temporal variables.   The strict upper bound in \eqref{sand} expresses this interval orientation as a bound on the radial rate of change.

 %%%%%%%%%%%%%%%%%%%%%%%%%%%
\section{Anisotropic Null Energy Conditions and Thermodynamic Stability}
\label{section5}

We now impose the NEC on the holographic RG flow.  The NEC implies that the contracted Ricci tensor with the null vectors $\xi^\mu$ satisfies: $R_{\m\n} \xi^\m \xi^\n\ge0$. Assuming that spatial isotropy in the theory is broken into two anisotropic planes, the NEC leads to two   functions $g_i(r)$ that are non-decreasing as a function of the radial coordinate $r$ along the RG flow. Application to the metric \eqref{metric1a} yields the following conditions \cite{Chu:2019uoh}:
\be \la{nec1a}
g_i'(r):=\prt{\prt{B'(r)-A_i'(r)}e^{K(r)+B(r)}}'\ge0\Rightarrow\prt{\prt{B'(r)-A_i'(r)}e^{d_1 A_1(r)+d_2 A_2(r)+B(r)}}'\ge0~,
\ee
where $K(r)$ is defined above \eqref{t1a}. The number of such non-decreasing functions $g_i$ corresponds to the number of rotational subgroups into which the original symmetry group $SO(d_1 + d_2)$ is broken. For the metric \eqref{metric1a}, we have $i = 1, 2$. The third NEC is
\be \la{nec3a}
N_3:=-d_1 A_1'(r)^2-d_2 A_2'(r)^2+B'(r) K'(r)-K''(r)\ge0~,
\ee
which implies the existence of a non-decreasing function 
\be
h(r):=-K'(r)e^{\frac{K(r)}{d_1+d_2}-B(r)}~,
\ee
with $h'(r)\ge0$ along the RG flow. These monotonicity statements refer to the chosen radial coordinate; its orientation relative to
the UV is specified separately.

We next consider the specific heat of a family of black-brane solutions of the same theory, with the same asymptotics and fixed boundary sources. The metric takes the form
\be\la{metric1t}
ds^2_{d+1}
=-e^{2B(r)}f(r)dt^2 +e^{2A_1(r)}d\vec{x}^2 +e^{2A_2(r)}d\vec{y}^2 +\frac{dr^2}{f(r)}~,
\ee
where $f(r_h)=0$ at a nonextremal horizon. Regularity of the Euclidean geometry gives the Hawking temperature
\be
T_H=\frac{1}{4\pi} \sqrt{(fe^{2B})'f'}\bigg|_{r_h} =\frac{e^{B(r_h)}|f'(r_h)|}{4\pi}~.
\ee
The thermal entropy density is proportional to the horizon
area. Suppressing its constant normalization, we write
\be
S_{th}=e^{K(r_h)}=e^{d_1 A_1(r_h)+d_2 A_2(r_h)}~.
\ee
We use $T_H$ to distinguish the temperature from the boundary
interval $T$. The stability condition considered here is
non-negative specific heat, $T_H\,dS_{th}/dT_H\ge0$; this does not establish stability against all other fluctuations, but suffices for our purposes.

If we take the exterior towards increasing $r$, so that $f'(r_h)>0$, and that the warp factors have no explicit
dependence on the family parameter $r_h$, the condition becomes
\be \la{stablea}
 \frac{1}{4 \pi} e^{B(r_h)-\prt{d_1 A_1(r_h)+d_2 A_2(r_h)}} \frac{\pp S_{th}}{\pp T_H}=\frac{d_1 A_1'(r_h)+d_2 A_2'(r_h)}{\pp_{r_h}f'(r_h)+f'(r_h) B'(r_h)}\ge 0~.
\ee
Here $f'(r_h)$ denotes the radial derivative of $f$ evaluated at the horizon, while $\pp_{r_h}f'(r_h)$ is its total derivative along the thermal family. This form applies to the scaling thermal families considered below.

Together, the directional NEC conditions \eqref{nec1a}, the radial NEC \eqref{nec3a}, and the specific-heat condition \eqref{stablea} constitute natural consistency constraints for well-defined theories and 
restrict the admissible backgrounds and scaling exponents considered below.

%%%%%%%%%%%%%%%%%%%%%%%%%%%
\section{An alternative c-Theorem from the Timelike Branch function $\cC$}
\label{section:C}

The functions $c$ and $c_T$ use the physical interval $T=t_{Im}-t_{Re}$, whose flow combines the two branches with opposite signs. Adiabaticity controls this signed contribution, but restricts the class of geometries. A natural question is then whether one can construct another physically motivated $c$-function whose monotonicity depends only on the condition $w\ge0$, valid for any RG flow without the need of adiabaticity. The answer is affirmative and follows directly from the analysis of Section~\ref{section3}. We now construct and briefly discuss such a $c$-function, $\cC$, whose monotonicity follows from the local effective-dimension condition alone.

The timelike branch supplies two intrinsic quantities: the finite imaginary area $S_{Im}$ and its conserved momentum $P=e^{\L_m}$. The imaginary area is finite under the first condition in \eqref{IRcond}, even when the branch time extent diverges. These quantities therefore allow a construction that requires neither the cancellation of divergent branch times nor a prescription fixing $c_{Re}$. The construction here is motivated by AdS$_3$, where the imaginary part of the timelike entanglement entropy is $\pi c/6$, independent of the interval; in higher dimensions $S_{Im}$ depends on $T$, and $\cC$ is the combination that becomes constant at fixed points.
 
%%%%%%%%%%%%%%%%%%%%%%%%%%%
\subsection{Definition and exact flow}

The normalized area and momentum, at a fixed point with the matching reference exponent $d_{x_r}=d_x$, scale as
$S_{Im}\propto T^{1-d_x}$ and $P\propto T^{-d_x}$. Their ratio thus has the scaling of a time interval. 
For a fixed reference exponent $d_x>0$, we define
\be\la{Teffdef}
T_{\rm eff}:=\frac{S_{Im}}{2}\Big(\frac{\pp S_{Im}}{\pp t_{Im}}\Big)^{-1}=\frac{S_{Im}}{2}e^{-\L_m}~,\qquad
\cC:=T_{\rm eff}^{d_x}\,\frac{\pp S_{Im}}{\pp t_{Im}}=\Big(\frac{S_{Im}}2\Big)^{d_x}e^{\L_m\prt{1-d_x}}~,
\ee
where we have used \eqref{derima}, evaluated at fixed IR cutoff and independent of it. Thus $\cC$ has the form of the branch function $c_{Im}$ in \eq{cfunctiona}, with the divergent branch time $t_{Im}$ replaced by the finite $T_{\rm eff}$, and equivalently the form of $c_T$ \eq{cf0Ta} with $T$ replaced by $T_{\rm eff}$.

Equations \eqref{s2a} and \eqref{t2} give
\be\la{Teffint}
T_{\rm eff}=\int_0^{t_{Im}/2}dt\, e^{2(\L-\L_m)}\Big|_{Im}
=\int_0^{\pi/2}d\theta\, \frac{\sin\theta}{\mu(\L_m+\ln\sin\theta)}~,
\ee
where we change coordinates $e^{\L-\L_m}=\sin\theta$, since $\L\le\L_m$ along the timelike branch, and $\mu=e^B|\L'|$, defined in \eq{muQ} is viewed as a function of $\L$. We again consider the natural assumptions of a smooth geometry with $\L'\neq0$, $\L$ increasing towards the UV and
$\L\to-\infty$ in the IR, a finite imaginary area, and differentiation under the integral sign. For differentiation under the integral sign it suffices that $1/d_{x_r}$ is bounded on the  branch, which holds for
the regular IR asymptotics considered here. The finite-area requirement is the first condition
in \eqref{IRcond}. At a specified turning point, the definition uses only the timelike branch and is independent of its joining to the spacelike branch.

The weight $\sin\theta/\mu$ in \eqref{Teffint} is positive. We write averages over it as
\be\la{avgdef}
\langle f\rangle:=\frac1{T_{\rm eff}}\int_0^{\pi/2}d\theta\,\frac{\sin\theta}{\mu}\,f~,
\ee
where $\langle1\rangle=1$, and $f$ and $\mu$ evaluated at $\L=\L_m+\ln\sin\theta$, by definition of $\theta$. Using $\pp_\L\ln\mu=1/d_{x_r}$, it follows that
\be\la{dTeff}
\frac{dT_{\rm eff}}{d\L_m}=-T_{\rm eff}\Big\langle\frac1{d_{x_r}}\Big\rangle~.
\ee
Hence, using \eq{Teffdef} and \eq{avgdef}
\be\la{Cflow}
\frac{d\cC}{dr_m}=d_x\L_m'\,\cC\,\langle w\rangle~.
\ee
The flow of $\cC$ is a positive average of $w$ over the timelike branch. Therefore $w\ge0$ between the IR and the turning point implies
\be\la{Cthm}
\frac{d\cC}{d\L_m}\ge0~ \Longleftrightarrow~ \L_m'\frac{d\cC}{dr_m}\ge0~.
\ee
Hence $\cC$ decreases from the UV to the IR. This uses neither adiabaticity nor $d_{x_r}>0$, and, for continuous $w\ge0$, $d\cC/d\L_m=0$ only when $w=0$ along the entire timelike branch between the IR and the turning point.

%%%%%%%%%%%%%%%%%%%%%%%%%%%
\subsection{Bound on the Flow Rate}

The upper bound \eqref{sand} has an exact counterpart for $\cC$. From \eqref{Cflow} and \eq{avgdef},
\be\la{sandC}
\frac{d\ln\cC}{d\L_m}=d_x\langle w\rangle=1-d_x\Big\langle\frac1{d_{x_r}}\Big\rangle~,
\ee
so if $w\ge0$ and $d_{x_r}>0$ between the IR and the turning point,
\be\la{Cbound}
0\le\frac{d\ln\cC}{d\L_m}<1~.
\ee
The upper bound is $dT_{\rm eff}/d\L_m<0$ by \eqref{dTeff}: $\cC$ decreases as $T_{\rm eff}$ increases. 
%%%%%%%%%%%%%%%%%%%%%%%%%%%
\subsection{Fixed points}

In an exactly scale-covariant geometry with the matching reference exponent, $w=0$ and $d\ln\mu/d\L=1/d_x$, so $\mu=\mu_*e^{\L/d_x}$ with constant fixed-point value  $\mu_*>0$. Then \eqref{Teffint} gives $T_{\rm eff}=J(d_x)/\mu(r_m)$ and
\be\la{Cfp}
\cC_*=\Big[\frac{J(d_x)}{\mu_*}\Big]^{d_x}~,\qquad
J(d_x):=\int_0^{\pi/2}d\theta\,(\sin\theta)^{1-1/d_x}=\frac{\sqrt\pi}{2}\frac{\Gamma(1-\frac1{2d_x})}{\Gamma(\frac32-\frac1{2d_x})}~,
\ee
convergent for $d_x>1/2$. At either asymptotic fixed point, if the fixed reference exponent $d_x$ matches the effective dimension, then $\cC$ approaches $\cC_*$. A single choice of $d_x$ gives finite
nonzero values at both endpoints only if their effective dimensions agree.

For example, for AdS$_{d+1}$ of radius $L$, $d_x=d-1$ and $\mu_*=(d-1)/L$. Restoring the entropy normalization,
\be\la{CAdS}
\cC_{AdS}=\frac{V_{d-2}}{4G_N}\Big[\frac{L}{d-1}J(d-1)\Big]^{d-1}\propto\frac{L^{d-1}}{G_N}~.
\ee
In $d=2$, with $c_{\rm CFT}=3L/2G_N$,
\be\la{C2d}
\cC_{AdS_3}=\frac{\pi L}{8G_N}=\frac{\pi c_{\rm CFT}}{12}=\frac12|{\rm Im}\,S_T|~,
\ee
in agreement with \cite{Doi:2023zaf}.

%%%%%%%%%%%%%%%%%%%%%%%%%%%
\subsection{Relativistic and anisotropic flows}

For Poincar\'e-invariant flows, $d_x=d-1$, $\L=(d-1)A$ and $B=A$, so the NEC gives
\be\la{Cpoinc}
w=-\frac{A''}{(d-1)A'^2}\ge0~.
\ee
Hence $\cC$ is monotonic along every such flow satisfying the assumptions above, including where $d_{x_r}<0$, and without adiabaticity.

For flows between AdS fixed points with the same entropy normalization,
\be\la{Cratio}
\frac{\cC_{UV}}{\cC_{IR}}=\Big(\frac{L_{UV}}{L_{IR}}\Big)^{d-1}\ge1~,
\ee
the ratio of the holographic central charges. In $d=2$, $\cC=S_{Im}/2$, so the imaginary part of the timelike entropy itself decreases from $\pi c_{UV}/6$ to $\pi c_{IR}/6$. Thus $\cC$ is monotonic under the same condition as the entanglement-entropy $c$-functions of \cite{Myers:2012ed,Chu:2019uoh}.

For anisotropic flows the construction applies separately to each direction, with $\L_{x_i}$ and $d_{x_i}$, and $\cC_{x_i}$ is monotonic whenever $w_i\ge0$ between the IR and the turning point. These directional conditions are discussed in Section~\ref{section6}.

 %%%%%%%%%%%%%%%%%%%%%%%%%%%
\section{Timelike $c$-functions in Holographic RG Flows}
\label{section6}

We apply the preceding results to the directional branch $c$-functions $c_x,c_y$ of \eqref{cfunctiona}, including briefly the backgrounds considered in \cite{Giataganas:2025div}.   Their monotonicity in the regulator limit is understood as in Section~\ref{sec:derivc}, with the positive normalization independent of the turning point. Equation \eqref{cxnormlimit} transfers the monotonicity criteria for $c_x,c_y$ to $c_T$. For interpolating flows, the local conditions $w_x,w_y\ge0$ guarantee monotonicity of $\cC_x,\cC_y$ under the assumptions of Section~\ref{section:C}; for the functions $c_x,c_y$ they are sufficient within the adiabatic class satisfying (H1)-(H3) of Section~\ref{sec:adiab}, with monotonicity understood in the normalized regulator limit and the IR regulator removed under the conditions of Section~\ref{sec:derivc}. 

%%%%%%%%%%%%%%%%%%%%%%%%%%%
\subsection{Poincar\'e-invariant Flows}

As a warm-up, and to demonstrate the holographic timelike $c$-function in the simplest setting, let us first consider Poincaré-invariant theories before turning to anisotropic cases. In this isotropic context we have $B(r)=A(r)$, and the RG flow is fully captured by the behavior of the conformal factor $A(r)$ where the bulk geometry is described by the metric \eqref{metric1a}. Such geometries arise naturally, for example, in Einstein–dilaton setups, where the scalar field has a nontrivial radial profile as in the  holographic RG flows of \cite{Girardello:1998pd, Freedman:1999gp}.

For Poincaré-invariant theories, there is a single $c$-function: $c_x = c_y = c$, with $d_x = d_y$, and the timelike $c$-function reduces to the form introduced in \cite{Giataganas:2025div}. From  \eqref{dxr1a}, the effective dimension is
\be \la{dxp}
d_{x_r}=\frac{\prt{d-1}A'^2}{A''+A'^2}~.
\ee
The NEC \eqref{nec3a} reduces to the simple condition $A''\le 0$. At saturation ($A''=0$), the effective dimension becomes background-independent and, wherever $d_{x_r}>0$, is the minimum
\begin{equation}
d_{x_r}=d-1~.
\end{equation}
For instance this value is appropriate to the UV fixed point of a CFT dual to asymptotically AdS space. For $d_x=d-1$ and $A'>0$, the adiabatic condition \eqref{adiabcond} becomes
\be\la{adiabpoinc}
A'A'''\ge A'^2A''+3A''^2~.
\ee
The evolution equation \eqref{c_final2a} for the $c$-function, with the endpoint correction dropped as in \eqref{c_tot2a} and the regulator removed in the normalized limit \eqref{cxnormlimit}, is
\be\la{poincare1a}
\frac{\pp c}{\pp r_m}\simeq 2e^{(d-1)A_m} ~T_{\infty}^{d_x-1} (d-1) A_m'\int_0^\frac{T}{2} d\tau \prt{1-\frac{d_x}{d_{x_r}}}~.
\ee
Let us examine the sign of the above derivative. The NEC: $A'' \le 0$, guarantees that $A'(r)$ is a non-increasing function of $r$ along the RG flow. To show this, notice that the UV boundary of the theory lies at $r\rightarrow \infty$ where $A_{UV}\rightarrow \infty$ and $A_{UV}'=A'_{min}$, and therefore $A_{min}'\ge0$. It follows that $A'\ge0$ for the entire RG flow. Thus the overall prefactor in \eqref{poincare1a} is manifestly non-negative. The sign of $\partial_{r_m}c$ is therefore controlled by the signed integral in \eq{poincare1a}. For adiabatic flows satisfying \eqref{adiabcond} the c-theorem is satisfied as long as the NEC holds and $d_x\le d_{x_r}$ for positive effective dimensions: the integrand is then non-negative and the $c$-function monotonically increases with $r_m$, or equivalently decreases toward the IR along the RG flow.

We can further illustrate the c-theorem explicitly with an example. Consider a theory with $A'' = 0$ at a fixed point, such as a CFT in the UV fixed point with a gravity dual that asymptotically approaches AdS. Then $d_{x_r}$ is bounded from below by its UV value $d_{x_r}^{(UV)}=d-1$ along the RG flow, since via \eqref{dxp} $A''\le0$ and we focus on positive effective dimensions (i.e. $A'^2 + A'' >0$). In this case we take $d_x :=d_{x_r}^{(UV)}= d - 1$ at the UV fixed point, and for adiabatic flows the c-theorem $\pp_{r_m} c\ge 0$ holds since
\be \la{drcmp}
\frac{\pp c}{\pp r_m}\simeq 2e^{(d-1)A_m} ~T_{\infty}^{d-2} (d-1) A_m'\int_0^\frac{T}{2} d\tau \prt{1-\frac{d-1}{d_{x_r}}}~,
\ee
and $d_{x_r} \ge d - 1:=d_{min}$ and $A_m' \ge 0$ due to the NEC.

This statement can also be made manifest by expressing the derivative \eq{drcmp} in terms of the metric variables:
\be
\frac{\pp c }{\pp r_m}\simeq 2e^{(d-1)A_m} ~T_{\infty}^{d-2} (d-1) A_m'\int_0^{T/2}d\tau\frac{\prt{-A''}}{A'^2}~,
\ee
where the NEC, $A''\le 0$, makes the integrand non-negative; this holds also where $d_{x_r}<0$. Hence, for any Poincaré-invariant theory with a conformal UV fixed point that satisfies the NEC and the adiabatic condition \eqref{adiabpoinc}, the timelike entanglement entropy $c$-function exhibits the expected monotonic behavior along the RG flow. The timelike branch function $\cC$ is monotonic under the NEC alone, Section~\ref{section:C}.

The NEC alone, however, does not guarantee monotonicity of $c$. For an Einstein-scalar flow approaching a regular AdS IR fixed point, the scalar backreaction determines the leading metric correction through $-A''\propto\|\phi'\|^2$, so an irrelevant mode of dimension $\Delta_{IR}>d$ gives $w\propto\mu^{2(\Delta_{IR}-d)}$. Although the NEC ensures $w\ge0$, the adiabatic condition fails near the IR when $\Delta_{IR}>d+\tfrac12$, and for $d+\tfrac12<\Delta_{IR}<\tfrac{3d}{2}$ the exact signed flow has
a negative UV-oriented derivative near the IR (Appendix~\ref{appA}). An explicit counterexample is the
supersymmetric flow of \cite{Freedman:1999gp} from $\mathcal N=4$ super Yang-Mills theory to the Leigh-Strassler fixed point: it satisfies the NEC, the normalized branch limit is not monotonic, whereas $\cC$ is monotonic. The numerical comparison is presented in Appendix~\ref{sec:fgpwsec}.

%%%%%%%%%%%%%%%%%%%%%%%%%%%
\subsection{Anisotropic RG flows}

We now turn to the more challenging case of anisotropic Lifshitz-like theories, where different spatial directions scale with distinct exponents and the metric functions take the form
\be\la{ansa1}
B(r)=z_0 r~,\quad A_1(r)=z_1 r~, \quad A_2(r)=z_2 r~,
\ee
with $z_i$ denoting the characteristic anisotropic Lifshitz-like exponents, which encode Lorentz-symmetry breaking and the degree of rotational symmetry breaking. In Poincaré coordinates, using the rescaling relations \eqref{rescalingsa}, one finds  $\a_i=\tilde{a_i}=z_i$, $\b=\tilde{\b}=z_0$ and $\delta=-\tilde{\delta}=-1$.  

From equation \eqref{dxresulta}, the effective dimensions  in the $x$- and $y$-directions are:
\be \la{dxlifaniso}
d_{x}=1+\frac{z_1(d_1-1)+z_2 d_2}{z_0}~\qquad \mbox{and}\qquad d_{y}=1+\frac{z_1 d_1+z_2 (d_2-1)}{z_0}~.
\ee
The parameters $z_i$ are subject to the NEC \eqref{nec1a} and \eqref{nec3a}, imposing nontrivial inequalities among them and restricting the space of consistent anisotropic RG flows. The effective dimensions characterize how the number of degrees of freedom scales in each anisotropic direction, and they play a central role in the monotonicity of the corresponding $c$-functions. In the following, we proceed by studying theories of different amounts of symmetry breaking and analyze the monotonicity of the corresponding timelike $c$-functions in each case. 

%%%%%%%%%%%%%%%%%%%%%%%%%%%
\subsubsection{Lifshitz-like Anisotropic Theories}
The simplest nontrivial anisotropic class arises when $z_0=z_1=z$ and $z_2=1$. The NECs \eqref{nec1a} and \eqref{nec3a} reduce to a single inequality:
\be\la{necanisoz} 
z\ge 1~.
\ee
Rotational symmetry is broken, so in general $c_x \neq c_y$ and $d_x \neq d_y$. From \eq{dxlifaniso} we obtain
\be \la{dxlifanisoa}
d_{x}=d_1+\frac{d_2}{z}~,\qquad d_{y}=1+d_1+\frac{d_2-1}{z}~.
\ee
The corresponding functions entering the c-flow equation are
\be 
\L_x(r)=(d_1 z+d_2)r~,\qquad \L_y(r)=\prt{\prt{d_1+1}z+d_2-1}r~.
\ee
For two nonempty spatial planes, the NEC
\eqref{necanisoz} ensures
\be\la{condaniso0}
\L_{\prt{x,y}}'>0~,\qquad d_{\prt{x,y}}>\frac12~.
\ee
Thus $\L_x,\L_y$ increase towards the UV and can tend to $-\infty$ as $r\to-\infty$.   Therefore, the orientation and IR convergence conditions hold.
The anisotropic theory characterized by a constant Lifshitz exponent $z$ is scale invariant, thus the derivative of the $c$-functions $c_x$ and $c_y$ is given by \eq{c_final2a} and for $d_{x_r}=d_x$ and $d_{y_r}=d_y$  yields
\be
\frac{\pp c_{(x,y)}}{\pp r_m}= 0~,
\ee 
as expected.  The matching functions $c_x,c_y$ and $\cC_x,\cC_y$ are constant under \eqref{IRclockprescription}, by \eqref{c_tot2a} and \eqref{Cflow}.

For nontrivial RG flows in which the theory interpolates between fixed points of different symmetry, i.e. different $z$, the NEC ensures $\L_{x,y}'>0$, Eq.~\eqref{condaniso0}, so that $\L_{x,y}$ increase towards the UV boundary at $r\to\infty$ and the radial criterion $\pp_{r_m}c\ge0$ is the UV-oriented one. The monotonicity of the $c$-functions is then governed by the local effective-dimension conditions
\[
d_{x_r}\ge d_{x}~,\qquad d_{y_r}\ge d_{y}~,
\]
which guarantee monotonicity of $\cC_x,\cC_y$ and, within the adiabatic class of Section~\ref{sec:adiab}, of $c_x,c_y$.

Let us illustrate these conditions in a convenient class of theories where the computation is tractable analytically. To describe an interpolation, specify a running extension of the metric. We consider a theory with a running Lifshitz exponent $z_r:=z_r(r)$ (we use the notation $z_r(r)$ for the scaling exponent along the RG flow, to avoid confusion with $z$, which is fixed and independent of $r$), with fixed-point values $z_{UV}$ and $z_{IR}$, and $B'=A_1'=z_r(r),$ $A_2'=1~.$

The parameters $d_x,d_y$ are fixed and evaluated at a fixed point, while $d_{x_r},d_{y_r}$ are their running counterparts at radial position $r$. Let $\L_x'=d_1z_r+d_2$ and $\L_y'=(d_1+1)z_r+d_2-1$.  
Choosing the IR fixed point to set $d_x,d_y$ in \eqref{dxlifanisoa}, so that $z:=z_{IR}$, the local functions read
\[
w_x=\frac{d_2\,(z-z_r)}{z\,d_x\L_x'}-\frac{d_1z_r'}{\L_x'{}^2}~,\qquad
w_y=\frac{(d_2-1)(z-z_r)}{z\,d_y\L_y'}-\frac{(d_1+1)z_r'}{\L_y'{}^2}~,
\]
so that for a sufficiently slowly varying exponent the flows \eqref{c_final2a} reduce to
\[
\frac{\pp c_x}{\pp r_m}\sim \L_{m,x}'\,d_2\int_0^{\frac T2}d\tau\,\frac{z-z_r}{z\L_x'}~,\qquad
\frac{\pp c_y}{\pp r_m}\sim \L_{m,y}'\,(d_2-1)\int_0^{\frac T2}d\tau\,\frac{z-z_r}{z\L_y'}~,
\]
and $z\ge z_r$ is the condition for non-negative integrands; while the $z_r'$ terms above are the exact corrections. Hence the conditions  
\be\la{zrcond}
1\le z_r\le z_{IR}~,\qquad z_r'\le0~,
\ee
give $w_x,w_y\ge0$ exactly. 
The radial NEC is $d_2(z_r-1)-d_1z_r'\ge0$, which follows from \eqref{zrcond}. The remaining nontrivial NEC is $z_r'+(z_r-1)((d_1+1)z_r+d_2)\ge0~,$ which bounds the rate at which $z_r$ decreases towards the UV. Adiabaticity is a separate restriction and for this running geometry, \eqref{adiabcond} reads
\be\la{adiabcondlif}
w_x'\le\left(z_r+\frac{d_1z_r'}{\L_x'}\right)w_x~,
\qquad
w_y'\le \left(z_r+\frac{(d_1+1)z_r'}{\L_y'}\right)w_y~.
\ee
These conditions involve $z_r''$ through $w_x',w_y'$ and do not follow from the NEC bound in general.

The local inequalities $w_x,w_y\ge0$ guarantee monotonicity of $\cC_x,\cC_y$. For the $c$-functions $c_x,c_y$, they are sufficient within the adiabatic class \eq{adiabcondlif}. Monotonicity is then established at every turning point in the normalized regulator limit. Note that slow variation of $z_r$ establishes adiabaticity away from the fixed points, where $w$ is of order one, but near the IR, where $w\to0$, slowness must be measured relative to the scale $\mu$. This can be understood as follows: away from the fixed points, where $w$ remains bounded away from zero, sufficiently small first and second derivatives of $z_r$ satisfy the displayed rate inequalities. Suppose the exponent approaches its IR value as $z_{IR}-z_r\propto e^{\nu r}$ for $r\to-\infty$, with $\nu>0$. Then $w\propto e^{\nu r}$ while $\mu\propto e^{z_{IR}r}$, so $w\propto\mu^{\nu/z_{IR}}$ and the adiabatic condition holds near the IR for $\nu<z_{IR}$ and fails for $\nu>z_{IR}$; at $\nu=z_{IR}$ subleading terms are needed to decide.

A simple, physically relevant scenario is a slowly and monotonically decreasing $z_r(r)$, with $z_{UV}\ge1$ at the UV boundary $r\to\infty$ and an IR value $z_{IR}>z_{UV}$. Monotonicity gives $1\le z_r\le z_{IR}$ and $z_r'\le0$ along the flow, so with the IR reference exponents, $z:=z_{IR}$, the conditions \eqref{zrcond} hold and $w_x,w_y\ge0$ exactly; hence $\cC_x,\cC_y$ are monotonic. The running NEC is satisfied for slow variation, where the $z_r'$ term is negligible and it reduces to $z_r\ge1$. For the branch functions the rate inequalities \eqref{adiabcondlif} must also hold throughout the flow: away from the fixed points they hold for sufficiently slow
variation relative to the local values of $w_x,w_y$, and near the IR they hold for $\nu<z_{IR}$, with the equality case determined by subleading terms. For regular asymptotic scaling endpoints, $\L_{x,y}'>0$ and
$d_x,d_y>1$ ensure the endpoint convergence required by
(H1)--(H3). Under the above rate conditions, $c_x,c_y$ and $c_T$ are therefore monotonic. When $z_{UV}\neq z_{IR}$ the endpoint dimensions \eqref{dxlifanisoa} differ, and the qualifications of Section~\ref{sec:dx} apply: with the IR reference exponents the functions are finite at the IR endpoint and behave as the powers stated there towards the UV, so that in directions with $d_{x}^{IR}<d_{x}^{UV}$ they diverge towards the UV.

%%%%%%%%%%%%%%%%%%%%%%%%%%%
\subsubsection{Lifshitz Theories}

The simplest spatially isotropic theory in the class \eqref{ansa1} is the Lifshitz theory with $z_0=z$ and $z_1=z_2=1$. This is a special case of the anisotropic geometries discussed above. The timelike $c$-function for this case has been briefly addressed in \cite{Giataganas:2025div}. The NEC again reduces to $z\ge1$, and the theory preserves spatial rotational symmetry, which ensures $c_x=c_y=c$ and $d_x=d_y$. From \eqref{dxresulta}, the effective dimension is
\be\la{dxlifanisolif}
d_x=1+\frac{d-2}{z}~.
\ee
The scaling function appearing in the flow equation is $\L(r)=(z+d-2)r$, and the NEC implies
\be\la{conslif0}
\L'(r)>0~,\qquad d_{x_r}=d_x>\frac12~.
\ee
Thus $\L$ increases towards the UV and tends to $-\infty$ in the IR and thus the IR convergence conditions hold. For a scale-invariant Lifshitz theory characterized
by a fixed exponent $z$, the matching reference exponent gives $w=0$, and \eqref{c_final2a} yields
$\pp_{r_m}c=0$, as expected, under the prescription \eqref{IRclockprescription}. The corresponding
function $\cC$ is also constant.

For a more general RG flow, the local effective-dimension condition is
\be  
d_{x_r}\ge d_x
\ee
where the running dimension is positive. This condition guarantees monotonicity of $\cC$
under the natural assumptions of Section~\ref{section:C}. For the branch function $c$, it is sufficient
within the adiabatic class satisfying (H1)-(H3) of Section~\ref{sec:adiab}.

We can again consider a Lifshitz exponent $z_r=z_r(r)$ varying along the RG flow, with fixed-point values $z_{UV}$ and $z_{IR}$. To specify the running geometry, we take $B'=z_r(r)$ and $A_1'=A_2'=1$.
The reference exponent $d_x$ is fixed at an endpoint, while $d_{x_r}$ varies along the geometry. Writing $\L'=z_r+d-2$, with $d_x=(z+d-2)/z$ at the reference fixed point we obtain
\be\la{wlifsplit}
w=\frac{(d-2)(z-z_r)}{d_x\,z\,\L'}-\frac{z_r'}{\L'^2}~,
\ee
so for $d>2$ and slowly varying $z_r$ the flow \eqref{c_final2a} reduces to
\be
\frac{\pp c}{\pp r_m}\sim\L_m'(d-2)\int_0^{\frac T2}d\tau\,\frac{z-z_r}{z\,\L'}~,
\ee
and $z\ge z_r$ is the condition for a non-negative integrand; the $z_r'$ term in \eqref{wlifsplit} is the exact correction. Thus,  the monotonicity condition $\pp_{r_m}c\ge0$ is obeyed whenever $z\ge z_r$, within the adiabatic class.

A particularly natural scenario is a slowly and monotonically decreasing $z_r(r)$, with the UV boundary at $r\to\infty$, so that $z:=z_{IR}=z_{\max}$. Choosing $d_x=d_x^{IR}$, the conditions $1\le z_r\le z_{IR}$ and $z_r'\le0$ hold everywhere along the flow and make both terms of \eqref{wlifsplit} non-negative, so $w\ge0$ exactly. The NEC \eqref{nec1a} requires $z_r'+(z_r-1)(z_r+d-1)\ge0,$ which must also be imposed, reducing for slow variation to $z_r\ge1$ when the derivative term is negligible relative
to the algebraic term; it fixes $z_{IR}\ge1$ but does not by itself enforce $z_r\le z_{IR}$, which here follows from the assumed monotonicity of $z_r$. The adiabatic condition \eqref{adiabcond} reads
\be\la{adiabcondhysca}
w'\le\Big(z_r+\frac{z_r'}{\L'}\Big)w~,
\ee
which holds for slow variation away from the fixed points and near the IR requires $\nu<z_{IR}$, while for $\nu=z_{IR}$ subleading terms decide, as in the anisotropic case.

Therefore, Lifshitz interpolations with a UV fixed point with $z_{UV}\ge1$, and a monotonically decreasing $z_r(r)$ that satisfy the running NEC and the adiabatic rate condition  \eqref{adiabcondhysca} obey the holographic timelike $c$-theorem under the natural holographic conditions (H1)--(H3): with the IR reference exponent, $\cC$ is monotonic by the local condition $w\ge0$ alone, while $c$ and $c_T$ are monotonic within this adiabatic class. The endpoint discussion of Section~\ref{sec:dx} applies when the reference exponent does not match both fixed points.

%%%%%%%%%%%%%%%%%%%%%%%%%%%
\subsubsection{Hyperscaling Violating, Lifshitz-like Anisotropic Theories}

We now study the timelike $c$-function in its most stringent setting: anisotropic geometries that break both Lorentz symmetry and scale invariance. A broad and physically important class of such backgrounds is provided by hyperscaling violating Lifshitz-like theories, characterized by a hyperscaling violation parameter $\theta$ and anisotropic Lifshitz exponents $z_i$. 

To maintain continuity with the previous sections, we adopt Poincaré coordinates in which the UV boundary is chosen to be at infinity for $\theta=0$ and positive $z_i>0 $. Starting from the metric \eq{metr1}, the exponents entering the polynomial scaling from \eq{metricpolyna} are 
\be\la{ansa2}
\b=z_0-\frac{\th}{d-1} ~,\quad \a_1=z_1-\frac{\th}{d-1}  ~, \quad \a_2=z_2-\frac{\th}{d-1}~,\quad \d=-1-\frac{\th}{d-1}~.
\ee 
After the appropriate radial coordinate transformation and rescaling of the spacetime coordinates, the metric functions in the domain-wall coordinate system \eqref{metric1a}, become
\be 
B(r)=\prt{1-\frac{d-1}{\th}z_0} \log r~,\qquad  A_i(r)=\prt{1-\frac{d-1}{\th}z_i} \log r~.
\ee
For the special case  $\th=0$ the above expressions are singular, but setting this value initially in the  Poincaré metric and performing a similar coordinate transformation, we obtain the anisotropic Lifshitz-like metric \eq{ansa1}. 
We keep all parameters free to allow direct applicability to diverse models and coordinate systems. The theory is anisotropic and the effective dimensions   $d_x$ and $d_y$ differ, where \eq{dxresulta} gives  
\be \la{dxlifaniso2}
d_{x}=1+\frac{(d_1-1)z_1+d_2 z_2 -\th}{z_0}~,\qquad d_{y}=1+\frac{d_1 z_1+ \prt{d_2-1}z_2-\th}{z_0}~,
\ee
and become equal whenever $z_1= z_2$, giving the special case of isotropic RG flows. 
The parameters $(z_i, \theta)$ are constrained by the three NEC conditions:
\bea 
&&(z_0-z_1) \prt{d_1 z_1 +d_2 z_2+z_0-\th}\ge0~,\quad (z_0-z_2) (d_1 z_1 +d_2 z_2+z_0-\th)\ge0~,\quad \\
&&(d-1) \left(d_1 z_1 (z_0-z_1)+d_2 z_2\prt{z_0-z_2}-\th z_0 \right)+\th^2\ge0~,
\eea
where the first two inequalities are multiplied by $(d-1)^2/(
\th^2 r^2)$ and the last one by $(d-1)/(\th^2 r^2)$, factors that have been trivially omitted since they are positive overall factors. These conditions restrict the region of parameter space where a physically sensible holographic theory exists. To interpret $\partial_{r_m}c$, we must identify the UV boundary in the coordinate system \eqref{metric1a}. Starting from Poincar\'e coordinates $r_p\in(0,\infty)$, the radial coordinate of \eqref{metric1a} is $r=|\delta+1|^{-1}r_p^{\delta+1}$,  so that the spatial metric elements scale as
$g_{ii}\propto r^{2\alpha_i/(\delta+1)}$.  Depending on the signs of $\alpha_i$ and $\delta$, the UV endpoint
sits at $r=\infty$ or $r=0$. We focus on regions where both spatial exponents $\alpha_i$ share a nonzero sign, so that both spatial warp factors grow towards the same UV endpoint.

To obtain an analytically tractable classification, we consider the common anisotropic subclass
\be\la{anisotr}
z:=z_1=z_0~,\qquad z_2=1~.
\ee
The NEC reduces to two independent inequalities:
\be
(z-1)\big(d_2-\theta+(d_1+1)z\big)\ge0~,
\qquad
(d-1)(z-1)d_2+\theta\big(\theta-(d-1)z\big)\ge0~.
\ee
The effective dimensions \eqref{dxlifaniso2} become
\be\la{dxlifanisoa3}
d_x=d_1+\frac{d_2-\theta}{z}~,\qquad
d_y=1+d_1+\frac{d_2-1-\theta}{z}~,
\ee
where $\theta$ shifts the spatial dimension on the $y$-plane. The corresponding radial functions are
\be
\L_x(r)=\frac{d-1}{\theta}(\theta-d_1z-d_2)\log r~,\qquad
\L_y(r)=\frac{d-1}{\theta}\big(\theta-(d_1+1)z-d_2+1\big)\log r~.
\ee
For the associated scaling thermal family, $S_{th}\propto T_H^{d_x}$, so non-negative specific heat requires
\be\la{dx1a}
d_x\ge0~.
\ee
Together with the first NEC inequality, this implies $z\ge1$ or $z<0$, and hence
\be\la{dy1a}
d_y=d_x+1-\frac1z\ge d_x\ge0~.
\ee
These non-negative bounds alone do not establish IR convergence, which requires $d_x,d_y>1/2$ in the scaling backgrounds.

Let $n:=d-1=d_1+d_2$, with both planes nonempty. Restricting to nonzero same-sign spatial exponents
and strictly positive specific heat, the NEC gives the following regions:
\begin{center}
\begin{tabular}{||c c c c||}
\hline
Subregime & Parameters & Allowed region & $\pp$ in \eq{metric1a}\\
\hline\hline
A) & $\alpha_i>0$, $\delta>-1$
  & $\theta<0$, $z\ge1$ & $\infty$\\
B) & $\alpha_i<0$, $\delta<-1$
  & $\theta>n$, $z<0$ & $\infty$\\
C) & $\alpha_i>0$, $\delta<-1$
  & $0<\theta<d_2$, $z\ge Z_C(\theta)$ & $0$\\
D) & $\alpha_i<0$, $\delta>-1$
  & excluded & -\\
\hline
\end{tabular}
\end{center}
Here
\be  
Z_C(\theta):=\frac{nd_2-\theta^2}{n(d_2-\theta)}>1~.
\ee
Subregime D) is empty in this subclass: $\delta>-1$ requires $\theta<0$, whereas $\alpha_2=1-\theta/n$ is then positive. This exclusion does not apply to arbitrary $z_i$.

The case $\theta=0$ was treated above; $z=0$ and boundaries with vanishing spatial exponents or
specific heat require separate treatment.

\begin{figure}[!t]
\centerline{\includegraphics[width=99mm]{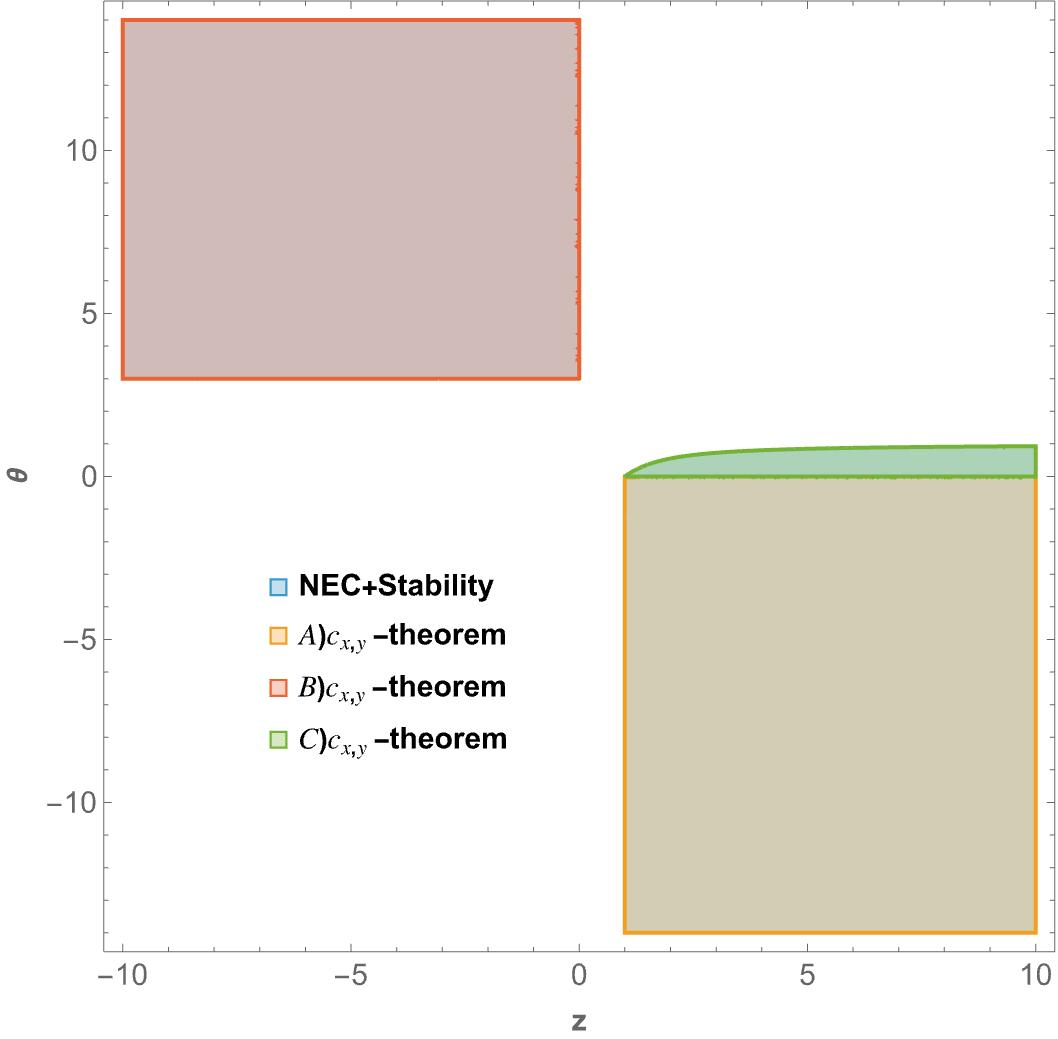}}
\caption{The parameter space of the anisotropic hyperscaling violation theory with $d_1=2 d_2=2$.  We consider the regimes where all spatial elements $g_{ii}$ diverge at the boundary. The shaded area of any color matches the parametric space allowed by the NEC and the thermodynamic stability conditions. In the parametric region A) where $\a_i>0$ and $\d>-1$ the c-theorem takes the form $\pp_{r_m}c\ge0$, and this covers the orange-shaded bottom right rectangle. In the parametric region B) where $\a_i<0$ and $\d<-1$ the c-theorem takes the form $\pp_{r_m}c\ge0$, and it covers the top left red-shaded region. In the parametric region C) where $\a_i>0$ and $\d<-1$ the c-theorem takes the form $\pp_{r_m}c\le0$, and it covers the thin green-shaded strip above region A).}
\label{fig:hysca_A2}
\end{figure}

In A) and B), $\L_x',\L_y'>0$, whereas in C), $\L_x',\L_y'<0$. Thus both $\L_x$ and $\L_y$
increase towards the UV and tend to $-\infty$ in the IR. The parameter bounds in these regions also imply $d_x,d_y>1$, so both IR conditions \eqref{IRcond} hold. The three regions are shown in Fig.~\ref{fig:hysca_A2}.

With the UV cutoff removed, the  directional branch flows are
\bea\la{c_final2ahx}
\frac{dc_x}{dr_m}\simeq 2e^{\L_{m,x}}T_{\infty,x}^{d_x-1}\L_{m,x}'\int_0^{T_x/2}d\tau_x \prt{1-\frac{d_x}{d_{x_r}}}  ~,
\eea
and
\bea\la{c_final2ahy}
\frac{dc_y}{dr_m}\simeq 2e^{\L_{m,y}}T_{\infty,y}^{d_y-1}\L_{m,y}'  \int_0^{T_y/2}d\tau_y\prt{1-\frac{d_y}{d_{y_r}}} ~.
\eea
The signed measures and endpoint corrections are those defined in Section~\ref{sec:derivc} for each
localization direction.

For the exact constant-exponent backgrounds, $d_{x_r}=d_x$ and $d_{y_r}=d_y$, so $w_x=w_y=0$. With matching reference exponents, the prescription \eqref{IRclockprescription} also gives $\varepsilon_{\mathcal R_x}
=\varepsilon_{\mathcal R_y}=0$. Consequently,
\be  
\frac{dc_x}{dr_m}=\frac{dc_y}{dr_m}=0~, \qquad \frac{d\cC_x}{dr_m}=\frac{d\cC_y}{dr_m}=0~.
\ee

For running interpolations, the orientation, convergence and NEC must be checked on the full running geometry. The conditions $w_x,w_y\ge0$ guarantee monotonicity of $\cC_x,\cC_y$ under the assumptions of Section~\ref{section:C}. For the branch functions $c_x,c_y$, these local conditions are sufficient within the adiabatic class, with monotonicity understood in the normalized regulator limit.

Thus a UV-oriented monotone has a non-negative radial derivative in A) and B) and a non-positive one in C). Since $\L_m'$ changes sign together with the location of the UV endpoint, this is the single criterion $dc/d\L_m\ge0$ in all three regions: the NEC and specific-heat constraints make the orientation of the flow follow the boundary automatically. For interpolating flows with endpoints in the same region of Fig.~\ref{fig:hysca_A2}, the timelike c-theorem therefore reads $\pp_{r_m}c_{x,y}\ge0$ in A) and B) and $\pp_{r_m}c_{x,y}\le0$ in C), for $\cC_{x,y}$ under $w_{x,y}\ge0$ and for $c_{x,y}$ in the adiabatic class.  The NEC and specific heat establish the parameter restrictions for the exact scaling backgrounds and fix the orientation of the flow so that the required direction of monotonicity is the same, $dc/d\L_m\ge0$, in every admissible region; along an interpolation, the local conditions $w_{x,y}\ge0$ and the directional adiabatic conditions are additional.  

Therefore, once the NEC and the specific-heat condition are imposed, the monotonicity condition along an interpolation between admissible backgrounds is the same in every region: $d_{x_r}\ge d_x$ and $d_{y_r}\ge d_y$ along the flow, which guarantees monotonicity of $\cC_{x,y}$, and, within the adiabatic class, of $c_{x,y}$. For interpolations with unequal endpoint dimensions, the endpoint qualifications of Section~\ref{sec:dx} apply.

%%%%%%%%%%%%%%%%%%%%%%%%%%%
\subsubsection{Hyperscaling Violating Theories}

Isotropic theories that break scale invariance in a controlled way are well known as hyperscaling-violating theories. They constitute a special subclass of the anisotropic hyperscaling-violating theories examined in the previous subsection. The $c$-function monotonicity was briefly discussed in \cite{Giataganas:2025div}; here we provide a more detailed analysis of the admissible parameter regimes and the conditions for monotonicity. These isotropic theories are characterized by two parameters, a hyperscaling violation exponent $\theta$, which quantifies the breaking of scale invariance, and a single Lifshitz exponent $z$, which measures the deviation from Lorentz invariance. They can be obtained as special, though not necessarily smooth, limits of the anisotropic hyperscaling violating theories. The most direct way to recover the isotropic theory from the anisotropic parent theory \eq{ansa2} is to set $z_1=z_2=1$ and $z_0=z$. Alternatively, in the subclass \eqref{anisotr}, shrinking the spatial plane spanned by $\vec{x}$ by taking $d_1\to0$, so that $d_2=d-1$, leaves the $y$-localized construction; the $x$-localized observable is then absent.

In Poincaré coordinates the exponents entering are
\be\la{ansa2h}
\b=z-\frac{\th}{d-1} ~,\qquad \a:=\a_i=1-\frac{\th}{d-1}  ~,\qquad \d=-1-\frac{\th}{d-1}~,
\ee
while in the coordinate system \eq{metric1a}, for $\theta\neq0$, the metric functions become
\be 
B(r)=\prt{1-\frac{d-1}{\th}z} \log r~,\qquad  A_i(r)=\prt{1-\frac{d-1}{\th}} \log r~.
\ee
The case $\theta=0$ reduces to the Lifshitz geometry discussed above. Since the theory is isotropic, the effective spatial dimensions coincide, $d_y = d_x$. From \eq{dxresulta}, we obtain
\be \la{dxlifaniso2h}
d_{x}=1+\frac{d-2 -\th}{z}~,
\ee
which can also be obtained from \eq{dxlifaniso2} upon setting $z_1 = z_2 = 1$ and $z_0=z$. This expression makes evident the role of the hyperscaling violation parameter $\theta$ in modifying the effective spatial dimension. Compared to the Lifshitz case \eq{dxlifanisolif}, the dimension entering the effective exponent is reduced from $d-2$ to $d-2-\theta$. This mirrors the role of $\theta$ in thermodynamics, where the thermal entropy scales as $S_{th}\propto T_H^{(d-1-\theta)/z}$.

The parameters $(z, \theta)$ are restricted by two NEC conditions \eq{nec1a} and \eq{nec3a}, whose nontrivial parts read
\bea 
(z-1) \prt{d-1+z-\th}\ge0~,\qquad  (d-1) \left((d-1)  (z-1)-\th z \right)+\th^2\ge0~,
\eea
together with the thermodynamic stability condition
\be 
\frac{d-\th-1}{z}\ge0~.
\ee
These ensure $d_x\ge0$ for any physically consistent hyperscaling violating theory.

To determine the sign of the derivative of the $c$-function, we must identify the location of the UV boundary in the coordinate system \eqref{metric1a}. Starting from Poincaré coordinates where $r_p\in(0,\infty)$, we transform to the coordinate system \eq{metric1a} via $r=|\d+1|^{-1} r_p^{\d+1}$ with $\d=\a-2$, taken positive, where the spatial metric components read $g_{ii}\propto r^{\frac{2\a}{\d+1}}$. Since only the sign of $\alpha$ and its position relative to $\alpha=1$ matter for the UV location, the possible cases divide into three subregimes. Writing $n:=d-1$, the NEC, strictly positive specific heat, $\theta\neq0$ and $\alpha\neq0$ give
\begin{center}
\begin{tabular}{||c c c c||} 
 \hline
Subregime & Spatial exponent & Allowed parameters & $\pp$ in \eq{metric1a}\\ [0.5ex] 
 \hline\hline
A) & $\a>1$ & $\theta<0$, $z\ge1$ & $\infty$ \\ 
 \hline
B) & $\a<0$ & $\theta>n$, $z<0$ & $\infty$ \\ \hline
C) & $0<\a<1$ & $0<\theta<n$, $z\ge1+\theta/n$ & $0$ \\
 \hline
\end{tabular}
\end{center}
In A) and B) the matching exponent satisfies $d_x>1$. In C), $d_x\ge1$ for $\theta\le n-1$, while for $n-1<\theta<n$, writing $\sigma:=n-\theta\in(0,1)$, the NEC bound $z\ge2-\sigma/n$ gives $d_x=1-(1-\sigma)/z\ge1-(1-\sigma)/(2-\sigma/n)>1/2$. Hence both IR conditions \eqref{IRcond} hold in all three regimes.

The evolution of the timelike $c$-function along the RG flow is given by \eqref{c_final2a},
\bea\la{c_hyscaa}
\frac{\pp c}{\pp r_m}\sim \L_m' \int_{0}^{\frac{T}{2}}d\tau\prt{1-\frac{d_x}{d_{x_r}}}~,\qquad \L_m'=-\frac{(d-1)(d+z-\th-2)}{\th r_m}~.
\eea
In A) and B) the NEC and the thermodynamic stability conditions imply $\L_m'>0$, with the boundary at infinity; in C) they impose the opposite sign, $\L_m'<0$, with the boundary at $r=0$. Thus in every subregime $\L$ increases towards the UV and tends to $-\infty$ in the IR. For the exact constant-exponent geometries, $d_{x_r}=d_x$, so the integrand vanishes and $\pp_{r_m}c=0$, as does $d\cC/dr_m$; these are the admissible fixed points. For interpolations between them, the sign of $\L_m'$ fixes the orientation: the c-theorem reads $\pp_{r_m}c\ge0$ in A) and B) and $\pp_{r_m}c\le0$ in C), i.e. $dc/d\L_m\ge0$ in all cases, and it follows from $w\ge0$ along the flow for $\cC$ under the assumptions of Section~\ref{section:C}, and for $c$ within the adiabatic class of Section~\ref{sec:adiab}.

In all physically admissible regions of the $(z,\theta)$ parameter space allowed by the NEC and thermodynamic stability, the orientation of the flow is therefore fixed automatically by the location of the UV boundary, in the way required by the c-theorem. This provides another consistency check of the general analysis: the NEC and thermodynamic stability select the admissible hyperscaling-violating scaling regimes and their orientation, while monotonicity along an interpolation follows from the effective-dimension condition, with adiabaticity needed for the  $c$-function.

%%%%%%%%%%%%%%%%%%%%%%%%%%%
\section{Discussions}\label{section7}

In this manuscript, we have studied the holographic timelike c-theorem in detail and subjected it to stringent tests in theories with reduced spacetime symmetries. We systematically extended the holographic formulation of the timelike $c$-function introduced in \cite{Giataganas:2025div} and we have made several advances. First, we extended the construction to theories that go far beyond Lorentz-invariant ones, including anisotropic theories with broken rotational symmetry, by introducing directional timelike $c$-functions $c_x,c_y$   \eq{cfunctiona} with their own reference exponents $d_x,d_y$, and we discussed in detail the application to Lifshitz-like and hyperscaling violating theories. Second, we derived the exact flow of the branch $c$-function at finite IR regulator and showed that it is a signed integral of a local effective-dimension condition \eq{localeff}, over the two branches, so that the local condition alone does not fix the sign; we identified the adiabatic class of flows, Section~\ref{sec:adiab}, in which it does, and we showed that the boundary $c$-function $c_T$ \eq{cf0Ta} has the same monotonicity. Third, we constructed a new timelike $c$-function $\cC$ \eq{Teffdef} from the finite imaginary area and the conserved momentum of the timelike branch alone; its exact flow is a positive weighted average of $w$, so $w\ge0$ suffices for its monotonicity for every flow, without adiabaticity. For instance, for Poincar\'e-invariant flows this condition is the NEC, between AdS fixed points $\cC$ reproduces the ratio of the holographic central charges, and in two dimensions $\cC$ is half the imaginary part of the timelike entanglement entropy, which therefore decreases from the UV to the IR. Fourth, we tested the theorems on explicit flows. Thus timelike RG flow irreversibility holds in Poincaré, Lifshitz-like, and hyperscaling-violating theories, both isotropic and genuinely anisotropic, with $\cC$ monotonic under the local condition and the introduced $c$-function monotonic in the adiabatic class.

This framework provides a generalization of c-theorems into regimes where traditional formulations based on spacelike entanglement entropy do not apply. The timelike formulation fills a conceptual gap, aligning with the physical expectation that RG flows in anisotropic theories should decrease the number of effective degrees of freedom, and that a $c$-function is expected to exist in such anisotropic theories. Our analysis suggests the existence of a timelike renormalization structure connecting entropy growth on the boundary with causal warping of the bulk geometry, and that timelike entanglement captures essential features of quantum degrees of freedom even in the absence of rotational, Lorentz, and scale invariance. 

In anisotropic settings, several distinct timelike $c$-functions naturally arise due to the directional breaking of rotational symmetry. For the classes of theories studied here, each directional function $\cC_{x},\cC_y$ is monotonic under its local condition, and the directional branch functions $c_x,c_y$ are monotonic within the adiabatic class. Accordingly, we adopt the most conservative approach and require the validity of a c-theorem independently in each direction. Consequently, any composite $c$-function constructed from these directional components that preserves their individual monotonicity is itself guaranteed to be monotonic. Nevertheless, it is conceivable that in highly anisotropic RG flows, effective degrees of freedom may redistribute heavily between different spatial directions, for instance, from $x$-modes to $y$-modes, such that one of the directional functions could exhibit local non-monotonic behavior, while an appropriately defined total averaged $c$-function still satisfies an overall c-theorem. Exploring whether a single covariant $c$-function can be defined in such highly anisotropic settings is an additional interesting direction.

Our formulation applies uniformly to both even and odd spacetime dimensions, as expected in a holographic construction. As a warm-up, we began the discussion with Poincaré-invariant theories and then progressively increased the complexity by studying theories with increasing amounts of symmetry breaking: anisotropic Lifshitz theories, followed by anisotropic hyperscaling-violating theories. In the exact scaling backgrounds the $c$-functions are constant, and the NEC and thermodynamic stability select the admissible parameter regions and fix the orientation of the flow, in agreement with the c-theorem. For running interpolations, a monotonically decreasing Lifshitz exponent towards the UV gives the directional effective-dimension inequalities with the IR reference exponents, so that $\cC_{x,y}$ are monotonic; monotonicity of the $c$-functions additionally requires the adiabatic conditions, whose leading IR condition holds for $\nu<z_{IR}$,
with $\nu$ the approach rate of the exponent and the equality case determined by subleading terms. Notice that this contrasts the standard entanglement entropy based $c$-functions, where it is known that RG flow can violate the monotonicity of entanglement entropy under RG transformations, so that the $c$-function based on the entanglement entropy does not capture fully the monotonicity of degrees of freedom in these theories \cite{Chu:2019uoh,Swingle:2013zla,Cremonini:2013ipa}. For Poincaré-invariant flows, $\cC$ is monotonic under the NEC, the same condition as for the spacelike entanglement $c$-functions \cite{Ryu:2006ef,Myers:2012ed,Chu:2019uoh}. 

The functions $c$ and $c_T$ are monotonic for adiabatic RG flows, while $\cC$ is monotonic under its local effective-dimension condition, without adiabaticity. The monotonicity boils down to the condition $w\ge0$, which is always compatible with the NEC: since $w=1/d_x-1/d_{x_r}$ \eq{localdx}, it holds along any flow for a reference exponent $d_x$ not exceeding the minimum of $d_{x_r}$ over the region where the running dimension is positive. This choice need not match either fixed point.  The content of the $c$-theorem lies in the matched case, in which $d_x$ equals the effective dimension of a fixed point so that the $c$-functions take finite fixed-point values; then $w\ge0$ requires that fixed-point dimension to be the minimum of $d_{x_r}$ along the flow, which the NEC guarantees for Poincar\'e-invariant flows and which for Lifshitz-like flows is a condition on the running exponents, such as $z_r\le z_{IR}$, and adiabaticity is a genuine additional restriction. Nevertheless, there exist flows, such as the one of Appendix \ref{sec:fgpwsec}, from $\mathcal N=4$ super Yang--Mills theory to the Leigh--Strassler fixed point, which do not satisfy our adiabaticity condition. This is an explicit example in which the pure timelike $\cC$ is monotonic, whereas $c$ and $c_T$ are not, since the relevant analysis holds for adiabatic flows.

All these results follow from the explicit expressions for the derivatives of the timelike $c$-functions in \eq{c_tot2ax}, \eq{c_tot2ay}, \eq{c_final2a} and \eq{Cflow}, showing how monotonicity is tied to simple geometric conditions involving radial derivatives of the metric, energy conditions and effective dimensions. The theories we have studied are the most general homogeneous spaces with monomial metric elements and are the most interesting and physically relevant due to the type of symmetries they possess. In the most general anisotropic backgrounds, where the metric elements $A_i(r)$ and $B(r)$ are arbitrary functions, the NEC and thermodynamic stability do not map one-to-one onto the directional conditions $w_i\ge0$, since the two involve different combinations of the derivatives of the metric functions and carry independent information away from scaling regimes. This is not unexpected, and it is in our view the correct state of affairs: for a generic bottom-up nonrelativistic background, the NEC and the specific-heat condition alone do not, and should not, guarantee the monotonicity of the $c$-functions considered here, since such fully arbitrary backgrounds typically cannot arise as solutions of physically motivated bulk actions. What our construction provides is a geometric criterion fixed by the metric alone, which does not require specifying the bulk matter dynamics and therefore applies to a broad class of backgrounds under the stated hypotheses; in all the examples studied, its monotonicity conditions are compatible with the nontrivial NEC and thermodynamic stability constraints, which select the admissible scaling regimes and the orientation of the flow. Whether the equations of motion of a specific bulk theory, for backgrounds that arise as its solutions, supply additional sufficient conditions for monotonicity is an open interesting and challenging question.

Our work suggests a broader conceptual lesson for renormalization group flows beyond Lorentz-invariant settings. When Lorentz symmetry is broken and space and time scale differently, RG irreversibility cannot, in general, be faithfully captured by observables defined purely on spacelike subregions. Instead, probes that are intrinsically sensitive to the temporal direction become appropriate. In this sense, timelike entanglement measures appear naturally suited to encode the loss of degrees of freedom along nonrelativistic RG flows, where conventional entanglement-based $c$-functions are known to fail. From this perspective, the monotonicity of the timelike $c$-functions constructed here reflects a structural feature of nonrelativistic quantum field theories: the RG flow itself intertwines energy scales with temporal structure.

A further observation arising from our work is the upper bound \eq{sand} on the rate of change of the timelike $c$-function along the RG flow, when $T$ is monotonically aligned with the RG direction. The bound states that the speed at which degrees of freedom are eliminated along the flow cannot exceed a value fixed by the bulk geometry. We showed that its content is the orientation of the boundary interval, and that $\cC$ obeys the analogous bound \eq{Cbound} as a consequence of the local conditions alone. From a field theoretic perspective, such a bound is suggestive of the local nature of Wilsonian renormalization group flow, which integrates out degrees of freedom gradually, within thin shells of energy, rather than eliminating all ultraviolet degrees of freedom at once; an arbitrarily rapid decrease of a $c$-function over a narrow range of scales might signal a breakdown of the assumptions underlying a smooth RG flow, such as locality or thermodynamic stability, or signal a phase transition. Whether the bound found here, or sharper rate bounds, admit such a field-theoretic interpretation is an open question. 

There are several compelling directions emerging from these results. One promising avenue is to explore nonequilibrium dynamics and real-time evolution in holographic quantum matter, which is distinct from the RG flow studied here. In holographic systems, excitations created at the boundary propagate causally into the bulk, potentially reducing the number of UV degrees of freedom accessible to boundary observers. This suggests a form of timelike irreversibility governed by entanglement growth and horizon formation. A timelike $c$-function may thus serve as a natural information-theoretic diagnostic of thermalization and scrambling, capturing how bulk causal structure enforces a unidirectional loss of short-distance information.  Moreover, certain excitations sourcing the flow can be viewed as propagating toward the horizon with the associated boundary spreading characterized by the butterfly velocity; therefore, the rate of change of the timelike $c$-function might encode aspects of scrambling dynamics. Such a relation is suggested also by entanglement entropy results: in global quenches of holographic field theories, the growth of entanglement entropy is known to be related to quantities characterizing chaos \cite{Mezei:2016zxg,Mezei:2016wfz}.

Studying timelike $c$-functions in dynamical spacetimes, such as black hole formation or Vaidya shells, could reveal universal transient laws or entanglement avalanche phenomena specific to timelike probes, as discussed in \cite{Liu:2013iza,Abajo-Arrastia:2010ajo}. Extensions of our analysis to higher-derivative theories would further test whether timelike monotonicity persists beyond Einstein gravity and whether additional restrictions on the couplings are needed. If monotonicity survives only in a restricted subset of higher-derivative couplings, this can be interpreted as a new information-theoretic constraint on consistent corrections to Einstein holography. A field-theoretic interpretation of $\cC$, in particular of the monotonic imaginary entropy in two-dimensional holographic flows, is another natural direction.

Finally, since many condensed matter systems are inherently anisotropic and nonrelativistic, timelike $c$-functions are natural candidates to provide a powerful diagnostic for classifying exotic critical points and phase transitions that lie beyond the conventional Lorentz-invariant paradigm and the Ginzburg-Landau framework. Traditional holographic $c$-functions have already yielded valuable insights in such contexts \cite{Baggioli:2020cld}, and the timelike formulation developed here has the potential to significantly enrich this program in the strongly anisotropic regimes.

\noindent \textbf{Acknowledgments.} This work is supported by the National Science and Technology Council (NSTC) of Taiwan with the Young Scholar Columbus Fellowship grant 114-2636-M-110-004.

%%%%%%%%%%%%%%%%%%%%%%%%%%%

\appendix
\section*{Appendices}
\addcontentsline{toc}{section}{Appendices}

\section{ Supporting Derivations of the Timelike Flow}
\label{appA}

We provide the regulated derivations and IR estimates used in the main text. Write $u:=\L-\L_m$ and $G:=e^{-B}/|\L'|=1/\mu$, so that $\pp_\L G=-G/d_{x_r}$.
All endpoint and differentiation assumptions are those stated in Sections~\ref{sec:derivc} and \ref{sec:adiab}.

At fixed radial cutoffs,
\be  
\frac{t_{Im}}2 =\int_{u_b}^{0}\frac{G\,du}{\sqrt{1-e^{2u}}},
\qquad
\frac{S_{Im}}2=e^{\L_m}\int_{u_b}^{0}\frac{Ge^{2u}\,du}{\sqrt{1-e^{2u}}}.
\ee
Let $D_m$ denote differentiation with respect to $\L_m$ at fixed radial endpoints. Since $D_mu_b=-1$,
\be\la{regid}
\begin{aligned}
D_m(S_{Im}/2)-e^{\L_m}D_m(t_{Im}/2)=\frac{S_{Im}}2 -e^{\L_m}\int_{u_b}^{0} \pp_uG\sqrt{1-e^{2u}}\,du -e^{\L_m}G_b\sqrt{1-e^{2u_b}}=0.
\end{aligned}
\ee
The last equality follows by integration by parts. The analogous calculation for the spacelike branch,
retaining both endpoint terms, gives $D_mS_{Re}+e^{\L_m}D_mt_{Re}=0$. Thus \eqref{derima} and \eqref{derrea} hold at finite radial cutoffs without separately evaluating the singular turning-point terms.

At fixed $\L_m$,  the prescription \eqref{IRclockprescription} gives $t_{Re}=\mathcal R e^{-\L_m/d_x}\to\infty$.
Under the stated IR conditions, this sends $\L_b\to-\infty$. Using $\pp_{\L_b}t_{Re}=-2G_b/\sqrt{1+e^{2u_b}}$, differentiation of \eqref{IRclockprescription} yields
\be\la{cutoffvelocity}
\frac{d\L_b}{d\L_m}=\frac{\sqrt{1+e^{2u_b}}}{2G_b}\left(\frac{t_{Re}}{d_x}+\left.\frac{\pp t_{Re}}{\pp\L_m}\right|_{r_b}\right).
\ee
Consequently,
\be\la{epsexplicit}
\varepsilon_{\mathcal R}=D(u_b)\left[G_b-\frac{\sqrt{1+e^{2u_b}}}{2}\left(
\frac{t_{Re}}{d_x}+\left.\frac{\pp t_{Re}}{\pp\L_m}\right|_{r_b}\right)\right],
\ee
where
\be  
D(u):=\frac1{\sqrt{1-e^{2u}}}-\frac1{\sqrt{1+e^{2u}}}\sim e^{2u}\qquad(u\to-\infty).
\ee
The endpoint condition \eqref{IRendpoint} gives $e^{2u_b}G_b\to0$. It also gives $e^{2u_b}t_{Re}\to0$: applying
l'H\^opital's rule to $\int_{\L_b}^{\L_0}G(v)\,dv/e^{-2\L_b}$ gives the derivative ratio $G_be^{2\L_b}/2\to0$.
Finally,
\be  
\left.\frac{\pp t_{Re}}{\pp\L_m}\right|_{r_b}=2\int_{\L_b}^{\infty}d\L\, \frac{Ge^{2(\L-\L_m)}}{[1+e^{2(\L-\L_m)}]^{3/2}}
\ee
has a finite limit by IR and UV convergence. Therefore $\varepsilon_{\mathcal R}\to0$, without adiabaticity. The same estimates remove the moving-endpoint contribution to $dS-e^{\L_m}dT$, giving $dS/dT=e^{\L_m}$ wherever $dT/d\L_m\neq0$.

For a regular scaling IR with $G\sim g_{IR}e^{-\L/d_*}$ and $d_*>1/2$, 
$t_{Re}\sim2d_*G_b$, and hence
\be\la{epsscaling}
\frac{d\L_b}{d\L_m}\to\frac{d_*}{d_x},
\qquad
\varepsilon_{\mathcal R}=\mathcal O(\mathcal R^{1-2d_*})\to0.
\ee
The condition $d_*>1/2$ is precisely the convergence condition for the imaginary area at a scaling IR.
For $d_x=d_*$ the leading coefficient vanishes. In an exactly scaling geometry with matched exponent
and the UV cutoff removed, the prescription fixes $u_b$, so $\varepsilon_{\mathcal R}=0$ at finite regulator.

For a matched regular scaling IR, suppose $w=a\mu^\kappa+o(\mu^\kappa)$, with $a,\kappa>0$, and the expansion holds after one logarithmic derivative. Then $Q=w/\mu\sim a\mu^{\kappa-1}$. Writing $\xi:=(\kappa-1)/d_*$, integration by parts with the kernel \eqref{Mdef} gives the leading signed integral
\be  
I:=\int_0^{T/2}d\tau\,w \sim-\xi Q_m\int_{-\infty}^{\infty}du\,M(u)e^{\xi u}.
\ee
For $0<\xi<1$, the integral is positive and convergent: $M\sim e^{2u}/2$ in the IR and $M\sim e^{-u}$ in
the UV. Under regular matching to the remaining geometry, this term dominates the contribution of order
$\mu(r_m)^{d_*}$. Thus $1<\kappa<1+d_*$ gives a negative UV-oriented derivative near the IR despite $w\ge0$.

For Poincar\'e-invariant scalar flows in canonical
normalization,
\be  
w=-\frac{A''}{(d-1)A'^2}=\frac{\beta^2}{2(d-1)^2}, \qquad \beta:=\frac{d\phi}{dA}.
\ee
An irrelevant IR mode has $\phi-\phi_{IR}\propto e^{(\Delta_{IR}-d)A}$, so $\kappa=2(\Delta_{IR}-d)$.
The negative-derivative window is therefore $d+1/2<\Delta_{IR}<3d/2$. 
The explicit supersymmetric example is discussed in Appendix~\ref{sec:fgpwsec}.

\section{ The flow from \texorpdfstring{$\mathcal N=4$}{N=4} to the Leigh-Strassler fixed point}
\label{sec:fgpwsec}

\begin{figure}[t]
\centering
\begin{minipage}[t]{0.49\textwidth}
  \centering
  \includegraphics[width=\linewidth]{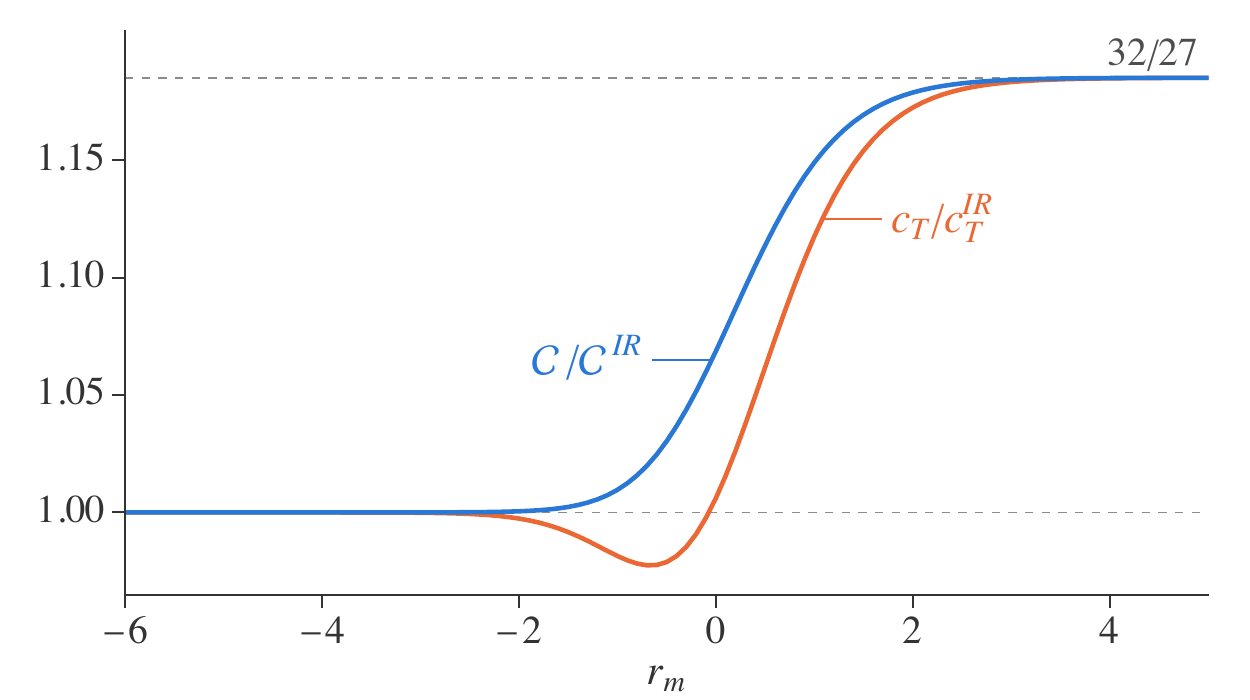}
  \captionof{figure}{The timelike function $\mathcal C$ remains monotonic along the  supersymmetric non-adiabatic flow of Appendix \ref{sec:fgpwsec}, with $L_{UV}=1$,  minimum of $A''$ at $r=0$, and $d_x=3$.
Both $c_T$ (orange) and $\cC$ (blue), normalized to their IR values, approach $32/27$ in the UV (upper dashed line). The NEC guarantees the monotonicity of $\cC$, as confirmed in the plot. The flow violates the adiabatic condition, so the monotonicity analysis of $c$ and $c_T$ does not apply, and indeed $c_T$ falls below its IR value, with a minimum $c_T/c_T^{IR}\simeq0.98$.}
\label{fig:fgpw}
\end{minipage}\hfill
\begin{minipage}[t]{0.49\textwidth}
  \centering
  \includegraphics[width=\linewidth]{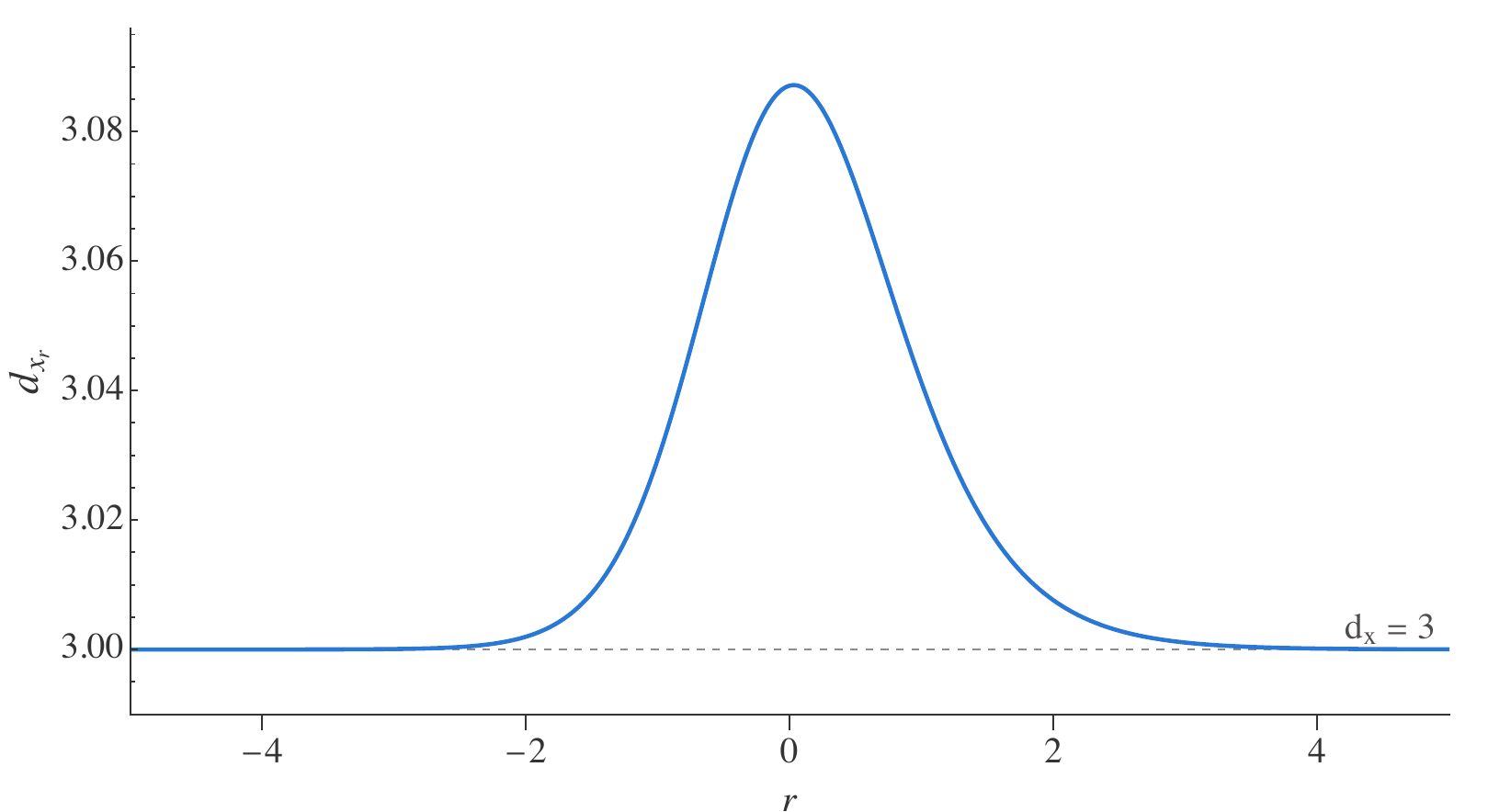}
  \captionof{figure}{The running effective dimension $d_{x_r}$ along the non-adiabatic flow of Appendix \ref{sec:fgpwsec}, as a function of the radial coordinate $r$. It equals $d_x=3$ at both fixed points and satisfies $d_{x_r}\ge d_x=3$ throughout, which is equivalent to the NEC $A''\le0$, and the local effective-dimension condition $w\ge0$ for this flow.}
  \label{fig:fgpw_dxr}
\end{minipage}
\end{figure}

The timelike-branch function $\mathcal C$ remains monotonic along the supersymmetric flow of \cite{Freedman:1999gp} to the Leigh–Strassler fixed point. The comparison with $c_T$ illustrates the stronger scope of the new theorem: this flow satisfies the NEC but lies outside the adiabatic class required for the monotonicity of $c_T$. 

The non-adiabatic supersymmetric flow has $(L_{UV}/L_{IR})^3=32/27$ and $\Delta_{IR}=3+\sqrt7$. Thus $\kappa_{IR}=2(\sqrt7-1)\simeq3.29$ lies in the negative-derivative window $1<\kappa_{IR}<4$ for $d_x=3$ derived in Appendix~\ref{appA}. The flow violates the adiabatic condition \eqref{adiabcond}  near the IR, so the monotonicity analysis of $c$ and $c_T$ does not apply, and the near-IR decrease of $c_T$ seen in Fig.~\ref{fig:fgpw} is precisely the behaviour predicted by the window. The non-monotonicity of $c_T$ is therefore not a failure of the construction but the expected behaviour outside the adiabatic class. The monotonicity of $\cC$, by contrast, requires only $w\ge0$, which is the NEC $A''\le0$ along the entire flow, so its theorem applies and $\cC$ is monotonic. Figure~\ref{fig:fgpw} compares $c_T$ and $\cC$, normalized to their IR values, with $L_{UV}=1$ and the minimum of $A''$ placed at $r=0$. Both UV/IR ratios approach $32/27$. Although $T(r_m)$ decreases monotonically towards the UV,  $c_T/c_T^{IR}$ reaches a minimum of about $0.98$.

The running effective dimension $d_{x_r}$ equals $d_x=3$ at both fixed points and satisfies $d_{x_r}\ge d_x=3$ throughout, which is equivalent to the NEC $A''\le0$, and the local effective-dimension condition $w\ge0$ for this flow. This behavior is presented in   Fig.~\ref{fig:fgpw_dxr}.

To obtain Figs.~\ref{fig:fgpw} and \ref{fig:fgpw_dxr}, we integrate the supersymmetric equations $\phi_j'=\frac g2\,\pp_{\phi_j}W$ and $A'=-\frac g3 W$, using the superpotential of \cite{Freedman:1999gp} with $g=2$, starting from the IR fixed point along its irrelevant direction with IR behaviour
$\phi_j-\phi_j^{IR}\propto e^{(\Delta_{IR}-4)A}$ and continuing with the exact AdS solutions beyond the integrated range. With $B=A$, $\Lambda=3A$ and $d_x=3$, we compute the branch times \eq{t2}, treating their common IR divergence together in the difference \eq{Tdef}, and the imaginary area \eq{s2a}, \eq{s3}, regularizing the turning point by the substitution $r=r_m-\sigma^2$ with $\sigma\ge0$ an auxiliary integration variable. From these, $c_T$ follows from \eq{cttt}, $\cC$ from \eq{Teffdef}, and $d_{x_r}$ from \eq{dxp} with $A''=-\frac{g^2}{6}|\nabla W|^2$ taken from the flow equations. The branch function $c$ has the monotonicity of $c_T$, \eqref{cbranchcomparison}.

\bibliography{timelike}

@article{Das:2025fcd,
    author = "Das, Rathindra Nath and Kundu, Arnab and Martins Costa, Matheus H. and Sarkar, Nemai Chandra",
    title = "{Temporal correlations and chaos from spacetime kernel}",
    eprint = "2512.06078",
    archivePrefix = "arXiv",
    primaryClass = "hep-th",
    month = "12",
    year = "2025"
}

@article{Conti:2025wwf, 
    author = "Conti, Andrea and Lozano, Yolanda and Rogdakis, Filippos and Rosen, Christopher",
    title = "{Defect entanglement entropy for superconformal monodromy defects}",
    eprint = "2511.22695",
    archivePrefix = "arXiv",
    primaryClass = "hep-th",
    month = "11",
    year = "2025"
}

@article{Maldacena:1997re,
    author = "Maldacena, Juan Martin",
    title = "{The Large $N$ limit of superconformal field theories and supergravity}",
    eprint = "hep-th/9711200",
    archivePrefix = "arXiv",
    reportNumber = "HUTP-97-A097, HUTP-98-A097",
    doi = "10.4310/ATMP.1998.v2.n2.a1",
    journal = "Adv. Theor. Math. Phys.",
    volume = "2",
    pages = "231--252",
    year = "1998"
}

@article{Witten:1998qj,
    author = "Witten, Edward",
    title = "{Anti de Sitter space and holography}",
    eprint = "hep-th/9802150",
    archivePrefix = "arXiv",
    reportNumber = "IASSNS-HEP-98-15",
    doi = "10.4310/ATMP.1998.v2.n2.a2",
    journal = "Adv. Theor. Math. Phys.",
    volume = "2",
    pages = "253--291",
    year = "1998"
}

@article{Jokela:2025qac,
    author = "Jokela, Niko and Kastikainen, Jani and Pen{\'\i}n, Jos{\'e} Manuel and Rodgers, Ronnie and Ruotsalainen, Helime",
    title = "{Entanglement C-functions of defects and interfaces in $\mathcal{N}=4$ supersymmetric Yang-Mills theory}",
    eprint = "2510.00123",
    archivePrefix = "arXiv",
    primaryClass = "hep-th",
    reportNumber = "HIP-2025-27/TH, NORDITA 2025-050",
    month = "9",
    year = "2025"
}

@article{Harper:2025lav,
    author = "Harper, Jonathan and Kawamoto, Taishi and Maeda, Ryota and Nakamura, Nanami and Takayanagi, Tadashi",
    title = "{Non-hermitian Density Matrices from Time-like Entanglement and Wormholes}",
    eprint = "2512.13800",
    archivePrefix = "arXiv",
    primaryClass = "hep-th",
    reportNumber = "YITP-25-186",
    month = "12",
    year = "2025"
}

@article{Mezei:2016zxg,
    author = "Mezei, M{\'a}rk",
    title = "{On entanglement spreading from holography}",
    eprint = "1612.00082",
    archivePrefix = "arXiv",
    primaryClass = "hep-th",
    doi = "10.1007/JHEP05(2017)064",
    journal = "JHEP",
    volume = "05",
    pages = "064",
    year = "2017"
}

@article{Mezei:2016wfz,
    author = "Mezei, M{\'a}rk and Stanford, Douglas",
    title = "{On entanglement spreading in chaotic systems}",
    eprint = "1608.05101",
    archivePrefix = "arXiv",
    primaryClass = "hep-th",
    doi = "10.1007/JHEP05(2017)065",
    journal = "JHEP",
    volume = "05",
    pages = "065",
    year = "2017"
}

@article{Gong:2025pnu,
    author = "Gong, XiangKun and Guo, Wu-zhong and Xu, Jin",
    title = "{Entanglement measures for causally connected subregions and holography}",
    eprint = "2508.05158",
    archivePrefix = "arXiv",
    primaryClass = "hep-th",
    month = "8",
    year = "2025"
}

@article{Guo:2024edr,
    author = "Guo, Wu-zhong and Jiang, Yao-zong and Xu, Jin",
    title = "{Pseudoentropy sum rule by analytical continuation of the superposition parameter}",
    eprint = "2405.09745",
    archivePrefix = "arXiv",
    primaryClass = "hep-th",
    doi = "10.1007/JHEP11(2024)069",
    journal = "JHEP",
    volume = "11",
    pages = "069",
    year = "2024"
}

@article{Nanda:2025tid,
    author = "Nanda, Kanhu Kishore and Narayan, K. and Porey, Somnath and Yadav, Gopal",
    title = "{dS extremal surfaces, replicas, boundary Renyi entropies in dS/CFT and time entanglement}",
    eprint = "2509.02775",
    archivePrefix = "arXiv",
    primaryClass = "hep-th",
    doi = "10.1007/JHEP11(2025)095",
    journal = "JHEP",
    volume = "11",
    pages = "095",
    year = "2025"
}

@article{Nunez:2025ppd,
    author = "Nunez, Carlos and Roychowdhury, Dibakar",
    title = "{Interpolating between spacelike and timelike entanglement via holography}",
    eprint = "2507.17805",
    archivePrefix = "arXiv",
    primaryClass = "hep-th",
    doi = "10.1103/x3zd-llsx",
    journal = "Phys. Rev. D",
    volume = "112",
    number = "8",
    pages = "",
    year = "2025"
}

@article{Nunez:2025puk,
    author = "Nunez, Carlos and Roychowdhury, Dibakar",
    title = "{Holographic timelike entanglement across dimensions}",
    eprint = "2508.13266",
    archivePrefix = "arXiv",
    primaryClass = "hep-th",
    doi = "10.1007/JHEP11(2025)100",
    journal = "JHEP",
    volume = "11",
    pages = "100",
    year = "2025"
}

@article{Xu:2024yvf,
    author = "Xu, Jin and Guo, Wu-zhong",
    title = "{Imaginary part of timelike entanglement entropy}",
    eprint = "2410.22684",
    archivePrefix = "arXiv",
    primaryClass = "hep-th",
    doi = "10.1007/JHEP02(2025)094",
    journal = "JHEP",
    volume = "02",
    pages = "094",
    year = "2025"
}

@article{Abajo-Arrastia:2010ajo,
    author = "Abajo-Arrastia, Javier and Aparicio, Joao and Lopez, Esperanza",
    title = "{Holographic Evolution of Entanglement Entropy}",
    eprint = "1006.4090",
    archivePrefix = "arXiv",
    primaryClass = "hep-th",
    doi = "10.1007/JHEP11(2010)149",
    journal = "JHEP",
    volume = "11",
    pages = "149",
    year = "2010"
}

@article{yang2014quantum,
  title={Quantum criticality of topological phase transitions in three-dimensional interacting electronic systems},
  author={Yang, Bohm-Jung and Moon, Eun-Gook and Isobe, Hiroki and Nagaosa, Naoto},
  journal={Nature Physics},
  volume={10},
  number={10},
  pages={774--778},
  year={2014},
  publisher={Nature Publishing Group UK London}
}

@article{isobe2016emergent,
  title={Emergent non-Fermi-liquid at the quantum critical point of a topological phase transition in two dimensions},
  author={Isobe, Hiroki and Yang, Bohm-Jung and Chubukov, Andrey and Schmalian, J{\"o}rg and Nagaosa, Naoto},
  journal={Physical review letters},
  volume={116},
  number={7},
  pages={076803},
  year={2016},
  publisher={APS}
}

@article{Liu:2013iza,
    author = "Liu, Hong and Suh, S. Josephine",
    title = "{Entanglement Tsunami: Universal Scaling in Holographic Thermalization}",
    eprint = "1305.7244",
    archivePrefix = "arXiv",
    primaryClass = "hep-th",
    reportNumber = "MIT-CTP/4475, MIT-CTP-4475",
    doi = "10.1103/PhysRevLett.112.011601",
    journal = "Phys. Rev. Lett.",
    volume = "112",
    pages = "011601",
    year = "2014"
}

@article{Baggioli:2024vza,
    author = "Baggioli, Matteo and Pujolas, Oriol and Wu, Xin-Meng",
    title = "{Holographic Lifshitz flows}",
    eprint = "2407.11552",
    archivePrefix = "arXiv",
    primaryClass = "hep-th",
    doi = "10.1007/JHEP09(2024)175",
    journal = "JHEP",
    volume = "09",
    pages = "175",
    year = "2024"
}

@article{Giataganas:2025ing,
    author = {Giataganas, Dimitrios and G{\"u}rsoy, Umut and Moran, Claire and Pedraza, Juan F. and Rodr{\'\i}guez Fern{\'a}ndez, David},
    title = "{Anisotropic critical points from holography}",
    eprint = "2509.15021",
    archivePrefix = "arXiv",
    primaryClass = "hep-th",
    reportNumber = "IFT-UAM/CSIC-25-98",
    month = "9",
    year = "2025"
}

@article{Cremonini:2014pca,
    author = "Cremonini, Sera and Dong, Xi and Rong, Junchen and Sun, Kai",
    title = "{Holographic RG flows with nematic IR phases}",
    eprint = "1412.8638",
    archivePrefix = "arXiv",
    primaryClass = "hep-th",
    reportNumber = "SU-ITP-15-02",
    doi = "10.1007/JHEP07(2015)082",
    journal = "JHEP",
    volume = "07",
    pages = "082",
    year = "2015"
}

@ARTICLE{1998Natur393550K,
       author = {Kivelson, S.~A. and Fradkin, E. and Emery, V.~J.},
        title = "{Electronic liquid-crystal phases of a doped Mott insulator}",
      journal = {},
         year = 1998,
        month = jun,
       volume = {393},
       number = {6685},
        pages = {550-553},
          doi = {10.1038/31177},
archivePrefix = {arXiv},
       eprint = {cond-mat/9707327},
 primaryClass = {cond-mat},
       adsurl = {https://ui.adsabs.harvard.edu/abs/1998Natur.393..550K}
}

@ARTICLE{2010ARCMP1153F,
       author = {{Fradkin}, Eduardo and {Kivelson}, Steven A. and {Lawler}, Michael J. and {Eisenstein}, James P. and {MacKenzie}, Andrew P.},
        title = "{Nematic Fermi Fluids in Condensed Matter Physics}",
      journal = {Annual Review of Condensed Matter Physics},
         year = 2010,
        month = apr,
       volume = {1},
        pages = {153-178},
          doi = {10.1146/annurev-conmatphys-070909-103925},
archivePrefix = {arXiv},
       eprint = {0910.4166},
 primaryClass = {cond-mat.str-el},
       adsurl = {https://ui.adsabs.harvard.edu/abs/2010ARCMP...1..153F}
}

@article{PhysRevB.89.155130,
  title = {Transport near the Ising-nematic quantum critical point of metals in two dimensions},
  author = {Hartnoll, Sean A. and Mahajan, Raghu and Punk, Matthias and Sachdev, Subir},
  journal = {Phys. Rev. B},
  volume = {89},
  issue = {15},
  pages = {155130},
  numpages = {15},
  year = {2014},
  month = {Apr},
  publisher = {American Physical Society},
  doi = {10.1103/PhysRevB.89.155130},
  url = {https://link.aps.org/doi/10.1103/PhysRevB.89.155130}
}

@article{PhysRevLett.88.137005,
  title = {Electrical Resistivity Anisotropy from Self-Organized One Dimensionality in High-Temperature Superconductors},
  author = {Ando, Yoichi and Segawa, Kouji and Komiya, Seiki and Lavrov, A. N.},
  journal = {Phys. Rev. Lett.},
  volume = {88},
  issue = {13},
  pages = {137005},
  numpages = {4},
  year = {2002},
  month = {Mar},
  publisher = {American Physical Society},
  doi = {10.1103/PhysRevLett.88.137005},
  url = {https://link.aps.org/doi/10.1103/PhysRevLett.88.137005}
}

@article{
doi:10.1126/science.1138584,
author = {Y. Kohsaka  and C. Taylor  and K. Fujita  and A. Schmidt  and C. Lupien  and T. Hanaguri  and M. Azuma  and M. Takano  and H. Eisaki  and H. Takagi  and S. Uchida  and J. C. Davis },
title = {An Intrinsic Bond-Centered Electronic Glass with Unidirectional Domains in Underdoped Cuprates},
journal = {Science},
volume = {315},
number = {5817},
pages = {1380-1385},
year = {2007},
doi = {10.1126/science.1138584},
URL = {https://www.science.org/doi/abs/10.1126/science.1138584},
eprint ={}
}

@article{
doi:10.1126/science.1152309,
author = {V. Hinkov  and D. Haug  and B. Fauqué  and P. Bourges  and Y. Sidis  and A. Ivanov  and C. Bernhard  and C. T. Lin  and B. Keimer },
title = {Electronic Liquid Crystal State in the High-Temperature Superconductor YBa${}_2$Cu${}_3$O${}_{6.45}$},
journal = {Science},
volume = {319},
number = {5863},
pages = {597-600},
year = {2008},
doi = {10.1126/science.1152309},
URL = {https://www.science.org/doi/abs/10.1126/science.1152309},
eprint ={}
}

@article{
doi:10.1126/science.1181083,
author = {T.-M. Chuang  and M. P. Allan  and Jinho Lee  and Yang Xie  and Ni Ni  and S. L. Bud’ko  and G. S. Boebinger  and P. C. Canfield  and J. C. Davis },
title = {Nematic Electronic Structure in the "Parent" State of the Iron-Based Superconductor $Ca(Fe_{1-x}Co_x)_2 As_2$},
journal = {Science},
volume = {327},
number = {5962},
pages = {181-184},
year = {2010},
doi = {10.1126/science.1181083},
URL = {https://www.science.org/doi/10.1126/science.1181083},
eprint = {}
}

@article{PhysRevB.59.8065,
  title = {Liquid-crystal phases of quantum Hall systems},
  author = {Fradkin, Eduardo and Kivelson, Steven A.},
  journal = {Phys. Rev. B},
  volume = {59},
  issue = {12},
  pages = {8065--8072},
  numpages = {0},
  year = {1999},
  month = {Mar},
  publisher = {American Physical Society},
  doi = {10.1103/PhysRevB.59.8065},
  url = {https://link.aps.org/doi/10.1103/PhysRevB.59.8065}
}

@article{Xia_2011,
   title={Evidence for a fractionally quantized Hall state with anisotropic longitudinal transport},
   volume={7},
   ISSN={1745-2481},
   url={http://dx.doi.org/10.1038/NPHYS2118},
   DOI={10.1038/nphys2118},
   number={11},
   journal={Nature Physics},
   publisher={Springer Science and Business Media LLC},
   author={Xia, Jing and Eisenstein, J. P. and Pfeiffer, L. N. and West, K. W.},
   year={2011},
   month=oct, pages={845–848} }

@article{Charmousis:2010zz,
    author = "Charmousis, Christos and Gouteraux, Blaise and Kim, Bom Soo and Kiritsis, Elias and Meyer, Rene",
    title = "{Effective Holographic Theories for low-temperature condensed matter systems}",
    eprint = "1005.4690",
    archivePrefix = "arXiv",
    primaryClass = "hep-th",
    reportNumber = "CCTP-2010-5, LPT-10-34",
    doi = "10.1007/JHEP11(2010)151",
    journal = "JHEP",
    volume = "11",
    pages = "151",
    year = "2010"
}

@article{Amoretti:2017axe,
    author = "Amoretti, Andrea and Are\'an, Daniel and Gout\'eraux, Blaise and Musso, Daniele",
    title = "{DC resistivity of quantum critical, charge density wave states from gauge-gravity duality}",
    eprint = "1712.07994",
    archivePrefix = "arXiv",
    primaryClass = "hep-th",
    reportNumber = "NORDITA-2017-139",
    doi = "10.1103/PhysRevLett.120.171603",
    journal = "Phys. Rev. Lett.",
    volume = "120",
    number = "17",
    pages = "171603",
    year = "2018"
}

@article{Nunez:2025gxq,
    author = "Nunez, Carlos and Roychowdhury, Dibakar",
    title = "{Time-like Entanglement Entropy: a top-down approach}",
    eprint = "2505.20388",
    archivePrefix = "arXiv",
    primaryClass = "hep-th",
    month = "5",
    year = "2025"
}

@article{Park:2018ebm,
    author = "Park, Chanyong and Ro, Daeho and Hun Lee, Jung",
    title = "{c-theorem of the entanglement entropy}",
    eprint = "1806.09072",
    archivePrefix = "arXiv",
    primaryClass = "hep-th",
    doi = "10.1007/JHEP11(2018)165",
    journal = "JHEP",
    volume = "11",
    pages = "165",
    year = "2018"
}

@article{Kolekar:2018chf,
    author = "Kolekar, Kedar S. and Narayan, K.",
    title = "{On AdS$_{2}$ holography from redux, renormalization group flows and c-functions}",
    eprint = "1810.12528",
    archivePrefix = "arXiv",
    primaryClass = "hep-th",
    doi = "10.1007/JHEP02(2019)039",
    journal = "JHEP",
    volume = "02",
    pages = "039",
    year = "2019"
}

@article{Giataganas:2025div,
    author = "Giataganas, Dimitrios",
    title = "{Holographic Timelike c-function}",
    eprint = "2505.20459",
    archivePrefix = "arXiv",
    primaryClass = "hep-th",
    month = "5",
    year = "2025"
}

@article{Mateos:2011ix,
    author = "Mateos, David and Trancanelli, Diego",
    title = "{The anisotropic N=4 super Yang-Mills plasma and its instabilities}",
    eprint = "1105.3472",
    archivePrefix = "arXiv",
    primaryClass = "hep-th",
    reportNumber = "ICCUB-11-142, MAD-TH-10-06",
    doi = "10.1103/PhysRevLett.107.101601",
    journal = "Phys. Rev. Lett.",
    volume = "107",
    pages = "101601",
    year = "2011"
}

@article{Mateos:2011tv,
    author = "Mateos, David and Trancanelli, Diego",
    title = "{Thermodynamics and Instabilities of a Strongly Coupled Anisotropic Plasma}",
    eprint = "1106.1637",
    archivePrefix = "arXiv",
    primaryClass = "hep-th",
    reportNumber = "ICCUB-11-145, MAD-TH-11-03",
    doi = "10.1007/JHEP07(2011)054",
    journal = "JHEP",
    volume = "07",
    pages = "054",
    year = "2011"
}

@article{Chu:2025sjv,
    author = "Chu, Chong-Sun and Parihar, Himanshu",
    title = "{Timelike entanglement entropy with gravitational anomalies}",
    eprint = "2504.19694",
    archivePrefix = "arXiv",
    primaryClass = "hep-th",
    month = "4",
    year = "2025"
}

@article{Freedman:1999gp,
    author = "Freedman, D. Z. and Gubser, S. S. and Pilch, K. and Warner, N. P.",
    title = "{Renormalization group flows from holography supersymmetry and a c theorem}",
    eprint = "hep-th/9904017",
    archivePrefix = "arXiv",
    reportNumber = "CERN-TH-99-86, HUTP-99-A015, USC-99-1, MIT-CTP-2846",
    doi = "10.4310/ATMP.1999.v3.n2.a7",
    journal = "Adv. Theor. Math. Phys.",
    volume = "3",
    pages = "363--417",
    year = "1999"
}

@article{Wilson:1974mb,
    author = "Wilson, Kenneth G.",
    title = "{The Renormalization Group: Critical Phenomena and the Kondo Problem}",
    reportNumber = "CLNS-296",
    doi = "10.1103/RevModPhys.47.773",
    journal = "Rev. Mod. Phys.",
    volume = "47",
    pages = "773",
    year = "1975"
}

@article{Girardello:1998pd,
    author = "Girardello, L. and Petrini, M. and Porrati, M. and Zaffaroni, A.",
    title = "{Novel local CFT and exact results on perturbations of N=4 superYang Mills from AdS dynamics}",
    eprint = "hep-th/9810126",
    archivePrefix = "arXiv",
    reportNumber = "CERN-TH-98-323, IFUM-633-FT, IMPERIAL-TP-98-99-4, NYU-TH-98-10-03",
    doi = "10.1088/1126-6708/1998/12/022",
    journal = "JHEP",
    volume = "12",
    pages = "022",
    year = "1998"
}

@article{Giataganas:2012zy,
    author = "Giataganas, Dimitrios",
    title = "{Probing strongly coupled anisotropic plasma}",
    eprint = "1202.4436",
    archivePrefix = "arXiv",
    primaryClass = "hep-th",
    reportNumber = "WITS-CTP-089",
    doi = "10.1007/JHEP07(2012)031",
    journal = "JHEP",
    volume = "07",
    pages = "031",
    year = "2012"
}

@article{Narayan:2022afv,
    author = "Narayan, K.",
    title = "{de Sitter space, extremal surfaces, and time entanglement}",
    eprint = "2210.12963",
    archivePrefix = "arXiv",
    primaryClass = "hep-th",
    doi = "10.1103/PhysRevD.107.126004",
    journal = "Phys. Rev. D",
    volume = "107",
    number = "12",
    pages = "126004",
    year = "2023"
}

@article{Jena:2024tly,
    author = "Jena, Siddhi Swarupa and Mahapatra, Subhash",
    title = "{A note on the holographic time-like entanglement entropy in Lifshitz theory}",
    eprint = "2410.00384",
    archivePrefix = "arXiv",
    primaryClass = "hep-th",
    doi = "10.1007/JHEP01(2025)055",
    journal = "JHEP",
    volume = "01",
    pages = "055",
    year = "2025"
}

@article{Grieninger:2023knz,
    author = "Grieninger, Sebastian and Ikeda, Kazuki and Kharzeev, Dmitri E.",
    title = "{Temporal entanglement entropy as a probe of renormalization group flow}",
    eprint = "2312.08534",
    archivePrefix = "arXiv",
    primaryClass = "hep-th",
    doi = "10.1007/JHEP05(2024)030",
    journal = "JHEP",
    volume = "05",
    pages = "030",
    year = "2024"
}

@article{Azeyanagi:2009pr,
    author = "Azeyanagi, Tatsuo and Li, Wei and Takayanagi, Tadashi",
    title = "{On String Theory Duals of Lifshitz-like Fixed Points}",
    eprint = "0905.0688",
    archivePrefix = "arXiv",
    primaryClass = "hep-th",
    reportNumber = "KUNS-2205, IPMU09-0055",
    doi = "10.1088/1126-6708/2009/06/084",
    journal = "JHEP",
    volume = "06",
    pages = "084",
    year = "2009"
}

@article{Afrasiar:2024lsi,
    author = "Afrasiar, Mir and Basak, Jaydeep Kumar and Giataganas, Dimitrios",
    title = "{Timelike entanglement entropy and phase transitions in non-conformal theories}",
    eprint = "2404.01393",
    archivePrefix = "arXiv",
    primaryClass = "hep-th",
    doi = "10.1007/JHEP07(2024)243",
    journal = "JHEP",
    volume = "07",
    pages = "243",
    year = "2024"
}

@article{Afrasiar:2024ldn,
    author = "Afrasiar, Mir and Basak, Jaydeep Kumar and Giataganas, Dimitrios",
    title = "{Holographic timelike entanglement entropy in non-relativistic theories}",
    eprint = "2411.18514",
    archivePrefix = "arXiv",
    primaryClass = "hep-th",
    doi = "10.1007/JHEP05(2025)205",
    journal = "JHEP",
    volume = "05",
    pages = "205",
    year = "2025"
}

@article{Zamolodchikov:1986gt,
    author = "Zamolodchikov, A. B.",
    title = "{Irreversibility of the Flux of the Renormalization Group in a 2D Field Theory}",
    journal = "JETP Lett.",
    volume = "43",
    pages = "730--732",
    year = "1986"
}

@article{Cardy:1988cwa,
    author = "Cardy, John L.",
    title = "{Is There a c Theorem in Four-Dimensions?}",
    doi = "10.1016/0370-2693(88)90054-8",
    journal = "Phys. Lett. B",
    volume = "215",
    pages = "749--752",
    year = "1988"
}

@article{Caputa:2024gve,
    author = "Caputa, Pawel and Chen, Bowen and Takayanagi, Tadashi and Tsuda, Takashi",
    title = "{Thermal pseudo-entropy}",
    eprint = "2411.08948",
    archivePrefix = "arXiv",
    primaryClass = "hep-th",
    reportNumber = "YITP-24-151",
    doi = "10.1007/JHEP01(2025)003",
    journal = "JHEP",
    volume = "01",
    pages = "003",
    year = "2025"
}

@article{Komargodski:2011vj,
    author = "Komargodski, Zohar and Schwimmer, Adam",
    title = "{On Renormalization Group Flows in Four Dimensions}",
    eprint = "1107.3987",
    archivePrefix = "arXiv",
    primaryClass = "hep-th",
    doi = "10.1007/JHEP12(2011)099",
    journal = "JHEP",
    volume = "12",
    pages = "099",
    year = "2011"
}

@article{Liu:2012eea,
    author = "Liu, Hong and Mezei, Mark",
    title = "{A Refinement of entanglement entropy and the number of degrees of freedom}",
    eprint = "1202.2070",
    archivePrefix = "arXiv",
    primaryClass = "hep-th",
    reportNumber = "MIT-CTP-4336",
    doi = "10.1007/JHEP04(2013)162",
    journal = "JHEP",
    volume = "04",
    pages = "162",
    year = "2013"
}

@article{Ryu:2006bv,
    author = "Ryu, Shinsei and Takayanagi, Tadashi",
    title = "{Holographic derivation of entanglement entropy from AdS/CFT}",
    eprint = "hep-th/0603001",
    archivePrefix = "arXiv",
    reportNumber = "NSF-KITP-06-11",
    doi = "10.1103/PhysRevLett.96.181602",
    journal = "Phys. Rev. Lett.",
    volume = "96",
    pages = "181602",
    year = "2006"
}

@article{Giataganas:2017koz,
	author = {Giataganas, Dimitrios and G\"ursoy, Umut and Pedraza, Juan F.},
	title = "{Strongly-coupled anisotropic gauge theories and holography}",
	eprint = "1708.05691",
	archivePrefix = "arXiv",
	primaryClass = "hep-th",
	reportNumber = "NCTS-TH/1712, NCTS-TH-1712",
	doi = "10.1103/PhysRevLett.121.121601",
	journal = "Phys. Rev. Lett.",
	volume = "121",
	number = "12",
	pages = "121601",
	year = "2018"
}

@article{Chu:2019uoh,
	author = "Chu, Chong-Sun and Giataganas, Dimitrios",
	title = "{$c$-Theorem for Anisotropic RG Flows from Holographic Entanglement Entropy}",
	eprint = "1906.09620",
	archivePrefix = "arXiv",
	primaryClass = "hep-th",
	reportNumber = "NCTS-TH/1904",
	doi = "10.1103/PhysRevD.101.046007",
	journal = "Phys. Rev. D",
	volume = "101",
	number = "4",
	pages = "046007",
	year = "2020"
}

@article{Myers:2010tj,
	author = "Myers, Robert C. and Sinha, Aninda",
	title = "{Holographic c-theorems in arbitrary dimensions}",
	eprint = "1011.5819",
	archivePrefix = "arXiv",
	primaryClass = "hep-th",
	doi = "10.1007/JHEP01(2011)125",
	journal = "JHEP",
	volume = "01",
	pages = "125",
	year = "2011"
}

@article{Casini:2011kv,
    author = "Casini, Horacio and Huerta, Marina and Myers, Robert C.",
    title = "{Towards a derivation of holographic entanglement entropy}",
    eprint = "1102.0440",
    archivePrefix = "arXiv",
    primaryClass = "hep-th",
    doi = "10.1007/JHEP05(2011)036",
    journal = "JHEP",
    volume = "05",
    pages = "036",
    year = "2011"
}

@article{Hung:2011ta,
    author = "Hung, Ling-Yan and Myers, Robert C. and Smolkin, Michael",
    title = "{Some Calculable Contributions to Holographic Entanglement Entropy}",
    eprint = "1105.6055",
    archivePrefix = "arXiv",
    primaryClass = "hep-th",
    doi = "10.1007/JHEP08(2011)039",
    journal = "JHEP",
    volume = "08",
    pages = "039",
    year = "2011"
}

@article{Cremonini:2013ipa,
    author = "Cremonini, Sera and Dong, Xi",
    title = "{Constraints on renormalization group flows from holographic entanglement entropy}",
    eprint = "1311.3307",
    archivePrefix = "arXiv",
    primaryClass = "hep-th",
    reportNumber = "SU-ITP-13-23",
    doi = "10.1103/PhysRevD.89.065041",
    journal = "Phys. Rev. D",
    volume = "89",
    number = "6",
    pages = "065041",
    year = "2014"
}

@article{Swingle:2013zla,
    author = "Swingle, Brian",
    title = "{Entanglement does not generally decrease under renormalization}",
    eprint = "1307.8117",
    archivePrefix = "arXiv",
    primaryClass = "cond-mat.stat-mech",
    doi = "10.1088/1742-5468/2014/10/P10041",
    journal = "J. Stat. Mech.",
    volume = "1410",
    number = "10",
    pages = "P10041",
    year = "2014"
}

@article{Ryu:2006ef,
	author = "Ryu, Shinsei and Takayanagi, Tadashi",
	title = "{Aspects of Holographic Entanglement Entropy}",
	eprint = "hep-th/0605073",
	archivePrefix = "arXiv",
	reportNumber = "NSF-KITP-06-31, KUNS-2021",
	doi = "10.1088/1126-6708/2006/08/045",
	journal = "JHEP",
	volume = "08",
	pages = "045",
	year = "2006"
}

@article{Doi:2022iyj,
	author = "Doi, Kazuki and Harper, Jonathan and Mollabashi, Ali and Takayanagi, Tadashi and Taki, Yusuke",
	title = "{Pseudoentropy in dS/CFT and Timelike Entanglement Entropy}",
	eprint = "2210.09457",
	archivePrefix = "arXiv",
	primaryClass = "hep-th",
	reportNumber = "YITP-22-121, IMPU22-0052",
	doi = "10.1103/PhysRevLett.130.031601",
	journal = "Phys. Rev. Lett.",
	volume = "130",
	number = "3",
	pages = "031601",
	year = "2023"
}

@article{Doi:2023zaf,
author = "Doi, Kazuki and Harper, Jonathan and Mollabashi, Ali and Takayanagi, Tadashi and Taki, Yusuke",
title = "{Timelike entanglement entropy}",
eprint = "2302.11695",
archivePrefix = "arXiv",
primaryClass = "hep-th",
reportNumber = "YITP-23-22",
doi = "10.1007/JHEP05(2023)052",
journal = "JHEP",
volume = "05",
pages = "052",
year = "2023"
}

@article{Basak:2023otu,
    author = "Basak, Jaydeep Kumar and Chakraborty, Adrita and Chu, Chong-Sun and Giataganas, Dimitrios and Parihar, Himanshu",
    title = "{Massless Lifshitz field theory for arbitrary z}",
    eprint = "2312.16284",
    archivePrefix = "arXiv",
    primaryClass = "hep-th",
    doi = "10.1007/JHEP05(2024)284",
    journal = "JHEP",
    volume = "05",
    pages = "284",
    year = "2024"
}

@article{Chu:2023zah, 
    author = "Chu, Chong-Sun and Parihar, Himanshu",
    title = "{Time-like entanglement entropy in AdS/BCFT}",
    eprint = "2304.10907",
    archivePrefix = "arXiv",
    primaryClass = "hep-th",
    doi = "10.1007/JHEP06(2023)173",
    journal = "JHEP",
    volume = "06",
    pages = "173",
    year = "2023"
}

@article{Narayan:2023zen,
    author = "Narayan, K.",
    title = "{Further remarks on de Sitter space, extremal surfaces, and time entanglement}",
    eprint = "2310.00320",
    archivePrefix = "arXiv",
    primaryClass = "hep-th",
    doi = "10.1103/PhysRevD.109.086009",
    journal = "Phys. Rev. D",
    volume = "109",
    number = "8",
    pages = "086009",
    year = "2024"
}

@article{Hoyos:2021vhl,
    author = "Hoyos, Carlos and Jokela, Niko and Pen\'\i{}n, Jos\'e Manuel and Ramallo, Alfonso V. and Tarr\'\i{}o, Javier",
    title = "{Risking your NEC}",
    eprint = "2104.11749",
    archivePrefix = "arXiv",
    primaryClass = "hep-th",
    reportNumber = "HIP-2021-16/TH",
    doi = "10.1007/JHEP10(2021)112",
    journal = "JHEP",
    volume = "10",
    pages = "112",
    year = "2021"
}

@article{Baggioli:2020cld,
    author = "Baggioli, Matteo and Giataganas, Dimitrios",
    title = "{Detecting Topological Quantum Phase Transitions via the c-Function}",
    eprint = "2007.07273",
    archivePrefix = "arXiv",
    primaryClass = "hep-th",
    reportNumber = "IFT-UAM/CSIC-20-96",
    doi = "10.1103/PhysRevD.103.026009",
    journal = "Phys. Rev. D",
    volume = "103",
    number = "2",
    pages = "026009",
    year = "2021"
}

@article{Casini:2006es,
    author = "Casini, H. and Huerta, M.",
    editor = "Sola, Joan",
    title = "{A c-theorem for the entanglement entropy}",
    eprint = "cond-mat/0610375",
    archivePrefix = "arXiv",
    doi = "10.1088/1751-8113/40/25/S57",
    journal = "J. Phys. A",
    volume = "40",
    pages = "7031--7036",
    year = "2007"
}

@article{Myers:2012ed,
    author = "Myers, Robert C. and Singh, Ajay",
    title = "{Comments on Holographic Entanglement Entropy and RG Flows}",
    eprint = "1202.2068",
    archivePrefix = "arXiv",
    primaryClass = "hep-th",
    doi = "10.1007/JHEP04(2012)122",
    journal = "JHEP",
    volume = "04",
    pages = "122",
    year = "2012"
}

@article{Myers:2010xs,
    author = "Myers, Robert C. and Sinha, Aninda",
    title = "{Seeing a c-theorem with holography}",
    eprint = "1006.1263",
    archivePrefix = "arXiv",
    primaryClass = "hep-th",
    doi = "10.1103/PhysRevD.82.046006",
    journal = "Phys. Rev. D",
    volume = "82",
    pages = "046006",
    year = "2010"
}
\end{document}